\documentclass[printer]{aa}  
\usepackage{graphicx}
\usepackage{amsmath}
\usepackage{amssymb}
\usepackage{color}
\usepackage[]{hyperref}

\usepackage{txfonts}
\def \cgssb {{\rm\,erg\,s^{-1}\,cm^{-2}\,arcsec^{-2}}}
\newcommand\Tstrut{\rule{0pt}{2.6ex}}         % = `top' strut
\newcommand\Bstrut{\rule[-1.4ex]{0pt}{0pt}}   % = `bottom' strut

\begin{document}

   \title{Discovery of intergalactic bridges connecting two faint $z\sim3$ quasars}

   \author{Fabrizio Arrigoni Battaia\inst{1,2}
   \and
   Aura Obreja\inst{3}
   \and
   J.~Xavier Prochaska\inst{4}
   \and
   Joseph~F. Hennawi\inst{5,6}
   \and
   Hadi Rahmani\inst{7}
   \and
   Eduardo~Ba{\~n}ados\inst{6}
   \and
   Emanuele~P. Farina\inst{6,1}
   \and
   Zheng~Cai\inst{4}
   \and
   Allison Man\inst{8}
          }

   \institute{Max-Planck-Institut f\"ur Astrophysik, Karl-Schwarzschild-Str 1, D-85748 Garching bei M\"unchen, Germany\\
              \email{arrigoni@mpa-garching.mpg.de}
         \and
	 European Southern Observatory, Karl-Schwarzschild-Str. 2, D-85748 Garching bei M\"unchen, Germany
	 \and
         University Observatory Munich, Scheinerstra{\ss}e 1, D-81679 Munich, Germany
         \and
         UCO/Lick Observatory, University of California, 1156 High Street, Santa Cruz, CA 95064, USA
         \and
         Department of Physics, Broida Hall, University of California, Santa Barbara, CA 93106-9530, USA
         \and
         Max-Planck-Institut f\"ur Astronomie, K\"onigstuhl 17, D-69117 Heidelberg, Germany
         \and
         GEPI, Observatoire de Paris, PSL Universit\'e, CNRS, 5 Place Jules Janssen, 92190 Meudon, France
         \and
         Dunlap Institute for Astronomy \& Astrophysics, 50 St. George Street, Toronto, ON M5S 3H4, Canada
             }

   \date{} 
 
  \abstract
  {We use the Multi-Unit Spectroscopic Explore (MUSE) on the Very Large Telescope to conduct a survey of $z\sim3$ physical quasar pairs at close separation ($<30\arcsec$) with a fast observation strategy (45 minutes on source).
  Our aim is twofold: (i) explore the Ly$\alpha$ glow around the faint-end of the quasar population; and (ii) take advantage of the combined illumination of a quasar pair to unveil large-scale intergalactic structures
  (if any) extending between the two quasars. In this work we report the results for the quasar pair SDSS~J113502.03-022110.9 - SDSS~J113502.50-022120.1 ($z=3.020, 3.008$; $i=21.84,22.15$), separated by $11.6\arcsec$ (or 89 projected kpc). 
  MUSE reveals filamentary Ly$\alpha$ structures extending between the two quasars with an average surface brightness of SB$_{\rm Ly\alpha}=1.8\times10^{-18}\cgssb$. Photoionization models of the constraints in the Ly$\alpha$, \ion{He}{II}$\lambda$1640, and \ion{C}{iv}$\lambda$1548 line emissions show that the emitting structures are intergalactic bridges with an extent between $\sim89$ (the quasars' projected distance) and up to $\sim600$~kpc. 
 Our models rule out the possibility that the structure extends for $\sim 2.9$~Mpc, i.e., the separation inferred from the uncertain systemic redshift difference of the quasars if the difference was only due to the Hubble flow. 
  At the current spatial resolution and surface brightness limit, the average projected width of an individual bridge is $\sim35$~kpc. We also detect a strong absorption in \ion{H}{i}, \ion{N}{v}, and \ion{C}{iv} along the background sight-line at higher $z$, which we interpret as due to at least two components of cool ($T\sim10^4$~K), metal enriched ($Z>0.3\, Z_{\odot}$), and relatively ionized circumgalactic or intergalactic gas surrounding the quasar pair. Two additional \ion{H}{i} absorbers are detected along both quasar sight-lines at $\sim -900$~and~$-2800$~km~s$^{-1}$ from the system, with the latter having associated \ion{C}{iv} absorption only along the foreground quasar sight-line. The absence of galaxies in the MUSE field of view at the redshifts of these two absorbers suggests that they trace large-scale structures or expanding shells in front of the quasar pair. Combining longer exposures and higher spectral resolution when targeting similar quasar pairs has the potential to  firmly constrain the physical properties of gas in large-scale intergalactic structures.} 


   \maketitle
%
%--------------------------------------------------------------------

\section{Introduction}

The current paradigm of structure formation predicts the presence of gaseous filaments connecting
galaxies (e.g., \citealt{White1987,Bond1996}), ultimately forming an intricate web known as the intergalactic medium (IGM; \citealt{meiksin09}).
Given the expected low densities for such gas ($n_{\rm H}\lesssim0.01$~cm$^{-2}$) and the budget of ionizing photons in the ultraviolet background (UVB; e.g., \citealt{hm12}), 
the direct observation of the IGM is predicted to be very challenging (surface brightness in Ly$\alpha$ emission predicted to be 
SB$_{\rm Ly\alpha}\sim10^{-19}-10^{-20}\cgssb$; \citealt{GW96,Bertone2012,Witstok2019}). Indeed, a direct detection of the IGM appears to 
be so far elusive even with current facilities (e.g., \citealt{Gallego2018,Wisotzki2018}), e.g., the Multi Unit Spectroscopic Explorer (MUSE; \citealt{Bacon2010}) and the Keck Cosmic Web Imager (KCWI; \citealt{Morrissey2012}).

It was however noticed early on that quasars could act as flashlights, possibly photoionizing the surrounding medium out to large distances. 
The ionized gas would then recombine emitting as main product Hydrogen Lyman-$\alpha$ (Ly$\alpha$) photons in copious amounts (e.g. \citealt{Rees1988, hr01}). 
This boosted glow (SB$_{\rm Ly\alpha}>10^{-19}\cgssb$) should then possibly be within reach of state-of-the-art instruments (\citealt{Cantalupo2005,kollmeier10}).

Following this idea, several works aimed for the Ly$\alpha$ emission from halos (nowadays known as the circumgalactic-medium, CGM; \citealt{TPW_2017}) out to intergalactic scales around individual high-$z$ quasars to constrain the physical properties of the diffuse gas phases (e.g., \citealt{HuCowie1987,heckman91a,Moller2000b,Weidinger04,Weidinger05,Christensen2006,hpk+09,cantalupo14,martin14a,hennawi+15,fab+16,Farina2017}).
At $z\sim3$, observations can now easily ($\sim 1$~hour on source) uncover the emission within 50 projected kpc, and reach an average maximum distance of $\sim 80$~projected kpc from the targeted quasar (\citealt{Borisova2016,FAB2019}).
This Ly$\alpha$ emission usually shows SB$_{\rm Ly\alpha} \sim4\times10^{-18}\cgssb$ with relatively quiescent line widths $\sigma_{\rm Ly\alpha}<400$~km~s$^{-1}$ (\citealt{FAB2019}), which are intriguingly similar to the velocity dispersion expected for 
halos hosting quasars at these redshifts ($\sigma=250$~km~s$^{-1}$ ; M$_{\rm DM}\sim10^{12.5}$~M$_{\odot}$; e.g., \citealt{white12}).
The uncertainties in the determination of the quasar systemic redshift, together with the haze possibly introduced by the Ly$\alpha$ radiative transfer still hamper, in most of the cases, a secure interpretation of the gas kinematics and/or configuration of the system as traced by the extended Ly$\alpha$ emission (\citealt{FAB2019}).
Notwithstanding these open issues, the Ly$\alpha$ nebulae are usually interpreted as physically associated with the targeted quasar, and tracing either the gravitational motions due to structure assembly (\citealt{Weidinger04,Weidinger05,FAB2018}), or the violent feedback of the central engine (\citealt{Cai2016}). 
Alternative interpretations explain the Ly$\alpha$ emission as not strictly associated with the targeted quasars, but due to structures along our line-of-sight to the quasar, like portions of the CGM of massive halos in the Hubble flow aligned along our line-of-sight (\citealt{Cantalupo2019}) or proto-galactic disks (\citealt{Martin2019}) illuminated by the quasar. 

Thanks to the aforementioned effort in the detection of the CGM around high-$z$ quasars, it starts to become evident that even around individual quasars it is extremely 
hard to detect diffuse emission at intergalactic distances ($>100$~kpc) unless additional companions (mostly active) are present in close proximity (\citealt{hennawi+15,FAB2019,FAB2018}), or 
much more sensitive observations are conducted. Dense environments seem to supply additional cool dense gas necessary for the detection of Ly$\alpha$ signal on very large scales (\citealt{hennawi+15,Cai2016,FAB2018}).
Further, the unification model for active galactic nuclei (AGN; e.g., \citealt{Anton93}) and evidences of anisotropic ionizing emission from high-redshift quasars (e.g., \citealt{QPQ2}) hint to the existence of shadowed regions around individual quasars. 
The presence of multiple quasars within the same structure thus increases the probability of large-scale gas to be illuminated by hard ionizing photons.
For these reasons, scientific teams have started to change their approach in unveiling IGM emission, passing from the targeting of individual quasars to short (\citealt{Cai2018}) or extremely long integrations ($>40$~hours; \citealt{Lusso2019}) of multiple high-redshift quasars, or overdensities hosting quasars (\citealt{Cai2016}).

Here, we report on our effort within this framework.
In particular, in 2015 we designed a survey of $z\sim3$ physically associated quasar pairs using the MUSE instrument on the Very Large Telescope (VLT) of the 
European Southern Observatory (ESO). We now have the first data of these observations, and 
here we present the results of the study of the first target.
Our work is structured as follows.
In Section~\ref{sec:selection} we explain how we selected the quasar pairs in our survey. Section~\ref{sec:obs} presents the observations and data reduction for the quasar pair here studied.
We highlight our results for the extended Ly$\alpha$ emission and for the detected absorptions in Section~\ref{sec:results}. Section~\ref{sec:pow_Lya} discusses the possible scenarios for the powering of the extended
Ly$\alpha$ emission, while Section~\ref{sec:model_abs} presents the results of the modeling of the absorbers. Finally, 
we summarize our findings in Section~\ref{sec:summary}. 

We adopt the cosmological parameters $H_{0}=70$~km~s$^{-1}$~Mpc$^{-1}$, $\Omega_{M}=0.3$, 
and $\Omega_{\Lambda}=0.7$, and therefore 1\arcsec corresponds to about 7.7~kpc at $z=3.020$ ($z_{\rm QSO2}$; details in Section~\ref{sec:selection}). All magnitudes are in the AB system (\citealt{Oke1974}), and all
distances are proper. 

\begin{figure*}
\centering
\includegraphics[width=0.9\textwidth]{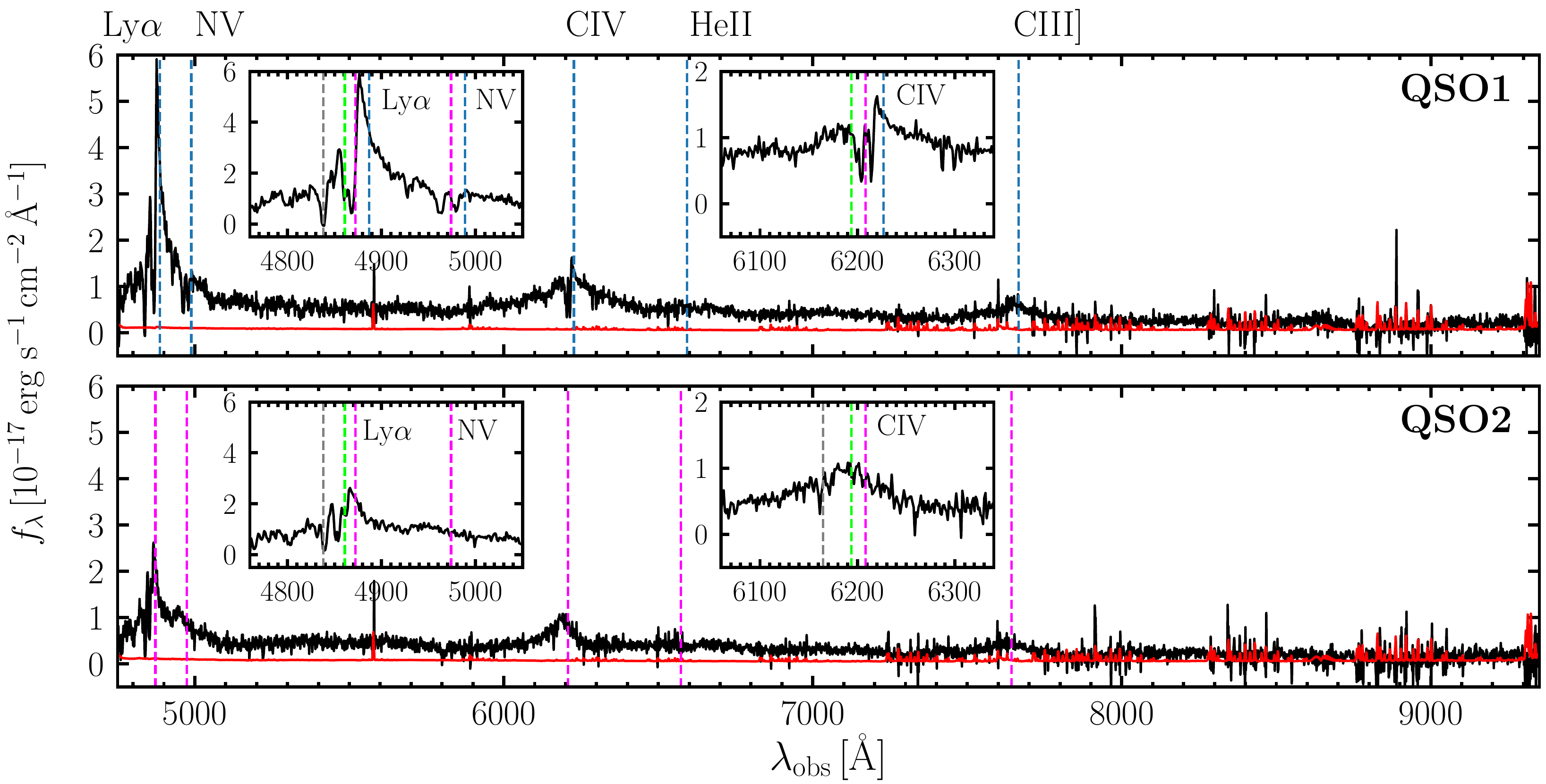}
\caption{1D spectra (black) for the two quasars of the pair, QSO1 (top) and QSO2 (bottom), 
as extracted from the MUSE data, using a circular aperture with radius 2\arcsec\ .
The red spectra indicate the error vectors.  The vertical dashed blue (magenta) lines indicate the position of important line emissions at the systemic redshift of QSO1 (QSO2).
For both objects, we show in the inset plots a zoomed version of the spectrum at the location of the Ly$\alpha$ and \ion{C}{iv} lines to highlight 
the presence of interesting absorptions. The vertical dashed gray and green lines indicate the location of an H Ly$\alpha$ absorptions present within both QSO1 and QSO2 spectra, 
and \ion{C}{iv} absorption along the QSO2 sight-line. The fit to these lines is shown in Section~\ref{sec:abs} and Figures~\ref{fig:ABSQSO1} and \ref{fig:ABSQSO2}.
Residuals due to frequent sky lines are evident at wavelengths $>7000$~\AA.}
\label{QSOpair_spectra}
\end{figure*}

%--------------------------------------------------------------------
\section{Selection of the quasar pairs}
\label{sec:selection}

The quasar pairs to be observed in our program have been selected from the twelfth data release of the Sloan Digital Sky Survey (SDSS) quasar catalog (\citealt{Paris2017}) using the following criteria:
\begin{itemize}
\item be at the lowest redshift for which the Ly$\alpha$ emission is detectable with MUSE, i.e. $3.0 \lesssim z < 3.9$, where sky lines are not dominant;
\item have a difference in redshift of $\Delta z \leq 0.03$ (corresponding to $\leq 2000$~km~s$^{-1}$). This small difference in redshift should ensure 
that the two quasars are physically associated (e.g., \citealt{Hennawi2006,QPQ1});
\item have a projected separation $\leq 0.5$~arcmin, so that both quasars sit within the MUSE field-of-view.
\item be well visible from VLT/ESO, i.e. Dec~$< 27$ degrees.
\end{itemize}

Importantly, in our selection we did not impose any constraint on the current luminosity of the quasars in the 
pair\footnote{The SDSS quasar catalog of \citet{Paris2017} includes quasars with $i$-mag down to 25.}. 
Our effort is thus complementary to the approach of \cite{Cai2018}, who selected pairs with at least one bright quasar ($g < 19$) visible from the Palomar and Keck sites.
The aforementioned criteria resulted in the selection of a total of 17 quasar pairs visible during the ESO semester P100. We however obtained data only on 7 of these targets due to weather conditions.

In this work we focus on the quasar pair SDSS~J113502.03-022110.9 - SDSS~J113502.50-022120.1 (henceforth QSO1 - QSO2), separated by $11.6\arcsec$ (or 89 kpc) 
and whose properties are summarized in Table~\ref{QSOpair}.
In particular, we double checked the redshift estimate of the SDSS catalog by using the known relation for the blueshift of the \ion{C}{iii}] line emission (\citealt{Shen2016}), and obtained consistent redshifts within the uncertainties\footnote{We remind that the work by \citet{Shen2016} do not cover quasars as faint as the one targeted in this work. For this case, the extrapolation of their relation to lower luminosities seems to give consistent results to the SDSS redshift pipeline.}. For completeness, we list in the table both redshifts, but we use the SDSS redshifts in the reminder of this work. 
The current redshift estimates place the two quasars at $\Delta v= 896 \pm 316 $~km~s$^{-1}$, which corresponds to a distance of $2.9\pm0.9$~physical Mpc if all the velocity difference is due to the Hubble flow. However, if we look at their spectra (e.g., Figure~\ref{QSOpair_spectra}), the observed Ly$\alpha$ emission peaks are only separated by $\Delta v= 598 \pm 98 $~km~s$^{-1}$ (or $1.9\pm0.3$~physical Mpc)\footnote{As the spectrum of QSO1 presents a strong absorber close to its Ly$\alpha$ line (Section~\ref{sec:results}), the velocity shift between the intrinsic Ly$\alpha$ peaks of the two quasars could be smaller.}.
These two quasars are $\approx3.8$ mag fainter than the average $M_{1450}=-27.12$ of the QSO MUSEUM sample of \citet{FAB2019}, and sit in a portion of sky with low galactic extinction $A_V=0.08$~mag (\citealt{sfd98})\footnote{The galactic extinction is reported to be in agreement within uncertainties ($A_V=0.07$~mag) when using \citet{Schlafly2011}.}.

\begin{table*}
\begin{center}
\caption{The targeted quasar pair}
\scalebox{1}{
\footnotesize
\setlength\tabcolsep{4pt}
\begin{tabular}{lcccccccc}
\hline
\hline
ID		&	SDSS name 		&	R.A.	&	Dec	&	$z_{\rm systemic}^{\rm a}$ 	        & $z_{\rm peakLy\alpha}^{\rm b}$  &   $i^{\rm c}$    &    $M_{1450}$   & Radio Flux$^{\rm d}$  \\
                &               		&     (J2000)   &     (J2000)   & 	SDSS (this work)		   	&			          & 	             & 	      	       & 	  (mJy/beam) \\ 
\hline
QSO1		&  SDSS~J113502.03-022110.9	&  11:35:02.030	& -02:21:10.93	&	$3.020\pm0.001$ ($3.019\pm0.003$)	&  3.011		          &  $21.84\pm0.02$  &    -23.44       & 	  <0.44       \\
QSO2		&  SDSS~J113502.50-022120.1	&  11:35:02.500	& -02:21:20.14	&	$3.008\pm0.001$ ($3.008\pm0.003$)	&  3.003		          &  $22.15\pm0.02$  &    -23.12       & 	  <0.44       \\ 
\hline
\hline
\end{tabular}
}
\flushleft{\scriptsize $^a$ Quasar systemic redshift from the SDSS catalog and, in brackets, from the peak of the \ion{C}{iii}] complex (i.e. \ion{C}{iii}] is a doublet $1906.7,1908.7$~\AA; and
\ion{Si}{iii}]$\lambda1892$ could be blended), after correcting for the expected shift (\citealt{Shen2016}). The intrinsic uncertainty on this 
correction is $\sim233$~km~s$^{-1}$ and dominates the error budget ($\Delta z \approx 0.003$).\\ 
$^b$ Redshift corresponding to the peak of the Ly$\alpha$ emission in the observed spectrum of 
each quasar.\\ 
$^c$ $i$ magnitude extracted from our data using the SDSS filter transmission curve and a circular aperture with a radius of $2\arcsec$. The SDSS $i$ magnitudes for the two quasars are $i_{\rm QSO1}=21.50\pm0.08$ and $i_{\rm QSO2}=22.05\pm0.12$.\\ 
$^d$ $3\times$ rms at $1.4$~GHz from the Very Large Array survey: Faint Images of the Radio Sky at Twenty-Centimeters (VLA FIRST; \citealt{Becker1994}).} 
\label{QSOpair}
\end{center}
\end{table*}

%log10(Lbol)
%45.5
%45.3
%Following Runnoe et al. 2012

%--------------------------------------------------------------------
\section{Observations and data reduction}
\label{sec:obs}

The quasar pair QSO1 - QSO2 was observed during UT 19 of February
2018 with clear sky conditions for the program 0100.A-0045(A) with the MUSE instrument
on the VLT 8.2m telescope YEPUN (UT4).
The observations consisted of 3 exposures of 880~s each, rotated with respect to each other by 90 degrees, 
and with a dither of a few arcseconds between them.
The data have been acquired with the nominal spectral range, and thus cover the wavelengths 4750.2 - 9350.2 \AA.

\begin{figure}
\centering
\includegraphics[width=0.9\columnwidth]{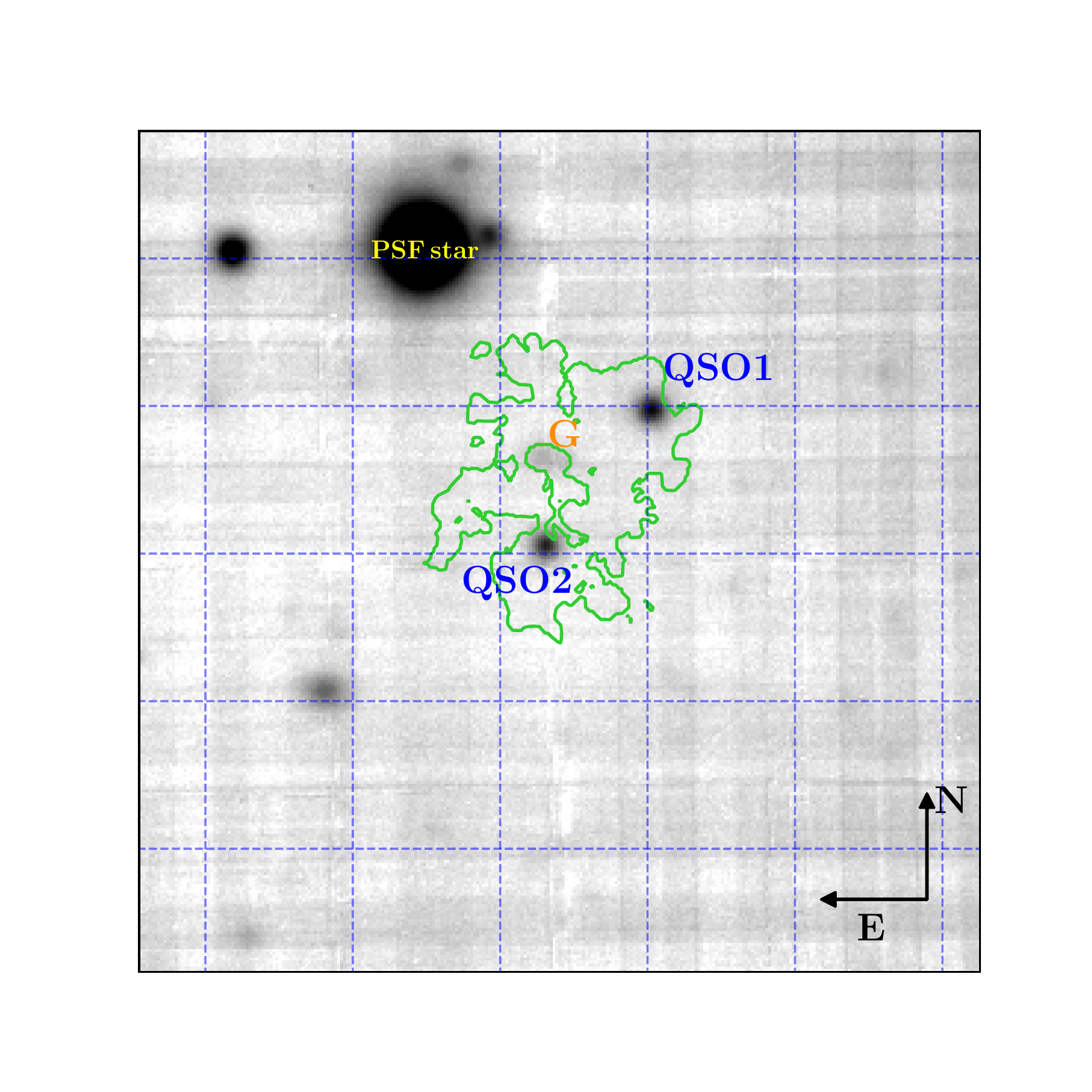}
\caption{White-light image of the observed 57\arcsec~$\times$~57\arcsec field of view.
We indicate the position of QSO1, QSO2, the star 2MASS J11350307-0220597
used to compute the point spread function of our data (Appendix~\ref{app:PSF}), and an interloper galaxy ``G'' (tentatively at $z=0.457\pm0.001$; Appendix~\ref{sec:int_gal}). Additionally, we indicate the $2\sigma$ isophote for the extended Ly$\alpha$ emission discovered around the quasar pair (Figure~\ref{fig:SB_Vel_map}).}
\label{whiteLight}
\end{figure}

The data were reduced using the MUSE pipeline recipes 
v2.2 (\citealt{Weilbacher2014}). In particular, each of the individual
exposures have been bias-subtracted, flat-fielded, twilight and illumination corrected, sky-subtracted, and wavelength
calibrated using the calibration data taken closest in time to the science frames. 
The flux calibration of each exposure has been obtained using a spectrophotometric standard star
observed during the same night of the science observing block. 
The individual exposures were then combined into a
single data cube. While we apply an initial sky subtraction using the MUSE pipeline, skyline residuals are further suppressed using
the software ZAP\footnote{\url{https://zap.readthedocs.io/en/latest/}} (\citealt{Soto2016})\footnote{We perform this step to search the data at large wavelength. At the location of the Ly$\alpha$ line there are no strong sky lines.}.
The seeing of the final combined data is measured from the star 2MASS J11350307-0220597 (see Appendix~\ref{app:PSF}), resulting in a
Moffat function with $\beta = 2.5$ and FWHM$= 1.66\arcsec$. 
The coadded spectrum of QSO1 and QSO2 as extracted from the final MUSE datacube are shown in Figure~\ref{QSOpair_spectra}.
Further we present in Figure~\ref{whiteLight} the white-light image of the observations field of view, obtained by collapsing the final
MUSE datacube.

The MUSE pipeline produces a variance datacube which is known to underestimate the true noise because it
neglects the correlated noise introduced during the resampling of the datacubes (e.g. \citealt{Borisova2016}). To
correct for this effect, we rescaled the variance cube layer by layer so that the average of 
each layer in the variance cube matches the average variance computed from each science layer after masking objects.

The final MUSE datacube has a $2\sigma$ surface brightness limit of SB$_{\rm Ly\alpha}=7\times10^{-19}\cgssb$ (in 1 arcsec$^2$ aperture) 
in a single channel (1.25\AA) at $\approx 4872$~\AA\ (Ly$\alpha$ at the redshift of QSO2). Given the stability of MUSE, further smoothing can allow us to push this sensitivity to lower levels (Section~\ref{sec:PSFsub}).

%--------------------------------------------------------------------
\subsection{Point spread function subtraction and extraction of the Ly$\alpha$ emission}
\label{sec:PSFsub}

Quasars easily outshine the radiation produced by the surrounding gas distribution, and their emission is smeared out to larger scales due to the seeing. 
For these reasons, the study of large-scale gas around quasars requires the subtraction of the unresolved 
quasar emission, as characterized by the point-spread-function (PSF) of the observations.
This problem has been empirically tackled in the literature by subtracting a wavelength-dependent PSF constructed from the data themselves
in several ways (e.g., \citealt{Moller2000a, Christensen2006, Husemann2014, Borisova2016}).
Given the presence of a bright star within the field-of-view of our observations, we were able to reconstruct the 
wavelength dependent PSF layer by layer at high signal to noise (S/N), as described in detail in the Appendix~\ref{app:PSF}.
The reconstructed layer-by-layer PSF was then subtracted at each quasar position out to a 5\arcsec\ radius after matching the quasar emission within a 1~arcsec$^2$ circle.
Before proceeding with the extraction of the Ly$\alpha$ signal, we removed all the continuum-detected sources from the datacube using the 
median-filtering routine {\tt contsubfits} in ZAP (\citealt{Soto2016}). 
We masked the location of very bright or extended continuum objects, like the star 2MASS J11350307-0220597 (Appendix~\ref{app:PSF})  and an interloper galaxy ``G'' tentatively at $z=0.457\pm0.001$ (more details in Appendix~\ref{sec:int_gal}) 
to avoid contamination from residuals.

\begin{figure*}
\centering
\includegraphics[width=1.0\textwidth]{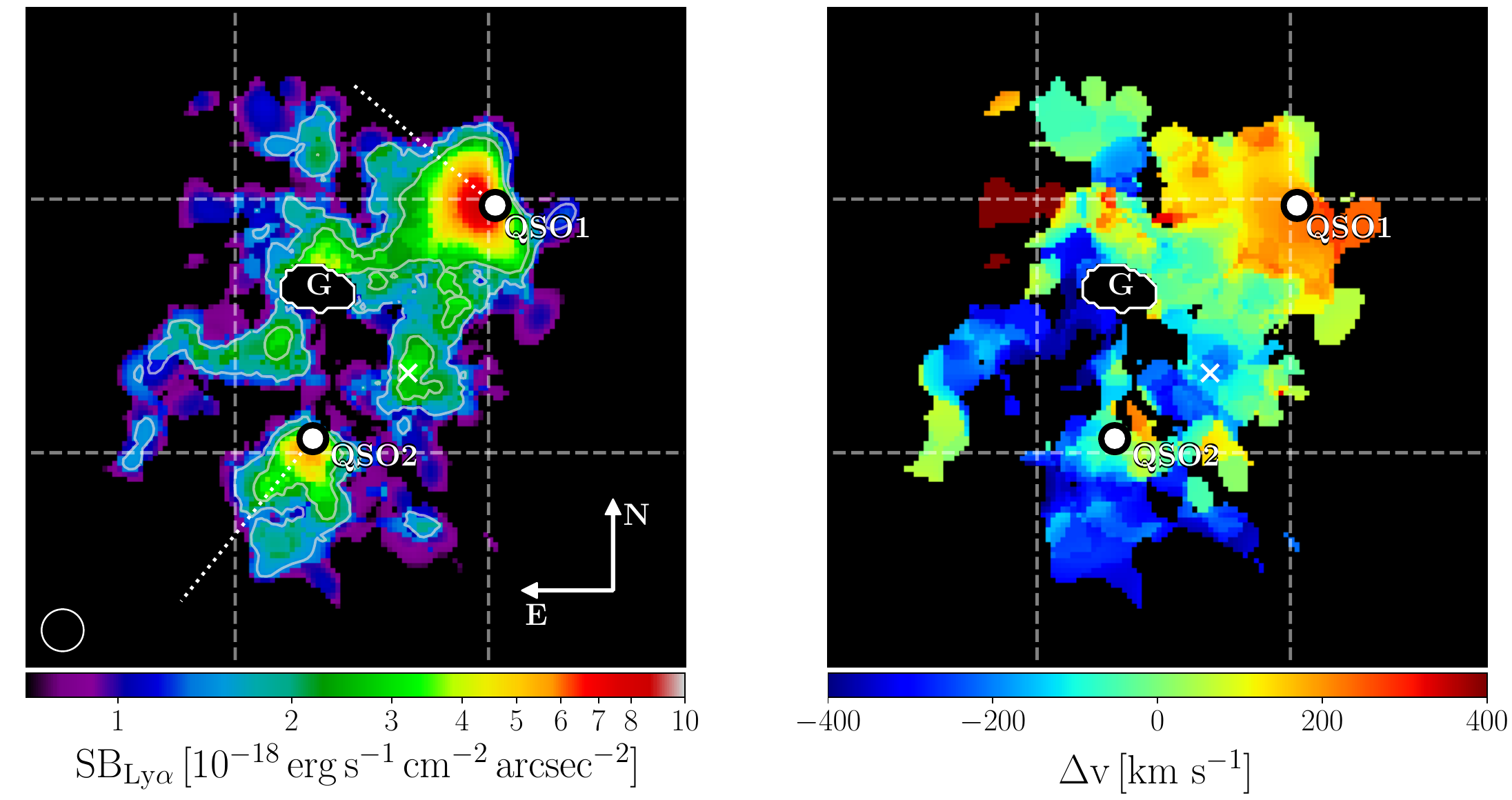}
\caption{The Ly$\alpha$ emission around the quasar pair in a field of view of about 200~kpc~$\times$~200~kpc (or 26\arcsec~$\times$~26\arcsec). 
Left: ``optimally extracted'' Ly$\alpha$ surface brightness map obtained after subtraction of the quasars point-spread-function (PSF) 
and continuum in the MUSE datacube (details in Section~\ref{sec:PSFsub}). 
To highlight the significance of the detected emission, we indicate the contours for S/N~$=3$ and $4$.
This image reveals Ly$\alpha$ bridges extending between the quasar pair. Right: flux-weighted velocity-shift map 
with respect to the systemic redshift of QSO2 obtained from the 
first order moment of the flux distribution. 
A velocity gradient between QSO1 and the portion of the nebula southern than QSO2 is evident 
(Figure~\ref{fig:Pseudoslit}). In both panels we indicate the position of the 
quasars QSO1 and QSO2  prior to PSF subtraction (white circles), and the masked interloper galaxy ``G'', tentatively at
$z=0.457\pm0.001$ (more details in Appendix~\ref{sec:int_gal}). Also, to guide the eye, we overlay a grid spaced by 10\arcsec (or 77~kpc). 
We also highlight the location of a bright knot (white cross) whose SB value is relevant for 
the discussion in Section~\ref{sec:qpq4}, the direction along which we trace the variations in SB$_{\rm Ly\alpha}$ in 
Section~\ref{sec:proj_dist}, and the seeing circle for these observations (bottom left corner).}
\label{fig:SB_Vel_map}
\end{figure*}

The Ly$\alpha$ signal was extracted from this final PSF and continuum subtracted datacube using custom routines written in the Python Programming Language\footnote{https://www.python.org/}.
First, we produced a sub-cube of the wavelength range where the extended Ly$\alpha$ line is expected, 
allowing for wide shifts of the line, i.e. $\pm7500$~km~s$^{-1}$ from the two quasar systemic redshifts.
This sub-cube covers 4750.2 - 5010 \AA. In the next step we smoothed the sub-cube with 
a Gaussian kernel of FWHM$=1.66\arcsec$, i.e. similar to the seeing.

We further marked all the regions in all the layers of the smoothed sub-cube above a S/N=2 \footnote{This threshold has been frequently used for detection of extended 
emission in MUSE data (e.g., \citealt{Borisova2016,FAB2019}).}, obtaining a segmentation map for each layer. 
Using these segmentation maps, we found the largest connected area with detection above S/N=2 to be 
of 2077 spaxels (or 83 arcsec$^2$), in the layer of 4872.7\AA\ (basically at the systemic redshift of QSO2).
We walk through the cube starting from this layer, first in the direction of increasing wavelength, 
and then in the direction of decreasing $\lambda$, to obtain a three-dimensional (3D) mask describing the Ly$\alpha$ emission within the datacube. As we move from layer to layer in either only increasing or only decreasing 
$\lambda$, we recursively attach to the detection area (defined from the previous processed layers) 
the regions of the new layer which have at least one voxel in common with it. 
We consider the union of the two areas obtained by walking the cube in increasing and decreasing $\lambda$
as the final detection area over the whole smoothed sub-cube. 
To avoid losing S/N$\rm>$2 regions at the central layers which are only slightly detached from the 
largest detection area, we found that one should start from an initial mask 
defined by the collapse of the three ``central'' segmentation maps, i.e. the combined segmentation map 
of the ``central'' layer (containing the largest connected detection area) together with the maps of the two adjacent layers. 
The selection of the largest detection on this collapsed layer and the percolation at larger and smaller wavelengths following the simple aforementioned constraints
allow us to obtain a 3D mask that can be used for the analysis of the extended emission. To avoid the inclusion of spurious signal, we restrict the mask to spaxels
with at least three 
layers along the wavelength direction.
The final 3D mask comprises 19253 voxels, extends for 21.25\AA\ (or 17 layers), and its flux-weighted center is at 4872.7~\AA, thus close to the systemic redshift of QSO2 (4872~\AA).

%--------------------------------------------------------------------

\section{Results}
\label{sec:results}

\subsection{Extended emission connecting the quasar pair}
\label{sec:em_res}

\begin{figure*}
\centering
\includegraphics[width=0.65\columnwidth]{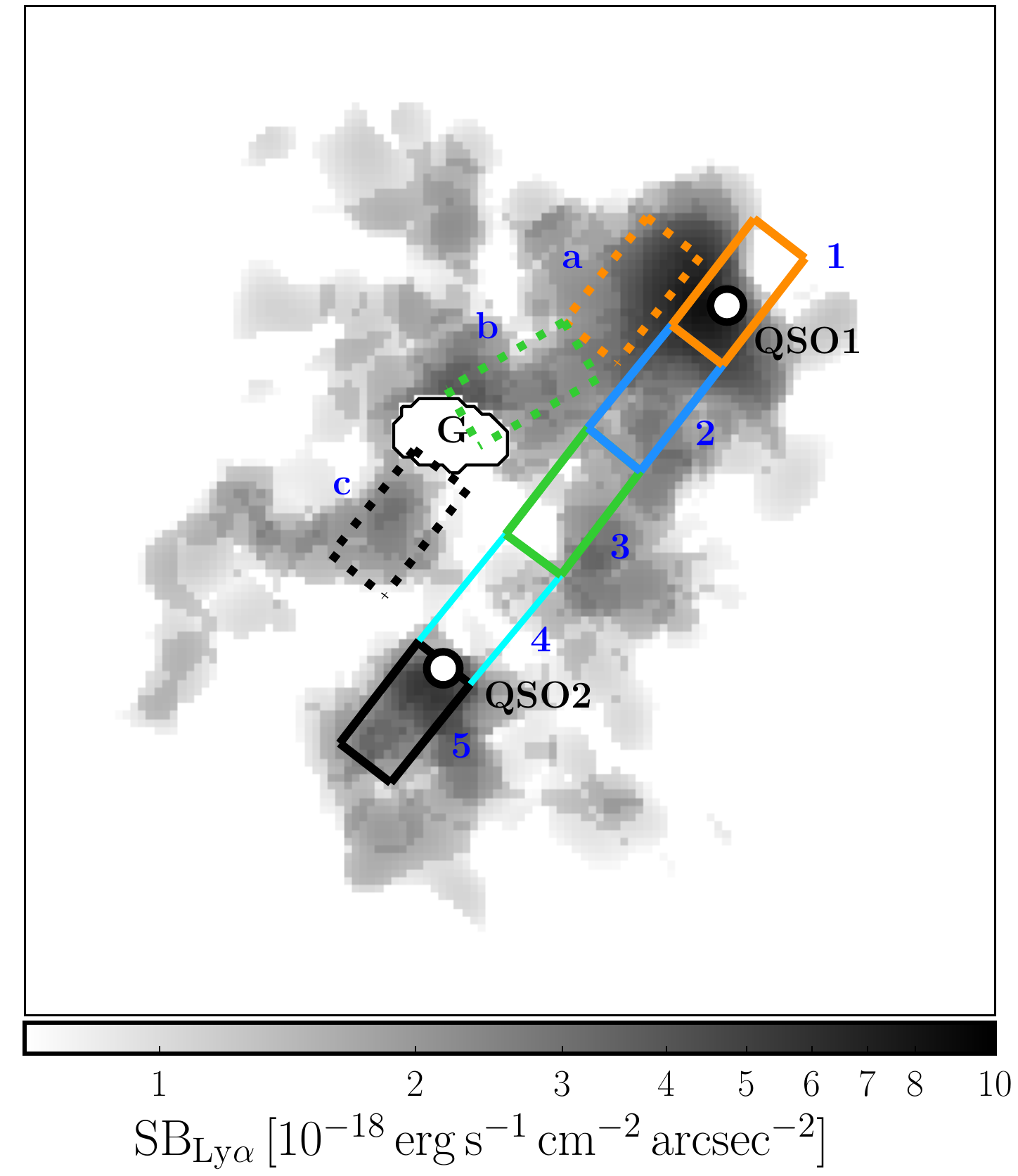}
\includegraphics[width=0.65\columnwidth]{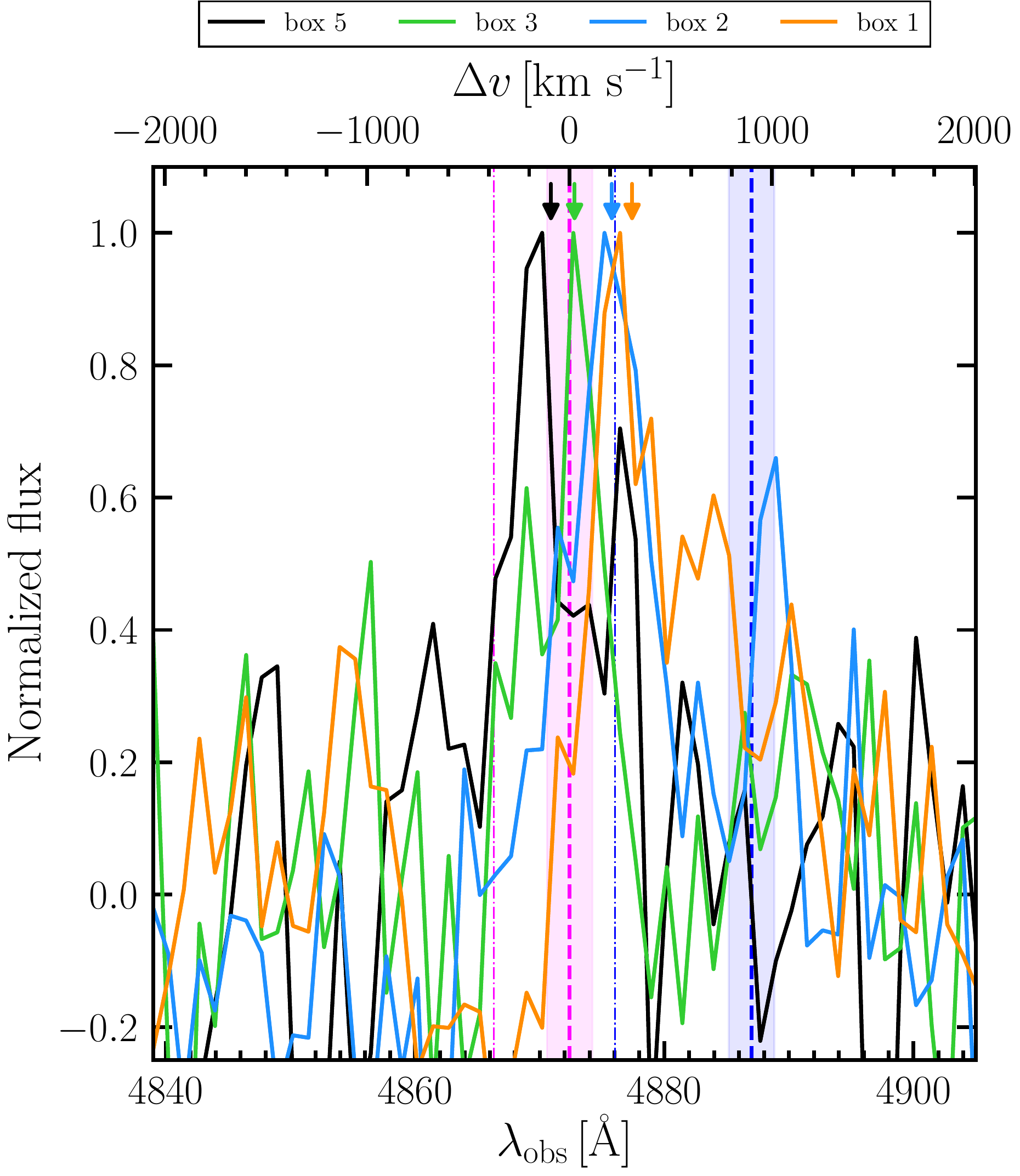}
\includegraphics[width=0.65\columnwidth]{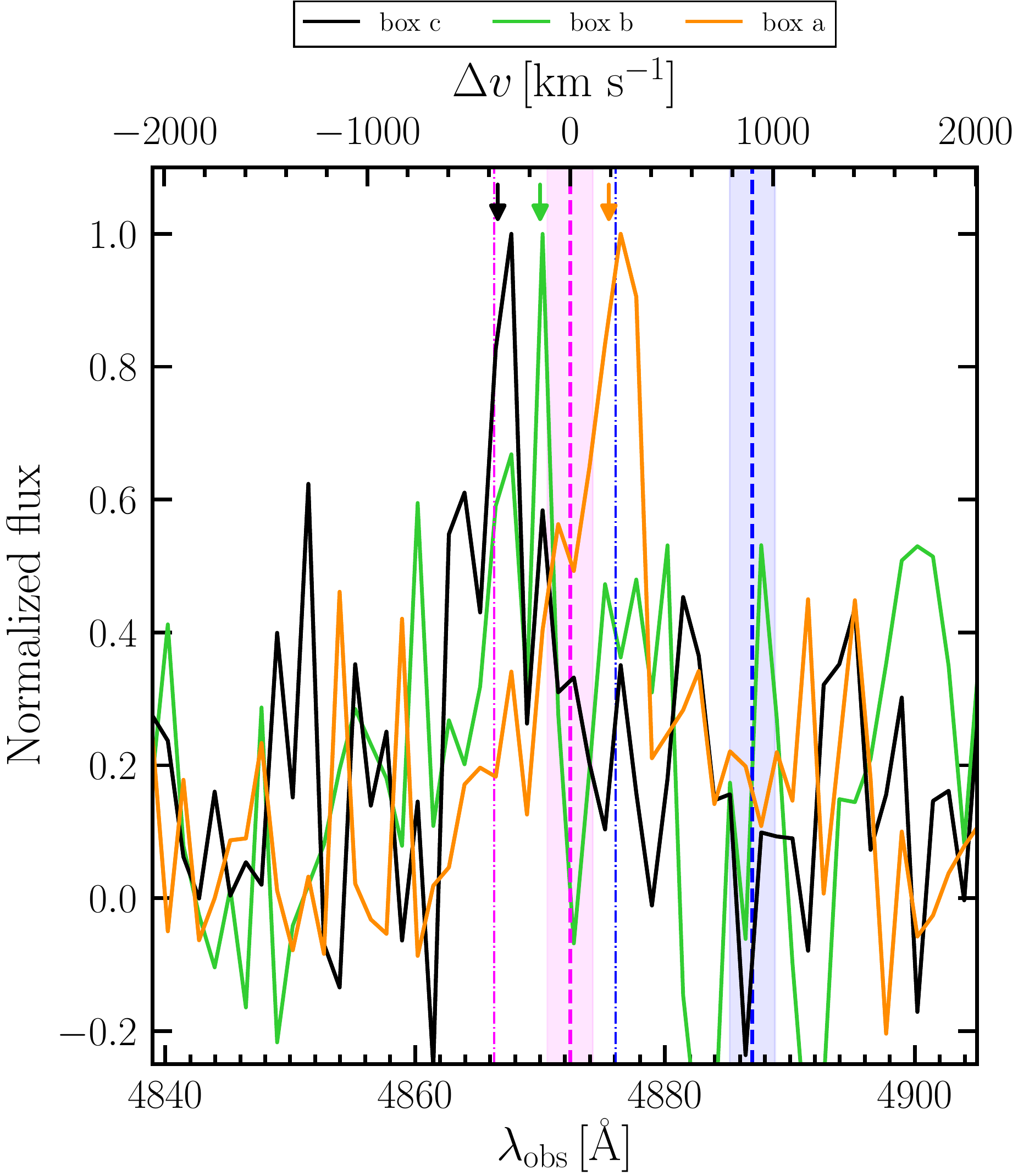}
\caption{Left: ``optimally extracted'' Ly$\alpha$ surface brightness map as in Figure~\ref{fig:SB_Vel_map} with 
the overlaid pseudoslits used to extract the spectra shown in the central and right panels. We assign an ID (blue) to each box of the pseudoslits.
Center: normalized spectra of the Ly$\alpha$ emission along the pseudoslit shown in the left panel with solid lines. Each spectrum is color-coded following the color of its box on the left (details in Section~\ref{sec:results}). 
The dashed (dotted-dashed) vertical lines show the systemic (peak of the Ly$\alpha$) redshifts for QSO1 (blue) and QSO2 (magenta).The respective shaded regions indicate the errors on 
the redshifts, as estimated by SDSS.  The velocity shifts $\Delta v$ are computed with respect to the systemic redshift of QSO2. 
We indicate with vertical arrows the flux-weighted centroid for each spectrum.
The flux-weighted velocity gradient of $\sim400$~km~s$^{-1}$ for the Ly$\alpha$ emission 
is in agreement with Figure~\ref{fig:SB_Vel_map}. 
The spectrum for box 4 (with no clear emission) is shown in Appendix~\ref{app:box2} (Figure~\ref{Box2_spectrum}).
Right: same as for the central panel, but for the second pseudoslit shown in the left panel with dotted lines. The flux-weighted velocity gradient of $\sim600$~km~s$^{-1}$ for the Ly$\alpha$ emission 
is in agreement with Figure~\ref{fig:SB_Vel_map}. The velocity gradients along the two pseudoslits are similarly increasing along the direction QSO2-QSO1.}
\label{fig:Pseudoslit}
\end{figure*}

We use the smoothed-cube masking described in the previous section to detect 
extended Ly$\alpha$ emission associated with the quasar pair. 
The optimally extracted SB map of this Ly$\alpha$ emission is computed
by integrating only the signal within the 3D mask as usually done in the literature (e.g., \citealt{Borisova2016}). 
Because of the irregular 3D morphology of the mask, each spaxel location of the optimally extracted SB map thus represents the signal integrated along a slightly different range in wavelength.
We show this SB 
map in the left panel 
of Figure~\ref{fig:SB_Vel_map}. The extended emission is detected at faint levels (average SB of ${\rm SB_{\rm Ly\alpha}}=1.8\times10^{-18}\cgssb$) on an area of 191~arcsec$^2$, covering the region between the two 
quasars\footnote{The area of 191~arcsec$^{2}$ corresponds to the whole Ly$\alpha$ nebula above S/N=2.}. 
At the positions of QSO1 and QSO2 the emission shows slightly higher levels with up to 
${\rm SB_{\rm Ly\alpha}}\sim8\times10^{-18}\cgssb$ in proximity to the brighter QSO1. 
The total luminosity of the extended emission is $L_{\rm Ly\alpha}=3.2\times10^{42}$~erg~s$^{-1}$.
The emitting structure shows a projected morphology reminiscent of intergalactic bridges or filaments connecting the two quasars. 
One bridge extends in the direction connecting the two quasars, while the second passes through the location of the interloper galaxy ``G'' (Appendix~\ref{sec:int_gal}). 
These structures have an average projected width of $\sim35$~kpc (or 4.5\arcsec) at the current spatial resolution and depth.
To enable the visualization of the significance of the detection and of the noise properties in our dataset, we overlay the S/N~$=3$, and $4$ contours on the optimally extracted SB map in Figure~\ref{fig:SB_Vel_map}, while in Appendix~\ref{app:NBandChi} we show a pseudo narrow-band image and a smoothed $\chi$ image of the central portion of the 
wavelength range covered by the 3D mask.

We also compute the first moment of the flux distribution within the 3D mask,
or in other words the flux-weighted velocity shift with respect to the systemic redshift of QSO2. 
We show the map for the shift in the right panel of Figure~\ref{fig:SB_Vel_map}.
The obtained shifts are in the range  $-400~{\rm km~s^{-1}} \lesssim  \Delta v \lesssim +400~{\rm km~s^{-1}}$ 
and show a gradient along the direction connecting QSO2 to QSO1. 
Specifically, starting from the southern regions close to QSO2, we see a shift of $\sim-200$~km~s$^{-1}$ 
which increases to $\sim 200$~km~s$^{-1}$ at the location of QSO1. The northern bridge, instead, shows 
a shift of $\sim -400$~km~s$^{-1}$ in the vicinity of QSO2 which similarly increases to $\sim 200$~km~s$^{-1}$ at the location of QSO1.

\begin{figure}
\centering
\includegraphics[width=0.9\columnwidth]{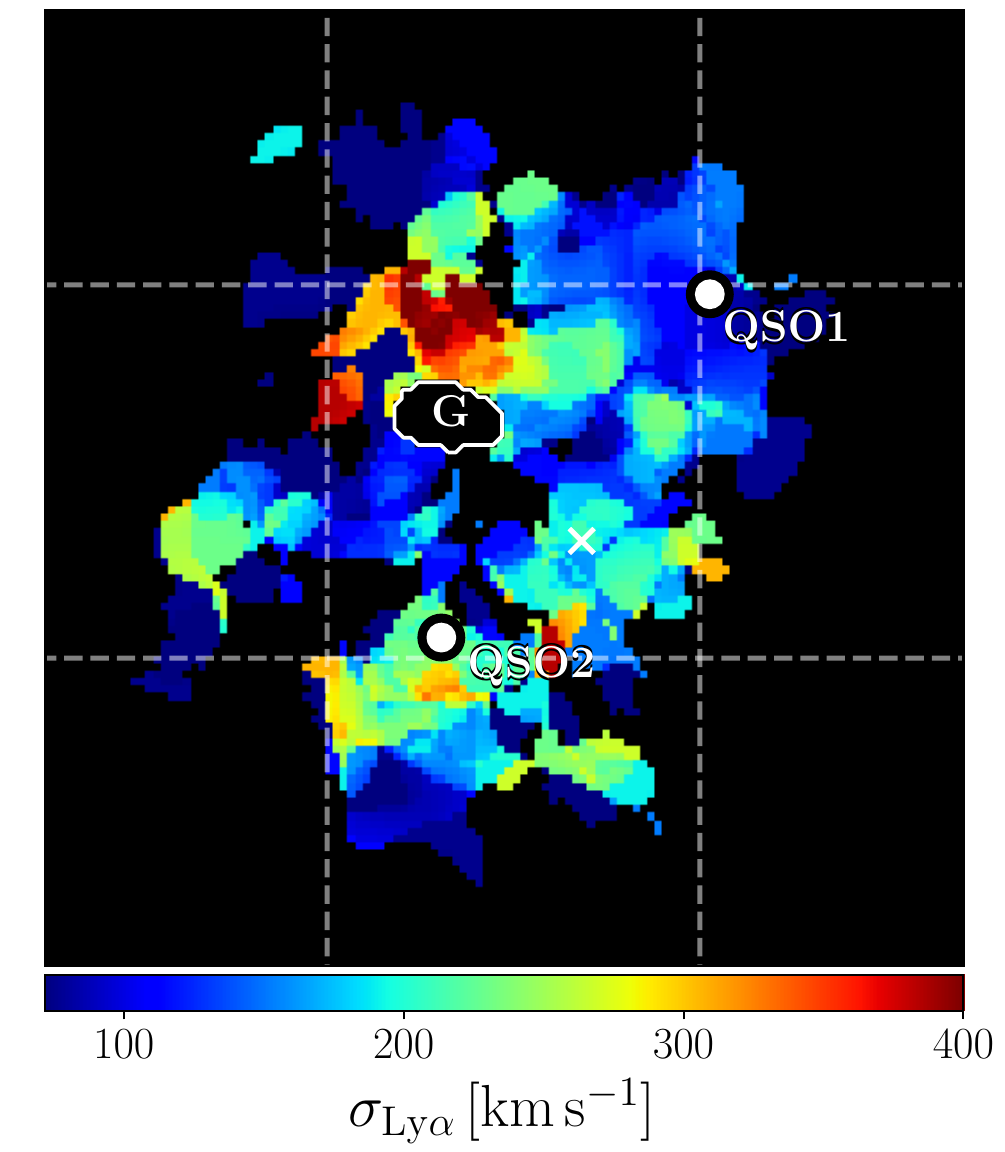}
\caption{The flux-weighted velocity-dispersion map obtained as the second order moment of the flux distribution 
within the 3D mask described in Section~\ref{sec:PSFsub}. 
The map shows the same field-of-view and uses the same symbols and nomenclature as in Figure~\ref{fig:SB_Vel_map}. 
Though quite noisy, as usually found around quasars, this map indicates that the emission is quiescent with an average velocity 
dispersion $\sigma_{\rm Ly\alpha}=162$~km~s$^{-1}$.}
\label{fig:sigma_map}
\end{figure}

To investigate these velocity gradients and visualize the line profile, we extract
spectra along two pseudoslits spanning the two bridges. 
In particular, for the bridge along the direction connecting the two quasars, we focus on obtaining spectra in five rectangular boxes. 
For this operation, we simply sum the fluxes layer by layer within each box, 
without using the aforementioned 3D mask. 
The rectangles have sides of $1\times$ and $2\times$ the seeing of our observations, i.e. $1.66\arcsec \times 3.32\arcsec$, 
and are placed as shown in the left panel of Figure~\ref{fig:Pseudoslit}, 
starting by centering the first region at the position of QSO1. 
The extracted 1D spectrum for each region\footnote{The spectrum for region 4 is shown in the appendix as it does not show a clear detection.} is shown in the central panel of Figure~\ref{fig:Pseudoslit}. 
Each spectrum is normalized at its peak to enable a better comparison of the line emission at the different locations.
This panel confirms the presence of a flux-weighted velocity gradient of about $400$~km~s$^{-1}$ along the direction QSO1-QSO2, though with slightly different values (-100, 300~km~s$^{-1}$) reflecting the uncertainties in this 
measure (vertical arrows in right panel of Figure~\ref{fig:Pseudoslit}). 
This gradient is smaller, but comparable with the velocity difference and location of the peaks of the quasars Ly$\alpha$ emission ($\Delta v= 598 \pm 98 $~km~s$^{-1}$; vertical dashed-dotted lines in the right panel 
of Figure~\ref{fig:Pseudoslit}). 
Similarly, we placed three boxes to cover the second bridge, starting with the first box ``a'' in vicinity of QSO1 (left panel of Figure~\ref{fig:Pseudoslit}). The normalized spectra of these three boxes are shown in the right panel 
of Figure~\ref{fig:Pseudoslit}, confirming the velocity gradient seen in the velocity map, from about $-400$~km~s$^{-1}$ (box ``c'') to $200$~km~s$^{-1}$ close to QSO1 (box ``a''). Also along this pseudoslit, the velocity gradient
roughly spans the velocity difference between the peaks of the quasars Ly$\alpha$ emission.

The similarity between the two velocity gradients along the two bridges is not surprising. Indeed, 
(i) the current observations are not extremely deep, leaving space for the presence of more diffuse gas (hence lower levels of Lyman-alpha emission)
connecting the currently observed bridges, with the observed gas being only the densest portion of the structure. (ii) Current cosmological simulations of structure formation
usually show multiple dense filamentary structures embedded in more diffuse intergalactic gas along the direction of massive halos, 
or multiple dense structures around massive interacting systems (e.g., \citealt{Rosdahl12,Mandelker2019}).

Further, along both bridges, the difference between the current quasars' systemic redshifts appear to be wider,  $\Delta v= 896 \pm 316 $~km~s$^{-1}$,  and seemingly less linked to the observed velocity 
difference within the extended Ly$\alpha$ emission. 
Nonetheless, the difference between the two quasars systemic redshifts could be due to the large uncertainties in those measurements (intrinsic uncertainties of 233~km~s$^{-1}$; Table~\ref{QSOpair}).
On top of this, the observed gradient and difference with respect to the uncertain quasars' systemic redshift could encode 
a mixture of radiative transfer effects, CGM kinematics and
intergalactic displacement along the line of sight. 
We discuss the possible configurations
of the system in Section~\ref{sec:pow_Lya}.

As a next step, we compute the flux-weighted velocity dispersion map as the
second moment of the flux distribution for the voxels encompassed by the 3D mask. 
Figure~\ref{fig:sigma_map} shows this map, which is clearly noisy due to the narrow spectral range 
of the detected emission. The obtained velocity dispersions are indeed relatively quiescent, 
with an average $\sigma_{\rm Ly\alpha} = 162$~km~s$^{-1}$ (or FWHM$=380$~km~s$^{-1}$)\footnote{The patch with high velocity dispersion ($\sigma_{\rm Ly\alpha}\sim400$~km~s$^{-1}$) 
slightly North than the interloper galaxy is at low S/N, and thus uncertain. We however include it when calculating the average value for $\sigma_{\rm Ly\alpha}$.}. 
This value is comparable, though lower than the average value observed around individual 
brighter $z\sim3$ quasars ($\sigma_{\rm Ly\alpha} = 265$~km~s$^{-1}$; \citealt{FAB2019}).

\begin{figure}
\centering
\includegraphics[width=0.85\columnwidth]{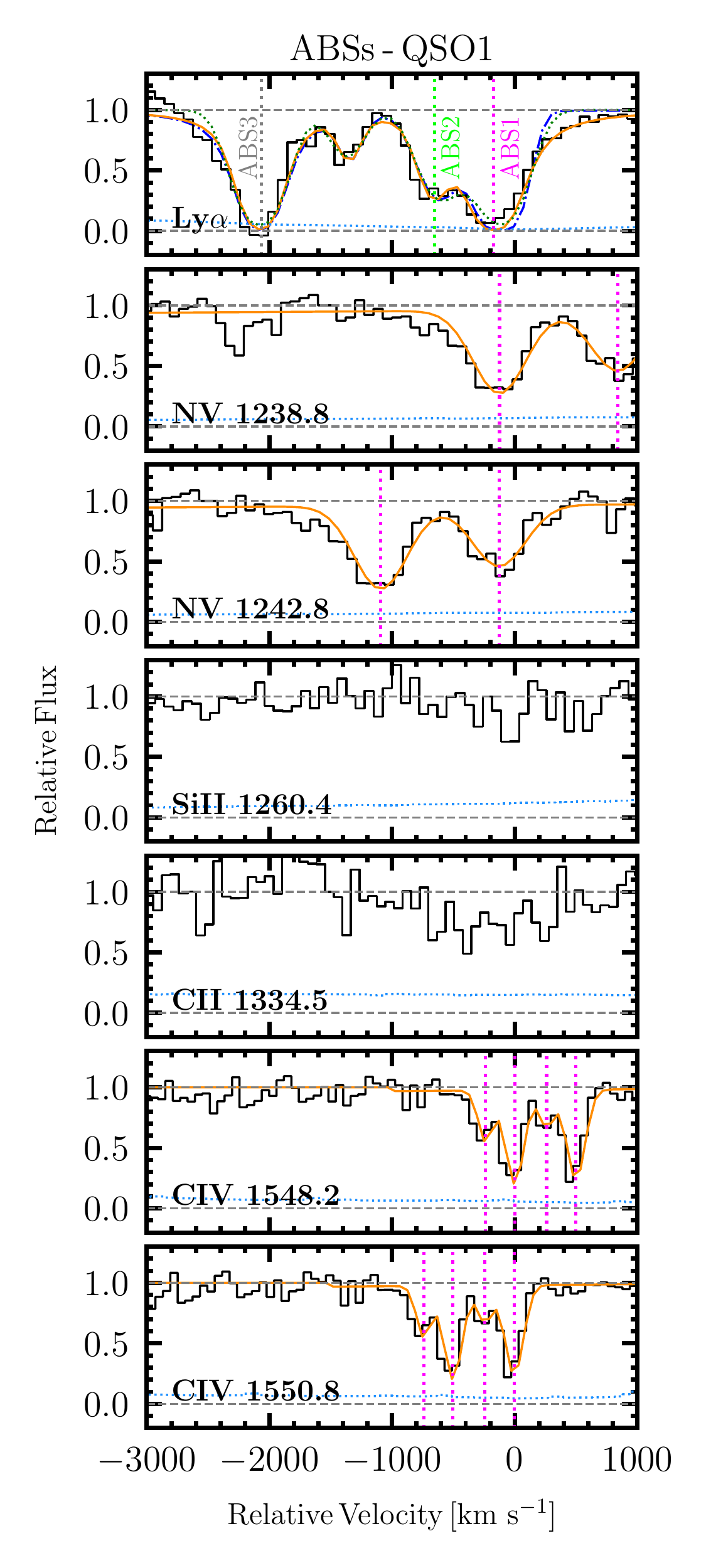}
\caption{The profiles of the absorbers along the QSO1 sight-line at the \ion{H}{i} Ly$\alpha$, \ion{N}{v}, 
\ion{Si}{ii}, \ion{C}{ii}, and \ion{C}{iv} lines. The black histograms show the continuum normalized data, 
while the orange lines are the sum of all the Gaussian components of the best fit. 
The locations of the absorbers considered in this analysis are highlighted with vertical dashed lines, 
i.e. ABS1 (magenta), ABS2 (lime), and ABS3 (gray).
The zero velocity is set to the redshift of the \ion{C}{IV} strongest component.
Table~\ref{Tab:ABS} gives all the fit parameters. For ABS1, we exclude the best fit solution for $N_{\rm HI}$ (orange) following physical arguments (Section~\ref{sec:model_abs}). We thus show the two extreme alternative fits 
 with $N_{\rm HI}$ and $b$ values that are favored by the current data, i.e.,  $b=100$~km~s$^{-1}$, log$(N_{\rm HI}/{\rm cm^{-2}})=17$ as a dashed-dotted blue line, and $b=200$~km~s$^{-1}$, log$(N_{\rm HI}/{\rm cm^{-2}})=15$ as a dotted green curve.}
\label{fig:ABSQSO1}
\end{figure}

\begin{figure}
\centering
\includegraphics[width=0.85\columnwidth]{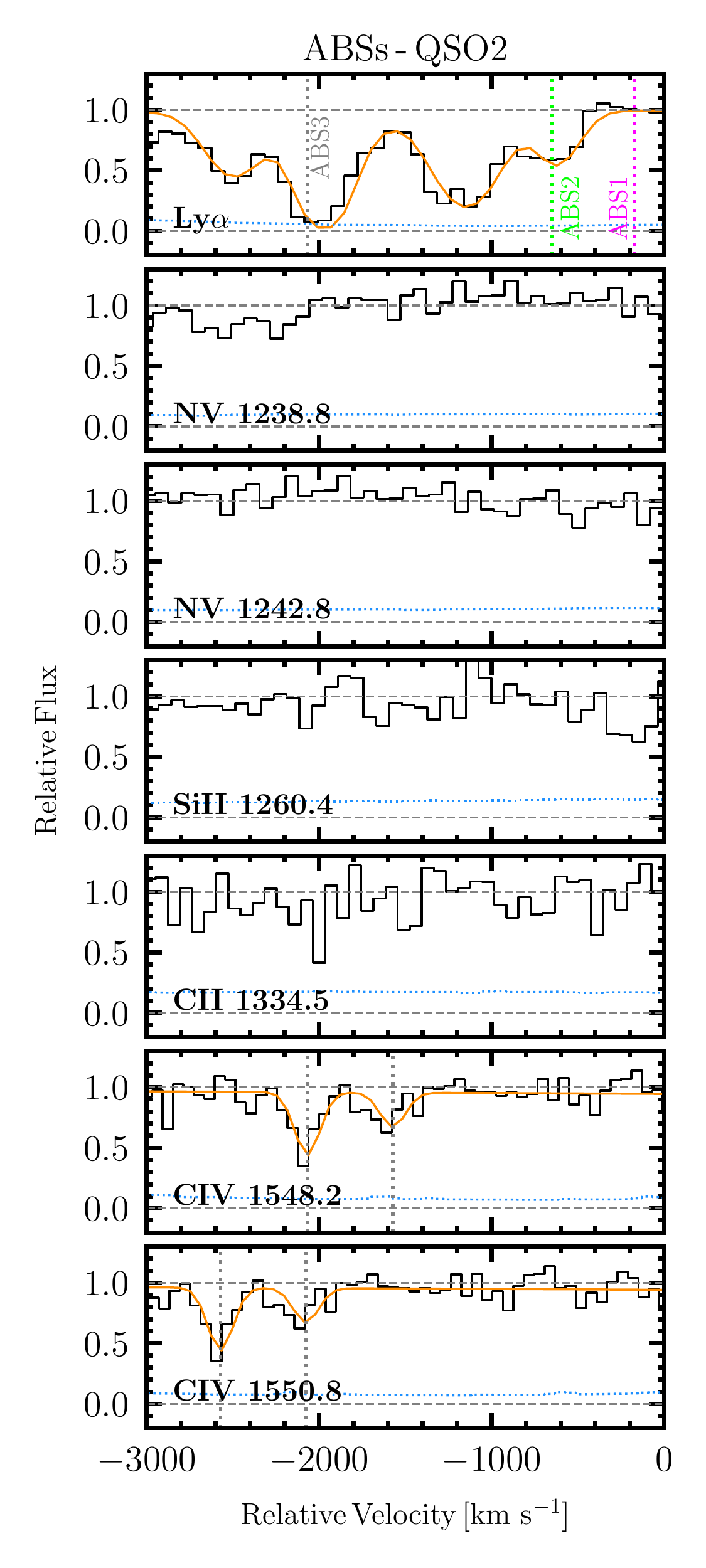}
\caption{The profiles of the absorbers along the QSO2 sight-line at the \ion{H}{i} Ly$\alpha$ line, 
\ion{N}{v}, \ion{Si}{ii}, \ion{C}{ii}, and \ion{C}{iv} lines. The black histograms show the continuum normalized data, 
while the orange lines are the sum of all the Gaussian components of the best fit. The location of the absorbers considered 
are highlighted with vertical dashed lines, i.e. ABS1 (magenta; not present along this sight-line), ABS2 (lime), and ABS3 (gray).
The zero velocity is set to the redshift of the \ion{C}{IV} strongest component of ABS1 along the QSO1 sight-line, as in 
Figure~\ref{fig:ABSQSO1}. Table~\ref{Tab:ABS} gives all the relevant fit parameters.}
\label{fig:ABSQSO2}
\end{figure}

Aside from the Ly$\alpha$ emission, we did not detect any other extended line emissions associated with the quasar pair down to the depth of the current observations.
In particular, we checked the \ion{C}{iv}$\lambda1549$ and \ion{He}{ii}$\lambda1640$ expected wavelengths as these two lines can give informations on metallicity, volume density $n_{\rm H}$, and speed of shocks (if any; \citealt{fab+15a}). Importantly, the 
\ion{He}{ii}/Ly$\alpha$ ratio is sensitive to $n_{\rm H}$ in a pure recombination scenario, with the ratio decreasing from the expected value of $0.34$ (at a temperature $T=2\times10^4$~K) if \ion{He}{ii} is not completely doubly ionized (i.e. at high enough densities; \citealt{fab+15b}). 
Here the observations achieved a $2\sigma$ surface brightness limit 
(in 1 arcsec$^2$ aperture) in a single channel (1.25\AA) of SB$_{\rm \ion{C}{iv}}=5.0\times10^{-19}\cgssb$ and SB$_{\rm \ion{He}{ii}}=4.6\times10^{-19}\cgssb$,  respectively for \ion{C}{iv} (at 6208.8~\AA) and \ion{He}{ii} 
(at 6573.5~\AA).
These slightly deeper sensitivities than at the location of the Ly$\alpha$ are due to the overall system efficiency of the facility which peaks at about 7000~\AA.
Considering the region where Ly$\alpha$ is detected (191~arcsec$^2$; Figure~\ref{fig:SB_Vel_map}), we obtain $5\sigma$ upper limits for \ion{C}{iv} and \ion{He}{ii} emissions in 5 channels maps, i.e. ${\rm SB}_{\rm \ion{C}{iv}}<2.3\times10^{-19}\cgssb$ and 
${\rm SB}_{\rm \ion{He}{ii}}<2.1\times10^{-19}\cgssb$. We show the maps at these wavelengths in Appendix~\ref{app:chiCIVHeII}. The average \ion{C}{iv}/Ly$\alpha$ and \ion{He}{ii}/Ly$\alpha$ ratios are thus constrained to be $<0.13$ ($5\sigma$) and 
$<0.12$ ($5\sigma$), respectively. Therefore, following \citealt{fab+15b}, if the Ly$\alpha$ emission is only due to recombination, the \ion{He}{ii} component cannot be fully doubly ionized given the observed low constraints, implying relatively high gas densities ($n_{\rm H}>0.1$~cm$^{-3}$). 
On the other hand, the low limit in the \ion{C}{iv}/Ly$\alpha$ ratio translates to metallicities likely lower than $Z_{\odot}$ unless the densities are extremely high, $n_{\rm H} \gg 1$~cm$^{-3}$ (\citealt{fab+15b}). These 
limits are similar and consistent with what has been usually found for extended Ly$\alpha$ nebulosities around quasars (e.g., \citealt{fab+15b,Borisova2016,FAB2018}) and in the so-called Lyman-Alpha Blobs 
(LAB; \citealt{Prescott2009,Prescott2013,fab+15a}) down to similar depths.   
We take into account the limits on these lines in Section~\ref{sec:pow_Lya}, where we discuss the possible system configurations.

\subsection{Gas absorption traced by the quasar pair}
\label{sec:abs}

As already shown in Figure~\ref{QSOpair_spectra}, we find various absorbers along the two quasar sight-lines.
Here we focus on reporting the properties of three absorbers (ABS1, ABS2, ABS3), while we discuss in detail their origin in Section~\ref{sec:model_abs}. In particular, we study a 
strong absorber (ABS1) along the QSO1 sight-line and close to the systemic redshift of QSO2, and two others (ABS2 and ABS3) 
found along both sight-lines to the two quasars. 

We analyze the absorption features, proceeding as follows.
We first model the continuum of each QSO by fitting low-order polynomials to spectral chunks that are free 
from absorption lines. After the continuum normalization, we 
model the Ly$\alpha$ absorption lines with Voigt profiles using \textsc{vpfit}\footnote{\url{http://www.ast.cam.ac.uk/rfc/vpfit.html}} v10.0. 
Given the spectral resolution of MUSE, we keep the Doppler $b$ parameter fixed at reasonable values while performing the fit of the Ly$\alpha$ absorption lines.
These absorptions are near the Ly$\alpha$ emission of the QSOs, and therefore 
the inferred column densities might be sensitive to the continuum placement. 
To take this into account we generate a few continuum models for each QSO and repeat the Voigt profile fitting. 
The dispersion in the resulted $N_{\rm HI}$ is incorporated in the quoted errors. 
We model each of the doublet absorption lines (\ion{C}{iv}$\lambda1548,1550$ and \ion{N}{v}$\lambda1238,1242$) 
using a double Gaussian profile. 
The sigma values of the two Gaussians of a doublet are tied to be the same, and the wavelength ratio is fixed at the 
value given by the atomic tables. 
We also allow the equivalent widths (EW) ratio of the two lines in a doublet to vary to take 
into account the possible saturation effect. 
We note that such models are not sensitive to continuum placements since the lines are reasonably narrow. 
We then obtain the lower limits of the column densities using the EWs of the lines and assuming the linear part 
of the curve-of-growth. 
For the non-detected transitions we use the S/N at the position of the lines to calculate the 
upper limit on the EWs. 
We further convert such limits to upper limits on column densities using the linear part of the curve-of-growth.
The fits performed along the two sight-lines are reported in Figures~\ref{fig:ABSQSO1} and \ref{fig:ABSQSO2}, 
while all the derived parameters are listed in Table~\ref{Tab:ABS}.

For ABS1 we estimate log($N_{\ion{H}{i}}/{\rm cm^{-2}})=15-17$, allowing the Doppler $b$ parameter 
to vary uniformly between $100$ and $200$~km~s$^{-1}$, with smaller $b$ at higher $N_{\ion{H}{i}}$. 
Allowing for even smaller $b$ parameters down to $50$~km~s$^{-1}$ increases the goodness of the fit ($\chi^2$ decreases from $\sim9$ to $\sim4$). 
However, these small $b$ values require very large column densities (log($N_{\ion{H}{i}}/{\rm cm^{-2}})>18$), 
which are disfavored by the lack of absorption at the location of low-ion transitions, 
like  \ion{Si}{ii}$\lambda1260$ or \ion{C}{ii}$\lambda1335$. As an additional test, we check for the presence of an associated Lyman limit system (LLS; log($N_{\ion{H}{i}}/{\rm cm^{-2}})>17.2$) 
by looking for its $912$~\AA\ break in the LRIS data of the Quasars Probing Quasars database (QPQ; \citealt{Findlay2018}). We find no clear evidence for the break, confirming that the values
log($N_{\ion{H}{i}}/{\rm cm^{-2}})=15-17$ are favored. For completeness, in Table~\ref{Tab:ABS} and in Figure~\ref{fig:ABSQSO1} 
we report examples for $b=50$ (solid orange line), $100$ (dashed-dotted blue line), and $200$~km~s$^{-1}$ (dotted green line). 

ABS1 has associated absorption in \ion{N}{v}$\lambda1240$ and \ion{C}{iv}$\lambda1549$. 
The absorption at the \ion{C}{iv} wavelength is best fit by two components (log($N_{\ion{C}{iv}}^{\rm strong}/{\rm cm^{-2}})>14.9$; log($N_{\ion{C}{iv}}^{\rm weak}/{\rm cm^{-2}})>14.5$), while the one at \ion{N}{v} 
can be fitted by a single Gaussian line (log($N_{\ion{N}{v}}/{\rm cm^{-2}})>15.5$) at the current spectral resolution.
The fit of the \ion{H}{I} absorption places ABS1 at $z=3.005\pm0.001$. This redshift is at $\Delta v=-230\pm140$~km~s$^{-1}$ from the systemic redshift of QSO2.
The two \ion{C}{iv} components show velocity shifts of $+173$~km~s$^{-1}$ and $-66$~km~s$^{-1}$ with respect to the \ion{H}{i} absorption, respectively for the strong and weak components.
The \ion{N}{v} is redshifted by $+45$~km~s$^{-1}$. These shifts justify the large $b$ parameter allowed during the fit of the \ion{H}{I} absorption.

The metal absorptions show relatively quiescent widths with the two \ion{C}{iv} components being characterized by
a velocity dispersion $\sigma^{\rm strong}=76\pm4$~km~s$^{-1}$ (or $0.39\pm0.02$~\AA) and $\sigma^{\rm weak}=58\pm10$~km~s$^{-1}$ (or $0.30\pm0.05$~\AA), and the \ion{N}{v} by $\sigma=215\pm10$~km~s$^{-1}$ (or $0.89\pm0.04$~\AA), after 
correcting for the MUSE spectral resolution.
The larger value for \ion{N}{v} could be partially due to the superposition of a second unresolved component.
However, if we assume that the two unresolved components share roughly the same $\sigma$, we would get $\sigma\sim150$~km~s$^{-1}$, still larger than \ion{C}{iv}.
Keeping in mind the large uncertainties in the determination of the two quasars redshifts, our fit overall suggests that the strong absorber ABS1 could be associated with QSO2. 
We cannot completely exclude that this absorber is due to intervening associated gas to QSO1 (at $\Delta v=-1120\pm140$~km~s$^{-1}$), but the narrowness of
its metal absorptions rules out the scenario in which this gas is outflowing at small distances from QSO1. 
Therefore, ABS1 is most likely produced by gas at least on CGM scales (around QSO1, QSO2 or the system comprising the two quasars; Section~\ref{sec:model_abs}).

ABS2 is placed at $\Delta v=-900$~km~s$^{-1}$ from QSO2, and, conversely, has log($N_{\ion{H}{i}}/{\rm cm^{-2}})\sim14$ 
along both sight-lines with no other absorption lines detected.  
This absorber is thus very similar to Ly$\alpha$ forest clouds (\citealt{meiksin09}).

The fit of ABS3, located at $\Delta v=-2800$~km~s$^{-1}$ from QSO2, shows high log($N_{\ion{H}{i}}/{\rm cm^{-2}})\sim19$ along both sight-lines ($b=50$~km~s$^{-1}$).
However, as for ABS1, the absence of low ion transitions possibly implies smaller values of $N_{\ion{H}{i}}$, i.e. log($N_{\ion{H}{i}}/{\rm cm^{-2}})\sim15-17$ ($b=200-100$~km~s$^{-1}$).
Also in this case we look for an associated 912~\AA\ break in the QPQ database (\citealt{Findlay2018}), finding no evidence for a LLS.  The values log($N_{\ion{H}{i}}/{\rm cm^{-2}})\sim15-17$ are thus favored also in this case.
Further, ABS3 shows strong \ion{C}{iv} absorption only towards QSO2 (log($N_{\ion{C}{iv}}/{\rm cm^{-2}})>14.7$). 
For both ABS2 and ABS3 we do not find associated galaxies at their corresponding redshifts (Section~\ref{sec:model_abs}).

%\begin{sidewaystable}
\begin{table*}
\begin{center}
\caption{The properties of the absorbers along the QSO1 and QSO2 sight-lines.}
\scalebox{1}{
\scriptsize
\setlength\tabcolsep{4pt}
\begin{tabular}{l | c | cc | cc | cc}
\hline
\hline
 & ID$_{\rm abs}^{\rm a}$ & ABS1$^{\rm b}$ (m) & ABS1-2\ion{C}{iv}  & ABS2 (l) & ABS2 (l) & ABS3$^{\rm l}$ (g) & ABS3$^{\rm l}$ (g) \Bstrut \\

 & ID$_{\rm s-l}^{\rm c}$                   & QSO1 & QSO1 & QSO1 & QSO2 & QSO1 & QSO2 \Tstrut\Bstrut \\
\hline
\ion{H}{I}                   & $z_{\rm abs}$ & $3.005\pm0.001$ & &  $2.99833\pm0.00026$ & $2.99901\pm0.00026$  & $2.98097\pm0.00021$ & $2.97952\pm0.00021$ \Tstrut \\
1215.670                  & log~($N/[{\rm cm^{-2}}])$         & $15-17$ & &   $14.6\pm0.2$ & $14.1\pm0.1$  & $15-17$ & $15-17$ \Tstrut \\
                                 & $b$ [km~s$^{-1}$]  & $200-100$ & &   50$^{\rm d}$ & 50$^{\rm d}$ & $200-100$ & $200-100$ \Tstrut\Bstrut \\
\hline
\ion{N}{V}$^{\rm g}$                  & $z_{\rm abs}$ & $3.00560\pm0.00015$ & & --- & --- & --- & --- \Tstrut \\
1242.804                  & log~($N/[{\rm cm^{-2}}])$         & $>15.5$ & & $< 14.2 $ & $< 14.2$ & $< 14.7$ & $< 14.7$ \Tstrut  \\
(1238.821)               & EW$_0^{\rm e}\ [{\rm \AA}]$   & 6.14$\pm$0.43 (4.5$\pm$0.48) & &  $< 0.37$ & $< 0.37$ & $< 1.1$ & $< 1.1$ \Tstrut \\
                                 & $\sigma_0^{\rm e}\ [{\rm \AA}]$               & 0.89$\pm$0.04 & & --- & --- & --- & --- \Tstrut\Bstrut  \\
\hline
\ion{C}{iv}$^{\rm g}$                  & $z_{\rm abs}$ & $3.00731\pm0.00005$ & $3.00412\pm0.00010$ &  --- & ---  & --- & $2.97967\pm0.00013$ \Tstrut \\
1550.770                  & log~($N/[{\rm cm^{-2}}])$         & $>14.9$ & $>14.5$ &  $< 14.1$ & $< 14.1$  & $< 14.1$ & $>14.7$ \Tstrut \\
(1548.195)                & EW$_0^{\rm e}\ [{\rm \AA}]$   & 3.02$\pm$0.27 (2.89$\pm$0.33) & 1.30$\pm$0.26 (0.94$\pm$0.24) &  $< 0.53$ & $< 0.53$  & $< 0.47$ & $1.99\pm0.36$ ($1.09\pm0.33$) \Tstrut \\
                                 & $\sigma_0^{\rm e}\ [{\rm \AA}]$               & 0.39$\pm$0.02 & 0.30$\pm$0.05 &  --- & ---  & --- & $0.38\pm0.05$ \Tstrut\Bstrut \\
\hline
\ion{C}{ii}                  & log~($N/[{\rm cm^{-2}}])$         & $<14.3$ & &  $<14.3$ & $<14.3$ & $<14.2$ & $<14.3$ \Tstrut \\
1334.532                  & EW$_0^{\rm e}\ [{\rm \AA}]$   & $<0.4^{\rm f}$ & &  $<0.4^{\rm f}$ & $<0.4^{\rm f}$ & $<0.3^{\rm f}$ & $<0.4^{\rm f}$ \Tstrut\Bstrut \\
\hline
\ion{Si}{ii}                  & log~($N/[{\rm cm^{-2}}])$         & $<13.2$  & &  $<13.2$ & $<13.2$ & $<13.0$ & $<13.2$ \Tstrut \\
1260.422                  & EW$_0^{\rm e}\ [{\rm \AA}]$   & $<0.24^{\rm f}$ & & $<0.24^{\rm f}$ & $<0.24^{\rm f}$ & $<0.18^{\rm f}$ & $<0.24^{\rm f}$ \Tstrut\Bstrut \\
\hline
\hline
\end{tabular}
}
\flushleft{\scriptsize $^{\rm a}$ For each absorber we report in brackets its color for the vertical dotted lines in Figures~\ref{fig:ABSQSO1} and \ref{fig:ABSQSO2}: (m) is magenta, (l) is lime, (g) is gray. \\ 
$^{\rm b}$ An alternative better fit ($\chi^2=3.6$ vs $\chi^2\sim9$ for the presented values) of the \ion{H}{I} absorption of this system could be done by fixing the Doppler $b$ parameter to $50$~km~s$^{-1}$. 
However, this alternative fit has a higher log$N_{\ion{H}{i}}=19.34$ which seems to be disfavored by the absence of low-ion transitions associated with this absorber and by the absence of a
clear Lyman limit break (Section~\ref{sec:abs}).\\
%In this case we get a much lower $N_{\ion{H}{i}}$ and a unrealistically high $b$: $z+{\rm abs}=3.00566\pm0.00024$, log$N_{\ion{H}{i}}=14.98\pm0.05$, $b=200\pm18$. This fit is not able to fit well the wing at positive velocities in Fig.~\ref{}{\bf add ref}, i.e. reduced $\chi^2=8.6$ in comparison to the reduced $\chi^2=3.6$ for the fit reported in the table.
$^{\rm c}$ This ID indicates the sight-line (s-l) on which the absorber is observed. \\
$^{\rm d}$  For this fit, the Doppler $b$ parameter is fixed to a value that is usually found in IGM and CGM absorbers (e.g., \citealt{meiksin09, QPQ8}). \\
$^{\rm e}$ Rest frame equivalent width $EW_0$ and rest frame $\sigma$ values. \\
$^{\rm f}$ $3\sigma$ upper limit from which we estimate the upper limit in column density assuming the linear part of the curve of growth. \\
$^{\rm g}$ In brackets we report the data for the line of the doublet at shorter wavelengths.\\
$^{\rm l}$ The values log$N_{\ion{H}{i}}=15 - 17$ are favored by the absence of low-ion transitions and the absence of a clear Lyman limit break (Section~\ref{sec:abs}). 
%For completeness, we find that a fit using $b=50$~km~s$^{-1}$ results in higher values: log$N_{\ion{H}{i}}=18.7\pm0.1$ and log$N_{\ion{H}{i}}=19.06\pm0.08$, respectively for the
%component along the QSO1 and QSO2 sightlines.
\\
} 
\label{Tab:ABS}
\end{center}
%\end{sidewaystable}
\end{table*}

Higher spectral resolution observations are required to firmly constrain the properties of all 
these three absorbers. Nevertheless, in Section~\ref{sec:model_abs} we show that already 
our current data allow us to roughly infer their nature.

\section{The powering of the extended Ly$\alpha$ emission}
\label{sec:pow_Lya}

Three powering mechanisms could be responsible for the extended Ly$\alpha$ emission detected around quasars: 
photoionization by the quasar (e.g., \citealt{heckman91a,hr01,Weidinger05}), scattering of Ly$\alpha$ photons from the quasar (e.g., \citealt{Dijkstra2008}), 
or shocks due to the quasar activity (e.g., \citealt{Cai2016}).
These mechanisms do not exclude each other, and could possibly act together. 
We explore in turn their contributions, if any, in the system studied here, 
using analytical considerations.
The modeling of these mechanisms in a cosmological context (e.g., \citealt{Gronke2017}) is beyond the scope of this work.

First, we focus on fast quasar winds. This phenomenon has been so far traced in emission only out to a few tens of kpc from the central engine (e.g., \citealt{Harrison2014}), and is usually manifested in emission lines with FWHM $\gtrsim 1000$~km~s$^{-1}$ and velocity shifts of at least few hundreds of km~s$^{-1}$ (e.g., \citealt{Mullaney2013,Cai2016}). 
The extended Ly$\alpha$ emission detected in our data differs substantially as it shows a relatively quiescent line profile with an average velocity dispersion $\sigma_{\rm Ly\alpha} = 162$~km~s$^{-1}$ (or FWHM$=380$~km~s$^{-1}$). Considering that this value is not corrected for the instrument spectral resolution and that resonant scattering of Ly$\alpha$ photons could broaden the line, it is safe
to say that fast winds do not play a major role in shaping the Ly$\alpha$ extended structure and its emission level 
that we observe. This is in agreement with what has been routinely found with short exposures for extended 
Ly$\alpha$ emission around $z\sim3$ quasars (\citealt{FAB2019}).

Secondly, we consider a photoionization scenario from both quasars.
Indeed, the case in which only one of the quasars shines on the gas seems to be ruled out by: (i) the 
higher Ly$\alpha$ fluxes in proximity of each quasar, and (ii) the absence of emission on large 
scales in the NW direction from QSO1 and SE direction from QSO2, where most likely only the contribution of one quasar 
(modulo its opening angle and presence of gas) is relevant. 
We thus explore the quasar pair photoionization scenario in the two limiting regimes for the recombination emission: 
optically thin ($N_{\rm HI}\ll 10^{17.2}$~cm$^{-2}$) or optically thick ($N_{\rm HI}\gg 10^{17.2}$~cm$^{-2}$) 
gas to the Lyman continuum photons. We do this in two steps. 
First, we show some expectations by following the model for cool gas around quasars introduced by \citet{qpq4}, 
and then we model the system using the photoionization code Cloudy (version 17.01), 
last described in \citet{Ferland2017}.

\subsection{Analytical estimates for the extended Ly$\alpha$ emission}
\label{sec:qpq4}

In the framework of \citet{qpq4}, the cool ($T\sim10^4$~K) gas is organized in clouds characterized by a 
single uniform hydrogen volume density $n_{\rm H}$, a cloud covering factor $f_{\rm C}$, and a hydrogen column density $N_{\rm H}$.  
Knowing these quantities and the luminosity of a quasar, one can estimate the $Ly\alpha$ emission at a distance $R$ from it.

Specifically, in the optically thick case, the Ly$\alpha$ SB scales with the luminosity of the central source and 
should decrease as $R^{-2}$ with increasing distance from a quasar (see \citealt{qpq4} for the derivation of the formula):

\begin{align}
\label{eqn:SB_thick}
{\rm SB}_{\rm Ly\alpha}^{\rm thick} & = 5.7\times10^{-17}\left( \frac{1+z}{4.014}\right)^{-4}\left( \frac{f_{\rm C}}{1.0}\right) \left( \frac{R}{50~{\rm kpc}}\right)^{-2}\\ \nonumber
& \times \left( \frac{L_{\rm \nu_{\rm LL}}}{7.6\times10^{29}~{\rm erg~s^{-1}~Hz^{-1}}} \right)~{\rm erg~s^{-1}~cm^{-2}~arcsec^{-2}}, \nonumber
\end{align}

\noindent where $L_{\nu_{\rm LL}}$ is the specific luminosity at the Lyman edge. 
To obtain this luminosity for the two quasars, we assume a spectral energy distribution (SED) which follows the form
$L_{\nu}=L_{\nu_{\rm LL}} (\nu/\nu_{\rm LL})^{\alpha_{\rm UV}}$ blueward of the Lyman limit $\nu_{\rm LL}$, 
and adopt a slope of $\alpha_{\rm UV}=-1.7$ (\citealt{lusso15}). 
As done in \citet{fab+15b}, $L_{\nu_{\rm LL}}$ is computed by integrating the \citet{lusso15} composite spectrum against the SDSS filter curve to give the correct $i$-band magnitude of the two quasars (as listed in Table~\ref{QSOpair}). We find $L_{\nu_{\rm LL}}^{\rm QSO1}=7.6\times10^{29}$~erg~s$^{-1}$~Hz$^{-1}$ and  $L_{\nu_{\rm LL}}^{\rm QSO2}=5.7\times10^{29}$~erg~s$^{-1}$~Hz$^{-1}$ for QSO1 and QSO2, respectively .  
The two quasars have a bolometric luminosity of $L_{\rm bol}^{\rm QSO1} = 1.5\times10^{46}$~erg~s$^{-1}$ and $L_{\rm bol}^{\rm QSO2} = 1.1\times10^{46}$~erg~s$^{-1}$, when using a standard quasar SED template as described in Section~\ref{sec:SEDs}.

We promptly demonstrate that the optically thick scenario is unlikely to be in place in this system.
We can indeed explore different configurations (e.g., different distances between the quasars), 
and add up the contribution to ${\rm SB}_{\rm Ly\alpha}^{\rm thick}$ given by equation~\ref{eqn:SB_thick} for each quasar. 
For this discussion, we focus on the region of the bridge indicated by a white cross in Figure~\ref{fig:SB_Vel_map}, 
which is characterized by SB$_{\rm Ly\alpha}=3.5\times10^{-18}\cgssb$, and assume in equation~\ref{eqn:SB_thick} 
an average redshift of 3.014 and $f_{\rm C}=1$, unless specified.
We first consider the case in which the redshift difference is mainly tracing peculiar velocities and thus the distance between the quasars and the region considered is roughly given by the projected distance, $R_{\rm x-QSO1}=57$~kpc and $R_{\rm x-QSO2}=31$~kpc, respectively.
Following equation~\ref{eqn:SB_thick}, the sum of the contributions due to the two quasars would then give ${\rm SB}_{\rm Ly\alpha}^{\rm thick}=1.5\times10^{-16}\cgssb$, which is  about $40\times$ higher than the observed value. Even if we consider a factor of two larger distances, we would obtain an  ${\rm SB}_{\rm Ly\alpha}^{\rm thick}$ $11\times$ higher than observed.
This can be reconciled by invoking a very low covering factor ($f_{\rm C}\sim 0.02-0.09$), obscuration of the quasars in the direction of the emitting gas, or larger distances between the two quasars and the observed gas.
Low covering factors for the emitting clouds are disfavored as the emission would have looked much clumpier than observed (e.g., \citealt{fab+15a}). Conversely, with the current dataset we cannot firmly verify if the 
two quasars are strongly obscured (by e.g. dust on small scales or their host galaxy) so that only few percent of their luminosity shines on the gas. However, we obtain a crude estimate of the intrinsic extinction E(B-V) 
affecting the two quasars by fitting their spectra with a reddened version of the expected power law of the composite SDSS spectrum ($\alpha_{\rm opt}=-0.46$, \citealt{VandenBerk2001}; Section~\ref{sec:SEDs}), after normalising it to the
continuum at 8200\AA. The power law is reddened using an SMC extinction curve (\citealt{Pei1992}), in which E(B-V) is a free parameter and $R_{V}=2.93$ is fixed. For both quasars we found E(B-V)$<0.06$, indicating
that the spectra of these quasars do not show significant intrinsic extinction along our line of sight. 
Nonetheless, the lack of strong obscuration and of dust has to be directly explored with follow-up observations e.g., in the near-infrared (e.g., \citealt{Banerji2015}) and submillimeter (e.g., \citealt{Venemans2017}) regimes.
We next explore larger distances.

The uncertain redshift difference between the two quasars could reflect their distance within the Hubble flow.
In this configuration, the zone considered for our estimates would then sit at much larger distances than previously considered.
If we assume the region to be at half way between the two quasar systemic redshifts, i.e.  $R_{\rm x-QSO1}= R_{\rm x-QSO2}=1.45$~Mpc, we obtain
${\rm SB}_{\rm Ly\alpha}^{\rm thick}=1.2\times10^{-19}\cgssb$. This value is about $30\times$ smaller than the observed SB.
Considering shorter distances given by the peak of the Ly$\alpha$ emission, i.e. $R_{\rm x-QSO1}= R_{\rm x-QSO2}=0.95$~Mpc, would only increase the SB to
${{\rm SB}_{\rm Ly\alpha}^{\rm thick}=2.8\times10^{-19}\cgssb}$. To match the value at the considered position, the two quasars should lie at a distance $R_{\rm x-QSO1}= R_{\rm x-QSO2}=267$~kpc, which would translate to a very small velocity or redshift difference, i.e. $\Delta v = 223$~km~s$^{-1}$ or $\Delta z = 0.002$ (comparable to the redshift error). Even in this configuration, a fully optically thick scenario is ruled out for small distances from each quasar (if they shine on the gas). 

We then focus on the optically thin case, which has been shown to only depend on the gas physical properties (e.g., $n_{\rm H}$, $N_{\rm H}$) provided the radiation is intense enough to keep the gas sufficiently ionized to be optically thin to the Lyman continuum photons (\citealt{qpq4}):

\begin{align}
\label{eqn:SB_thin}
{\rm SB}_{\rm Ly\alpha}^{\rm thin} & = 1.8\times10^{-18}\left( \frac{1+z}{4.014}\right)^{-4}\left( \frac{f_{\rm C}}{1.0}\right) \left( \frac{n_{\rm H}}{0.24~{\rm cm^{-3}}}\right) \\ \nonumber
& \times \left( \frac{N_{\rm H}}{10^{20.5}~{\rm cm^{-2}}} \right)~{\rm erg~s^{-1}~cm^{-2}~arcsec^{-2}}. \nonumber
\end{align}

As shown in equation~\ref{eqn:SB_thin}, if we assume the median $N_{\rm H}$ value from absorption studies of 
quasar halos (log$N_{\rm H}=20.5$; \citealt{QPQ8}) and a plausible $n_{\rm H}$ for CGM gas\footnote{Because of its location, the CGM gas is expected to have densities ranging from interstellar gas densities ($n_{\rm H}\sim10^{-2}-10^4$~cm$^{-3}$; e.g., \citealt{Draine2011,KlessenGlover2016}) to IGM densities.}, 
the optically thin scenario can match the observed average ${\rm SB}_{\rm Ly\alpha}$. 
This first order calculation holds only if the two quasars are able to keep the gas ionized enough to be optically thin to the ionizing radiation.
As we demonstrate in the next section, this is not the case for distances $R\gtrsim100$~kpc, and so a fully optically thin scenario holds only if the system extent is similar to or slightly larger than the observed projected distance.

\subsection{Photoionization models for the extended Ly$\alpha$ emission}

In the following sections, we construct photoionization models assuming different configurations of the 
quasar pair to test which one is more likely given the constraints on the different extended line emissions reported in 
Section~\ref{sec:results}. Specifically, we will base our investigation on the estimates presented in the previous section, and thus focus
on three configurations: (i) the quasars sits at a separation similar to the projected distance, (ii) the quasars are within the Hubble flow with a separation of $2.9$~Mpc, and 
(iii) the quasars are placed at an intermediate distance between the two aforementioned cases.
Before describing the Cloudy calculation, we first describe the parametrization of the two quasars spectral energy distributions (SEDs), and discuss how we consider the impact of 
resonant scattering.

\subsubsection{The assumed SED for the two quasars}
\label{sec:SEDs}

\begin{figure}
\centering
\includegraphics[width=1.0\columnwidth]{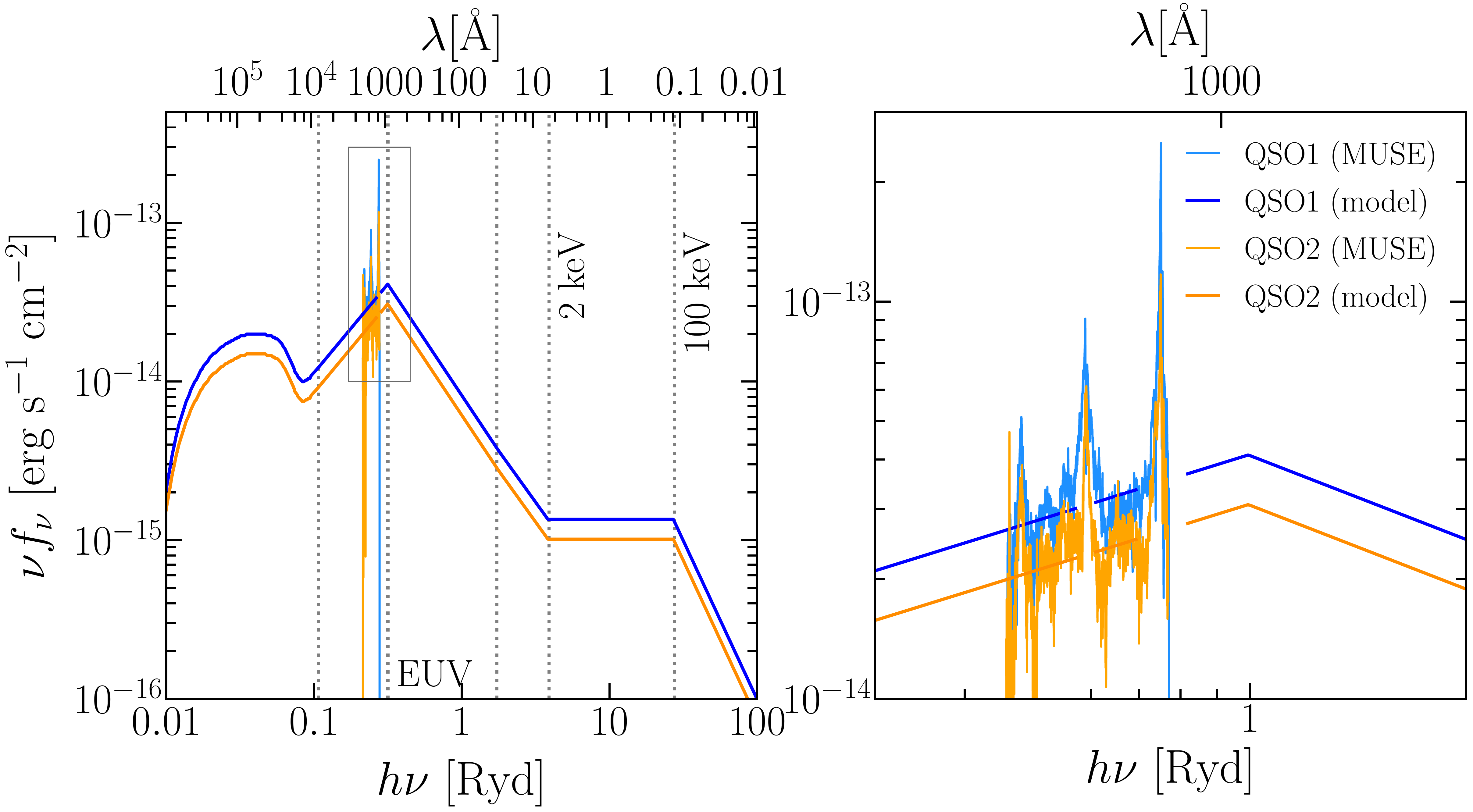}
\caption{The spectral energy distribution (SED) of QSO1 (blue) and QSO2 (orange), 
used as incident radiation in the Cloudy calculations. We compare the models with the available MUSE data (lighter color for each quasar). 
In the left panel, the vertical lines indicate the energies used to define the different power-laws (section~\ref{sec:SEDs}).
The right panel is a zoomed in version of the box highlighted in the left panel. 
The emission from the wavelength ranges around the \ion{C}{iv} and Ly$\alpha$ line locations are masked 
to prevent contributions from resonantly scattered photons, as done in \citealt{fab+15b}.}
\label{fig:SED}
\end{figure}

For the quasars' SEDs we adopt the same assumptions as in \citet{fab+15b} because we do not have complete coverage of the quasars' spectra.
The only exception to the modeling of \citet{fab+15b}, is the assumption of the simple power-laws measured by \citet{VandenBerk2001} and \citet{lusso15} in the rest-frame optical and UV, respectively.
In the mid-IR part of the SED we assume the composite spectra by \citet{Richards2006}.
In Figure~\ref{fig:SED} we show the shape of the assumed SED for both QSO1 and QSO2, together with their MUSE spectra. 
We provide below a summary of the power-laws used to parametrized {the SEDs}:

\begin{equation}
f_\nu \propto \begin{cases} \nu^{\alpha_{\rm opt}}, & \text{if } 0.11\,{\rm Ryd} \leq h\nu\leq 1 \,{\rm Ryd} \\
\nu^{\alpha_{\rm EUV}}, & \text{if } 1\,{\rm Ryd} \leq h\nu\leq 30 \,{\rm Ryd} \\ 
\nu^{\alpha}, & \text{if } 30 \, {\rm Ryd}\leq h\nu<2 \, {\rm keV} \\
\nu^{\alpha_{\rm X}}, & \text{if } 2 \, {\rm keV}\leq h\nu<100 \, {\rm keV} \\
\nu^{\alpha_{\rm HX}}, & \text{if } h\nu\geq100 \, {\rm keV},
\end{cases}
\end{equation}

\noindent where $\alpha_{\rm opt}=-0.46$ (\citealt{VandenBerk2001}),  $\alpha_{\rm EUV}=-1.7$ (\citealt{lusso15}), $\alpha=-1.65$ (i.e., obtained to match an $\alpha_{\rm OX}=-1.5$; \citealt{Strateva2005}), $\alpha_{\rm X}=-1$, and $\alpha_{\rm HX}=-2$.
These assumptions are regarded as standard in photoionization modeling of active galactic nuclei (AGN; e.g., \citealt{Baskin2014}).

\subsubsection{Approximating the impact of resonant scattering}
\label{sec:t_scatt}

Because of the large optical depth at line center (e.g., \citealt{GW96}), Ly$\alpha$ photon propagation should be affected by substantial resonant
scattering under most astrophysical conditions. Even at very close separation from a quasar, the gas can be found to be optically thick to the Ly$\alpha$
transition (i.e., $N_{\ion{H}{i}}\gtrsim10^{14}$~cm$^{-2}$; e.g., \citealt{Gallimore1999}). Hence, a Ly$\alpha$ photon typically experiences a large number of scatterings before escaping 
the system or clouds in which it starts to interact (e.g., \citealt{Neufeld_1990, Dijkstra2006}). 

However, it is usually found and assumed that the scattered Ly$\alpha$ line photons from the quasar do not 
contribute significantly to the SB$_{\rm Ly\alpha}$ surrounding quasars on large scales, i.e. $\gtrsim 100$~kpc 
(e.g., \citealt{qpq4,cantalupo14,fab+15b}). 
Indeed, the quasar's Ly$\alpha$ photons very efficiently diffuse in velocity space. 
Consequently, the vast majority of these photons escape the system at very small scales ($\lesssim 10$~kpc),
without propagating to larger distances (e.g., \citealt{Dijkstra2006, Verhamme2006}).

In this work we thus use a twofold approach. 
First, we neglect the contribution due to resonant scattering in the Cloudy calculations, such that we can 
mimic the expected negligible contribution of scattering on large scales and have ``clean'' predictions.
To achieve this, we mask the quasars' input SEDs at the Ly$\alpha$ line location as done in \citet{fab+15b} 
(Figure~\ref{fig:SED}). This method still allows us to account for the scattered Ly$\alpha$ photons arising 
from the diffuse continuum produced by the gas itself, which, however, appear to be negligible in our calculations\footnote{This contribution depends on the broadening of the line due to turbulence. 
We assume turbulent motions of 50~km~s$^{-1}$ to account for the typical equivalent widths seen for optically thick 
absorbers in quasar spectra, i.e., $\sim 1-2\AA$ (\citealt{qpq6}). Our results are not sensitive to this parameter.}. 
Second, we introduce an approximate estimate for the contribution from resonant scattering of quasar  Ly$\alpha$ photons, 
which is found to be more relevant on small scales.
To compute this estimate, we need to know: (i) the fraction of the quasar's Ly$\alpha$ photons seen 
by a parcel of gas in the nebula, or the probability that the quasar's Ly$\alpha$ photons scatters in the direction of 
a portion of the nebula, and (ii) the probability of scattering and escaping the nebula in the direction of the observer.
For each photon, both these probabilities can be written as $P= W({\rm cos(\theta)})e^{-\tau_{\rm esc}}$, 
and are thus governed by the phase function $W({\rm cos(\theta)})$ (or angular redistribution function, 
which parametrizes the probability of a photon to be scattered in a certain direction) and by the optical depth for 
the Ly$\alpha$ photons $\tau_{\rm esc}\sim N_{\rm HI}\sigma_{\alpha}(\nu, T)$, where $\sigma_{\alpha}(\nu, T)$ is the cross section for the Ly$\alpha$ scattering (e.g., \citealt{Stenflo1980, Laursen2009, Dijkstra2017}). 
For simplicity, we assume: (i) $W({\rm cos(\theta)})\sim0.5$ as it 
corresponds to the most probable value of  $cos(\theta)$, 
and (ii) a similar optical depth between quasar and nebula, and nebula and observer. 
Also, as the cross section depends on the gas motions, we assume the gas to be in
infall towards the quasars with a velocity of $200$~km~s$^{-1}$ as shown in cosmological simulations (e.g., \citealt{Goerdt2015}). We then compute the estimate for the SB$_{\rm Ly\alpha}$ due to scattering as
\begin{equation}
{\rm SB}_{\rm Ly\alpha}^{\rm scatt; QSO} =  \frac{f_{\rm conv}}{4 \pi (1+z)^{4}}R^{-2} \int_{\lambda=4865\AA}^{4888\AA} L_{\rm Lya; QSO}(\lambda) P(\lambda,N_{\rm HI}, T)^2{\rm d\lambda}
\label{eqn:scatt}
\end{equation}

\noindent where $f_{\rm conv}$ is the conversion from steradians to arcsec$^2$, $R$ is the distance from the quasar, and $L_{\rm Lya; QSO}(\lambda)$ is the observed quasar luminosity spectrum. We convolve this with the aforementioned probability to observe a quasar Ly$\alpha$ photon after scattering,  $P(\lambda, N_{\rm HI}, T)^2$, and use the observed wavelength range [4865,4888] in which we see extended Ly$\alpha$ emission (e.g., Figure~\ref{fig:Pseudoslit}). As reference, if we integrate the quasars' spectra in this range without applying the probability we get $L_{\rm Lya; QSO1}=1.17\times10^{43}$~erg~s$^{-1}$ and $L_{\rm Lya; QSO2}=6.74\times10^{42}$~erg~s$^{-1}$ for QSO1 and QSO2, respectively. We note that these luminosities are similar to the luminosity of the extended structure ($L_{\rm Ly\alpha}=3.2\times10^{42}$~erg~s$^{-1}$). The use of the observed spectrum $L_{\rm Lya; QSO}(\lambda)$ is conservative because a non-negligible fraction of the quasars' photons could have been absorbed in the system and along our line-of-sight before reaching the observer. 
We use the $N_{\rm HI}$ and $T$ of the Cloudy calculations in the formula of P.\\

This treatment is very crude and has to be regarded as an indicative reference, given that 
we use a fixed set of parameters for $\theta$, the relative gas velocity, and $P$.
Only a Monte Carlo simulation of Ly$\alpha$ radiative transfer applied to cosmological simulations of quasar pairs 
could properly handle this problem and give more detailed insights. 
However, Monte Carlo simulations of Ly$\alpha$ radiative transfer are beyond the scope of this work, 
and, in any case, none of our results should depend strongly on this effect given the large extent of the system studied.

We stress that similar considerations also apply to the resonant \ion{C}{iv} line (e.g., \citealt{Berg2019}). 
However, in this work we neglect the contribution of resonant scattering to the \ion{C}{iv} line, 
since this process should be less efficient for the \ion{C}{iv} photons than for the Ly$\alpha$,
due to the much lower abundance of metals. 
Taking into account that resonant scattering is important mainly at small distances from the quasars, 
neglecting this effect does not affect the main results of this work.

\subsubsection{Photoionization models for a single quasar}
\label{sec:sing_qso}

To have a reference for the subsequent modeling of the quasar pair, we first show the results of photoionization of gas illuminated by a single faint quasar, QSO1.
On top of the assumption for the quasar SED and for the resonant scattering already presented, we select the model parameter grid for this visualization as follow.
We assume (i) a standard plane-parallel geometry for the slab, (ii) a fixed volume density $n_{\rm H}=1,0.1,0.01$~cm$^{-3}$ whose values should encompass possible values in the quasar CGM, 
(iii) a fixed metallicity $Z=0.1\, Z_{\odot}$ close to the value seen in absorption studies around $z\sim2$ quasars ($\sim 0.3\, Z_{\odot}$; \citealt{QPQ8}), and (iv)
we stop our calculations when a total Hydrogen column density $N_{\rm H}=10^{20.5}$~cm$^{-2}$ is reached. This value is the median $N_{\rm H}$ estimated for absorbers around $z\sim2$ quasars out to an impact parameter of $300$~kpc (\citealt{QPQ8}).
We then place the slab of gas at increasing distance from the quasar to show how this would affect the predicted Ly$\alpha$ emission. Specifically, we place the slab at 30 different distances spaced in logarithmic bins between 20 and 1500~kpc.

\begin{figure}
\centering
\includegraphics[width=0.98\columnwidth]{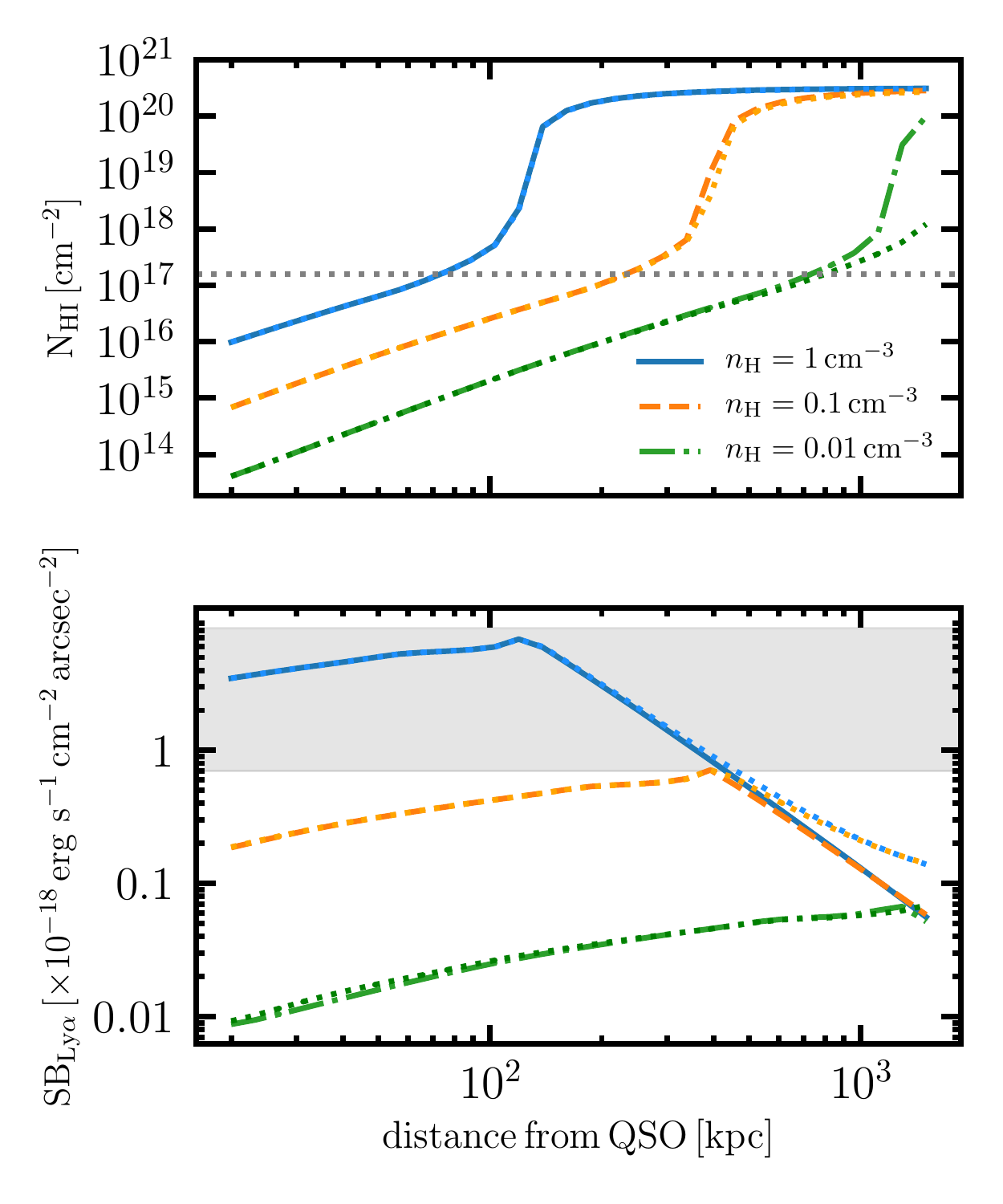}
\caption{Cloudy predictions for plane parallel slabs with log$(N_{\rm H}/{\rm cm^{-2}})=20.5$, 
photoionized by a single quasar with the characteristic SED of QSO1.
The slabs -- characterized by uniform $n_{\rm H}$ -- are placed at increasing distance from the quasar. 
{\it Top:} column density of \ion{H}{i} as a function of distance from the quasar. The horizontal dotted line indicates the threshold 
between the optically thin and thick regimes. For $n_{\rm H}=0.1$~cm$^{-3}$ the quasar is able to keep the gas ionized 
up to $\sim 200$~kpc.
{\it Bottom:} predicted SB$_{\rm Ly\alpha}$ as a function of distance from the quasar. The dotted lines indicate Cloudy models which 
take into account the presence of the $z=3$ UVB by \citet{hm12}. The gray shaded region shows the range of observed SB$_{\rm Ly\alpha}$.}
\label{fig:Cloudy_QSO1}
\end{figure}

Figure~\ref{fig:Cloudy_QSO1} shows the results of this calculation for the $N_{\rm HI}$ (top panel) and the SB$_{\rm Ly\alpha}$ (lower panel) as a function of distance from QSO1. 
The two regimes described in Section~\ref{sec:qpq4}, optically thin and optically thick to the ionizing radiation, are readily evident (the dotted gray line in the top panel indicates $N_{\rm HI}=10^{17.2}$~cm$^{-2}$).
A slab can be optically thin further away from the quasar than a denser slab, following the relation 

\begin{equation}
R_{\rm n_{\rm H}^{\rm smaller}}=\sqrt{n_{\rm H}^{\rm larger}/n_{\rm H}^{\rm smaller}}R_{\rm n_{\rm H}^{\rm larger}}.
\label{eq_R}
\end{equation}

This can be easily obtained by comparing the number of ionizing photons at the two different distances or, in other words, by finding at which distance the ionization parameter 
U\footnote{The ionization parameter is defined to be the ratio of the number density of ionizing photons to hydrogen atoms, 
$U \equiv \Phi_{\rm LL}/cn_{\rm H}$. The number of ionizing photons depends on the distance from the ionizing source as $\Phi_{\rm LL}\propto R^{-2}$.} is the same for models with different densities. 
We note that the $N_{\rm HI}$ saturates to the total gas content on short distances after the models transition 
from optically thin to optically thick.

\begin{figure*}
\centering
\includegraphics[width=0.95\textwidth]{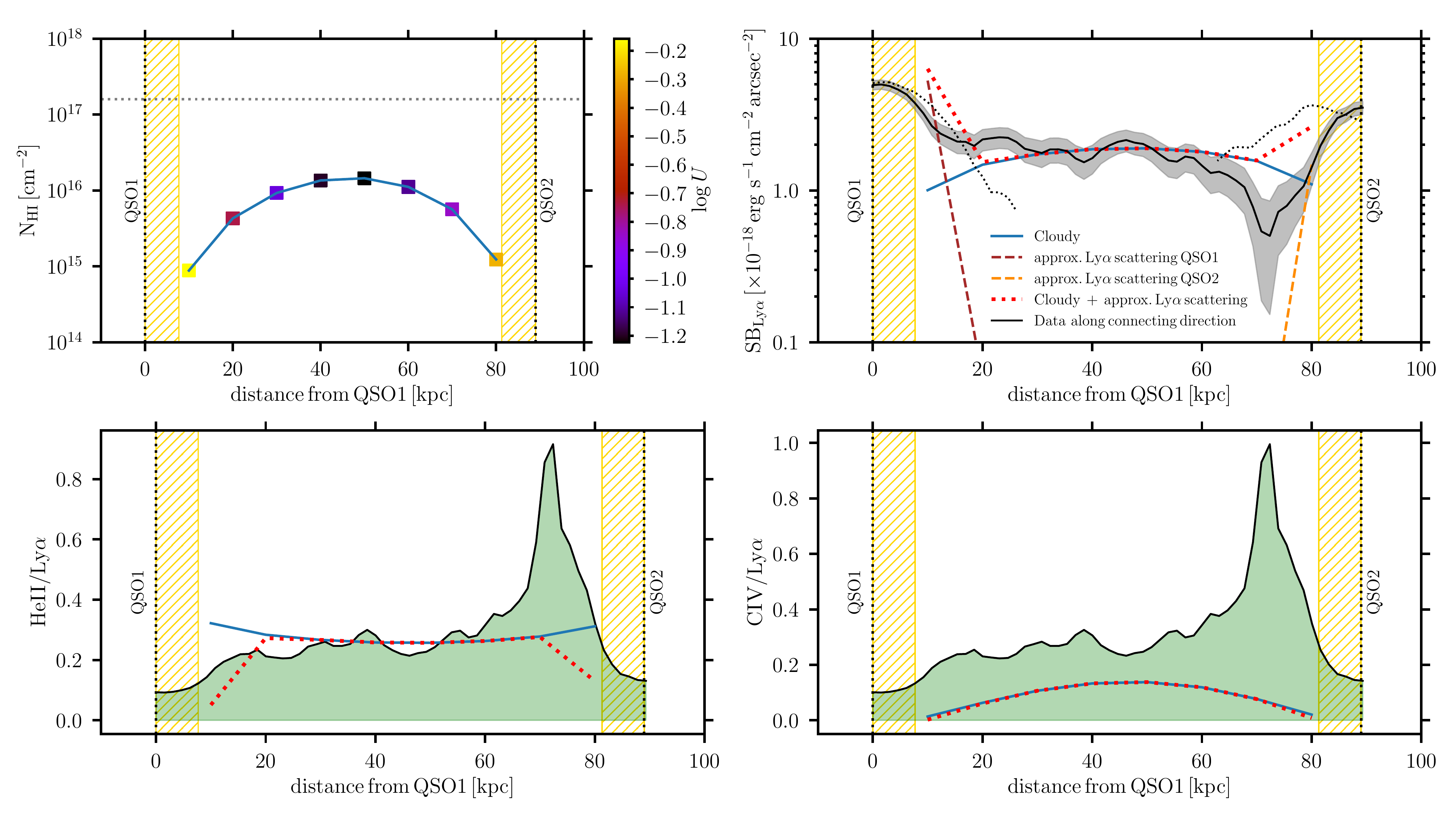}
\caption{Cloudy predictions for plane parallel slabs with total 
log$(N_{\rm H}/{\rm cm^{-2}})=20.5$ and $n_{\rm H}=0.5$~cm$^{-3}$, illuminated by the quasar pair QSO1 and QSO2 placed 
at a separation equal to their observed projected distance (89~kpc).
{\it Top left:} column density of \ion{H}{i}, N$_{\ion{H}{i}}$, as a function of distance. 
For each model the data-points are color-coded by their respective ionization parameter.
In this scenario all the models are optically thin to the ionizing radiation, i.e. N$_{\ion{H}{i}}<10^{17.2}$~cm$^{-2}$.
{\it Top right:} comparison of the observed (black line with shaded $1\sigma$ error) and predicted SB$_{\rm Ly\alpha}$ (blue line). 
The brown and dashed lines indicate the contribution due to scattering of Ly$\alpha$ photons from the quasars, 
as explained in Section~\ref{sec:t_scatt}. The red dotted line indicates the total SB$_{\rm Ly\alpha}$
summing up the Cloudy prediction and the scattering contribution. 
The thin black dotted lines are the observed SB$_{\rm Ly\alpha}$ along the directions  NE for QSO1 and SE for QSO2, at angles 52 and 142 
degrees East from North, respectively (details in Section~\ref{sec:proj_dist}).
{\it Bottom left:} comparison of the observed (black line is the $2\sigma$ upper limit) and the predicted (blue line) \ion{He}{ii}/Ly$\alpha$ ratio as function of the distance from the quasars.
The green shaded area represents the parameter space allowed by the observations. 
{\it Bottom right:} comparison of the observed (black line is the $2\sigma$ upper limit) and the predicted (blue line) \ion{C}{iv}/Ly$\alpha$ ratio as function of the distance from the quasars.
The green shaded area represents the parameter space allowed by the observations. 
The vertical dotted lines in each panel indicate the position of the two quasars, while the striped yellow regions 
represent the zones used to normalize the quasar PSF, characterized by large uncertainties and, therefore, not considered in the analysis. 
The dotted red line represents the ratio corrected for the presence of Ly$\alpha$ scattering.} 
\label{fig:Cloudy_projDist}
\end{figure*}

The prediction for the optically thin regime does not follow exactly the aforementioned relation SB$_{\rm Ly\alpha}\propto N_{\rm H} n_{\rm H}$ as Cloudy takes into account both temperature changes of the recombination coefficients and the contribution to the Ly$\alpha$ emission from cooling. 
Both these phenomenon increase with distance from the quasar as the temperature drops 
increasing the recombination efficiency (e.g., \citealt{StoreyHummer1995}) and 
collisional coefficients ($T\sim10^{4.2}$~K; e.g., \citealt{Raymond1976,Wiersma2009}). 
In the optically thick regime, SB$_{\rm Ly\alpha}\propto L_{\nu_{\rm LL}}$, scaling with the
distance following $R^{-2}$, as expected. For $>100$~kpc, the presence of additional ionizing photons from the metagalactic ultraviolet background (UVB; e.g., \citealt{hm12}) introduces mild differences in the predicted SB$_{\rm Ly\alpha}$, and very slight changes in the ionized fraction. This is illustrated by the deviation from the predicted  $R^{-2}$ relation towards higher SB$_{\rm Ly\alpha}$ of the dotted curves, 
which show the Cloudy models run with the UVB from \citet{hm12} at $z=3$. 
We do not show the scattering contribution here since it seems irrelevant at these distances 
(e.g., the dashed lines in Figure~\ref{fig:Cloudy_projDist}).
From Figure~\ref{fig:Cloudy_QSO1}, it is already clear that relatively dense gas ($n_{\rm H}>0.1$~cm$^{-3}$) is needed to produce the high levels of SB$_{\rm Ly\alpha}$ detected around 
the observed quasar pair in the short exposures with MUSE. 
\citet{heckman91a,cantalupo14, hennawi+15, fab+15b} have already shown that such dense gas is needed to explain the emission 
around individual quasars.

\subsubsection{Photoionization models for a quasar pair at the observed projected distance}
\label{sec:proj_dist}

In this section we present the modeling of a photoionization scenario in which the two quasars sit at a separation
similar to the observed projected distance (89~kpc), and both illuminate the gas responsible for the extended Ly$\alpha$ emission. In this framework, the two quasars are likely in a merger phase which would explain the observed velocity shift 
of the Ly$\alpha$ emission and the uncertain difference in velocities between the quasar systemics. 
Large peculiar velocities are thus in play.

Our model grid covers distances [10,80] kpc from each quasar in steps of 10~kpc. 
This is achieved by normalizing each quasar spectrum at different values of $f_{\nu_{\rm LL}}$ 
depending on its distance from the slab.
To be conservative, we do not consider distances smaller than 10 kpc, because of uncertainties due to the quasars' PSF subtraction, and because in such close 
proximity to the quasars we expect density variations and effects due to, e.g., the interstellar medium of the host galaxy.
For simplicity, we assume (i) a plane parallel geometry, (ii) a fixed volume density $n_{\rm H}=0.5$~cm$^{-3}$, (iii) $Z=0.1\, Z_{\odot}$, and (iv) we stop our calculations when $N_{\rm H}=10^{20.5}$~cm$^{-2}$ is reached.

In Figure~\ref{fig:Cloudy_projDist} we show the prediction of this set of models in terms of:
$N_{\rm HI}$ (top-left), SB$_{\rm Ly\alpha}$  (top-right), and line ratios \ion{He}{ii}/Ly$\alpha$ (bottom-left) 
and \ion{C}{iv}/Ly$\alpha$ (bottom-right). For the observational data, we
extract the average Ly$\alpha$ emission along the direction connecting the two quasars, using a slit with width $2\times$ the seeing of our observations (solid black line).
It is important to note that, of course, in proximity of the quasars there are variations in the SB$_{\rm Ly\alpha}$ depending on the direction along which we place the slit. 
To appreciate the difference in Ly$\alpha$ profiles along different directions close to the two quasars, we show how the SB$_{\rm Ly\alpha}$
behaves along the NE direction from QSO1 and the SE from QSO2 at angles of 52 and 142 degrees East of 
North (black dotted lines in top-right plot). These two directions have been chosen because they are perpendicular 
and parallel, respectively, to the direction connecting the two quasars (142 degrees East of North).
For the ratios, we divide the $2\sigma$ SB limits per layer at the \ion{He}{ii} and \ion{C}{iv} locations (Section~\ref{sec:em_res}) by the aforementioned SB$_{\rm Ly\alpha}$ within the two quasars. The allowed parameter space is 
indicated by the green shaded region. 

\begin{figure*}
\centering
\includegraphics[width=0.95\textwidth]{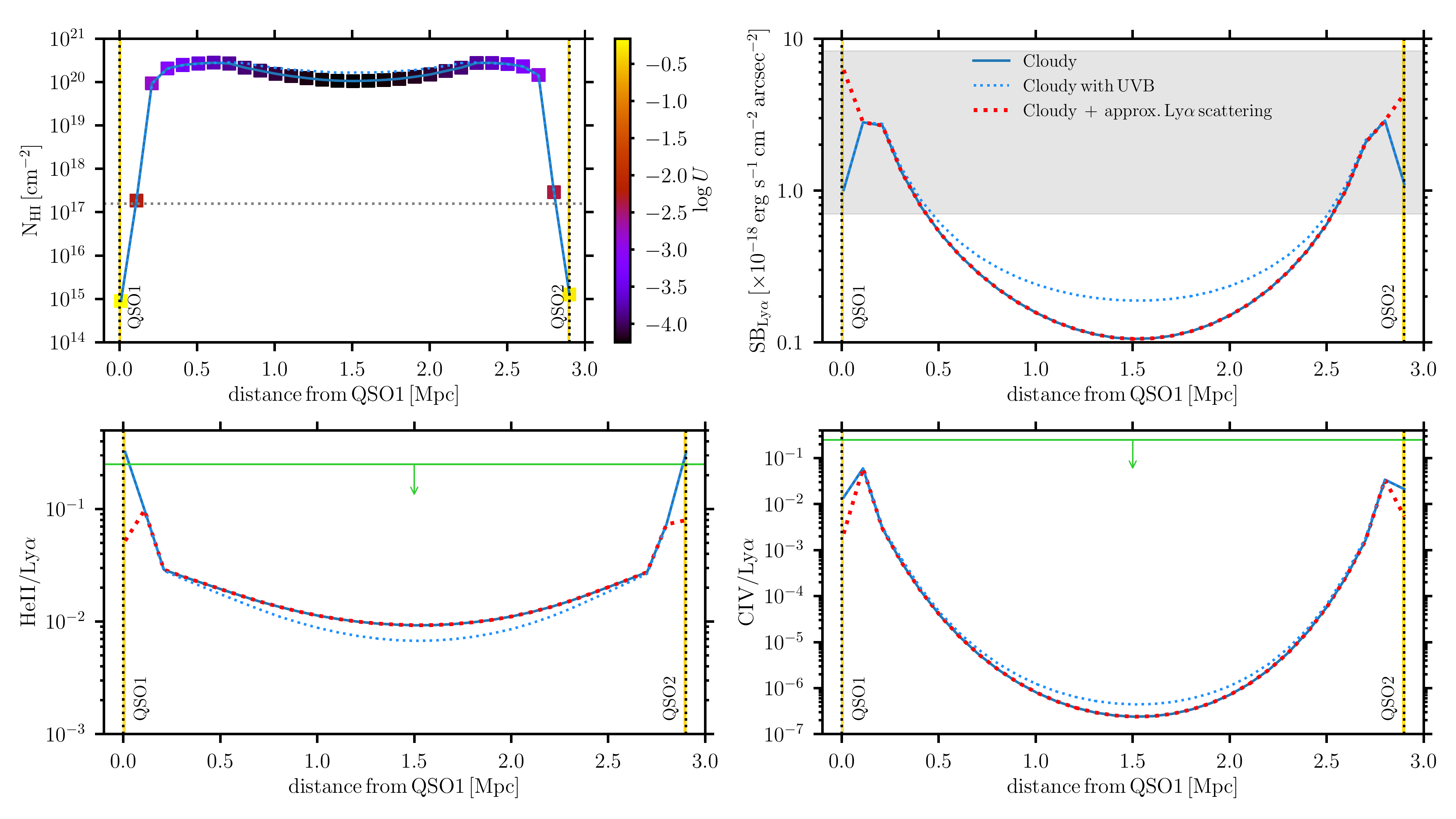}
\caption{Cloudy prediction for plane parallel slabs with total log$(N_{\rm H}/{\rm cm^{-2}})=20.5$ 
and $n_{\rm H}=0.5$~cm$^{-3}$, illuminated by the quasar pair QSO1 and QSO2, placed at a separation of 2.9~Mpc, 
as derived from their systemic redshifts.
{\it Top left:} column density of \ion{H}{i}, N$_{\ion{H}{i}}$, as a function of distance. 
The data-points are color-coded by the ionization parameter.
In this scenario most of the models are optically thick to the ionizing radiation, 
i.e. N$_{\ion{H}{i}}\gg10^{17.2}$~cm$^{-2}$, with only the CGM regions of the two quasars characterized by optically thin gas.
{\it Top right:}  Predicted SB$_{\rm Ly\alpha}$ for the Cloudy models without (solid blue line) and with (dotted blue line) the UVB. 
The red dotted line indicates the total SB$_{\rm Ly\alpha}$, summing up the Cloudy prediction and the scattering contribution 
estimated following Section~\ref{sec:t_scatt}. The gray shaded region shows the range of observed SB$_{\rm Ly\alpha}$.
{\it Bottom left:} Predicted \ion{He}{ii}/Ly$\alpha$ ratio as function of the distance from the quasars without 
(solid blue line) and with (dotted blue line) the UVB.
{\it Bottom right:} Predicted \ion{C}{iv}/Ly$\alpha$ ratio as function the distance from the quasars without 
(solid blue line) and with (dotted blue line) the UVB.
In each of the bottom panels, the dotted red lines represent the ratios corrected for the presence of Ly$\alpha$
scattering as modeled in Section~\ref{sec:t_scatt}, while the green horizontal line indicates the local $2\sigma$ upper limit on each ratio (Figure~\ref{fig:Cloudy_projDist}).
In all four panels, the vertical dotted lines indicate the position of the two quasars, while the striped yellow 
regions show the zones used to normalize the quasars PSFs.}
\label{fig:Cloudy_HubDist}
\end{figure*}

Figure~\ref{fig:Cloudy_projDist} shows that our model grid can reproduce the roughly flat SB$_{\rm Ly\alpha}$ 
and the line ratios observed, with the \ion{He}{ii}/Ly$\alpha$ ratio possibly starting to show some tension 
with our simple modeling.
In this configuration, the gas emitting Ly$\alpha$ emission is highly ionized and thus optically thin to the ionizing radiation. This is due to the relatively high ionization parameter U at each location, log$U> -1.2$. 
The presence of Ly$\alpha$ resonant scattering appears to be non-negligible on scales $R<20$~kpc from the quasar, 
and could help in explaining the low \ion{He}{ii}/Ly$\alpha$ ratios observed at these locations. 
Finally, we stress that the assumption of a constant $N_{\rm H}$ value along all of the emitting bridge, implies a 
total mass of cool ($T\sim10^4$~K) gas of $M_{\rm cool}=f_{\rm C} A N_{\rm H} m_{\rm p}/X=3.9\times10^{10}$~M$_{\odot}$, 
where $A$ is the area covered by the bridge, $m_{\rm p}$ is the proton mass, and $X=0.76$ is the Hydrogen mass fraction (\citealt{qpq4}). 
If the two quasars are hosted by a halo of $M=10^{12.5}$~M$_{\odot}$ (average halo hosting quasars at these redshifts; \citealt{white12}), the detected cool gas mass would represent 9.4\% 
of the total gas mass within the halo, after
removing the mass expected to be in stars, $M_{*}=(8.3\pm2.8)\times10^{10}$~M$_{\odot}$ (\citealt{Moster2018}). Once taken into account that our observations are not sensitive to very diffuse gas ($n_{\rm H}\sim10^{-2}$~cm$^{-3}$), 
this estimate seems surprisingly close to the 
fraction of cool gas seen in similar massive halos in current cosmological simulations (15\%; e.g., \citealt{cantalupo14, FAB2018}).

\subsubsection{Photoionization models for a quasar pair within the Hubble flow}
\label{sec:hub_dist}

Now, we assume a photoionization scenario in which the two quasars sit at their systemic redshifts, 
and thus at a distance of $2.9$~Mpc.
As this stretched configuration would place the filamentary emission along our line of sight, 
we again assume that both quasars illuminate the gas responsible for the observed extended Ly$\alpha$ emission.
Because of the finite speed of light, this assumption requires that QSO2 has been active for at least $18.9$~Myr, while QSO1 for $9.4$~Myr. These values seem reasonable given the current estimates for quasars' lifetimes
(e.g., \citealt{Martini2004, Eilers2017, Schmidt2018, Khrykin2019})\footnote{Considering
the faint luminosity of the two quasars, a shorter quasar lifetime will only affect strongly 
distances of $\sim100$~kpc or smaller (Figure~\ref{fig:Cloudy_QSO1}), i.e. the extent of the highly ionized region will be accordingly smaller.}.
In this framework, the two quasar halos are not yet strongly interacting and the velocity shift of the Ly$\alpha$ line (Figure~\ref{fig:SB_Vel_map})
would be a mixture of complex radiative transfer effects and velocities tracing the Hubble flow.

Our model grid covers distances [10, 2890] kpc from each quasar in steps of about 100~kpc. As in Section~\ref{sec:proj_dist}, this is achieved by normalizing each quasar spectrum at different 
values of $f_{\nu_{\rm LL}}$ depending on its distance from the slab.
For simplicity, we assume (i) a plane parallel geometry, (ii) a fixed volume density $n_{\rm H}=0.5$~cm$^{-3}$, (iii) $Z=0.1\, Z_{\odot}$, and (iv) we stop our calculations when $N_{\rm H}=10^{20.5}$~cm$^{-2}$ is reached.
These $N_{\rm H}$, $n_{\rm H}$, and $Z$ are likely too high for the average cloud in the IGM (e.g., \citealt{meiksin09}), but here we are interested in conservatively high values which should produce the highest signal observable as the models remain
optically thick (Section~\ref{sec:qpq4} and \citealt{qpq4}). As we explore large distances from the quasars, 
we also run models with the UVB of \citet{hm12}.

In Figure~\ref{fig:Cloudy_HubDist} we show the predictions of this set of models 
in the same observables as in Figure~\ref{fig:Cloudy_projDist}.
We plot the predictions of the ``clean'' Cloudy models (blue solid line), the Cloudy models with the UVB as additional source (blue dotted line), and the Cloudy models plus our approximated contribution of the Ly$\alpha$ scattering (red dotted line). 
As expected from the single source model presented in Section~\ref{sec:sing_qso}, 
the Ly$\alpha$ emission shows its maximum levels SB$_{\rm Ly\alpha}\approx3\times10^{-18}\cgssb$ at the transition between 
the optically thin and thick regimes ($R\sim100$~kpc from each quasar), 
with the expected decline as $R^{-2}$ in the optically thick regime, reaching the minimum (SB$_{\rm Ly\alpha}\approx10^{-19}\cgssb$) at the half distance between the quasars. The contribution of ionizing photons from the UVB almost precisely doubles the Ly$\alpha$ emission at this location. In this ``Hubble-flow'' scenario, the \ion{He}{ii} and \ion{C}{iv} line emissions will be extremely faint and already at the limit of current facilities capabilities for close 
separations ($R\sim100$~kpc) from each quasar. In this regard, it is interesting to note that our approximate treatment of scattering creates a region with peak \ion{He}{ii}/Ly$\alpha$ at a distance of about $100$~kpc from the quasar. 
This effect remains to be verified with detailed radiative transfer simulations.
Also, the ratio \ion{C}{iv}/Ly$\alpha$ peaks at the same location as \ion{He}{ii}/Ly$\alpha$. 
Its trend, however, is not mainly driven by the assumption on the Ly$\alpha$ scattering, 
but by the higher excitation of Carbon on smaller scales.

In this configuration, it is difficult to directly compare our photoionization models with the observations as complex projection effects can drastically change the predicted curves. For this reason, we do not attempt to plot our data in Figure~\ref{fig:Cloudy_HubDist}, but we only show the observed range of SB$_{\rm Ly\alpha}$ and the local $2\sigma$ upper limit on the ratios \ion{He}{ii}/Ly$\alpha$ and \ion{C}{iv}/Ly$\alpha$. 
Nevertheless, from our models, it is clear that the observed emission would be dominated by gas at small distances from the two quasars, i.e. in their CGM. In this scenario, we would thus expect to see two nebulae sitting at the systemic redshift of the quasars, and thus to find at least some overlapping emission showing double peaks, with each peak sitting at the systemic of the two quasars or at the redshift of the Ly$\alpha$ peak of the two quasars. 
We inspect our data for such a signature, finding a signal at the systemic of QSO1 only in close proximity to its location ($2\arcsec$ or 10 projected kpc) and along the direction connecting the two quasars (within box 2), as shown by the blue line in Figure~\ref{fig:Pseudoslit}. 
This signature is very concentrated spatially ($<1\arcsec$), and for this reason we suspect that it is due 
to a compact object.
Also, there is tentative evidence for  a double peak in close proximity of QSO2 (black line in Figure~\ref{fig:Pseudoslit}). This double peak is also extremely localized and could be due to radiative transfer effects at this location. 
We thus conclude that there are no obvious signatures of a superposition of two nebulae at different redshifts.

Finally, we stress that, in this framework, the direction of the discovered bridges of Ly$\alpha$, stretching between the two quasars, would be due to a very improbable chance alignment of dense structures in the two distinct CGMs. 
This alignment is quite unlikely also because of the absence of additional extended emission in 
other directions. We thus argue that this scenario is not able to reproduce the observations.

\subsubsection{Photoionization models for a quasar pair at an intermediate distance}
\label{sec:int_dist}

\begin{figure*}
\centering
\includegraphics[width=0.95\textwidth]{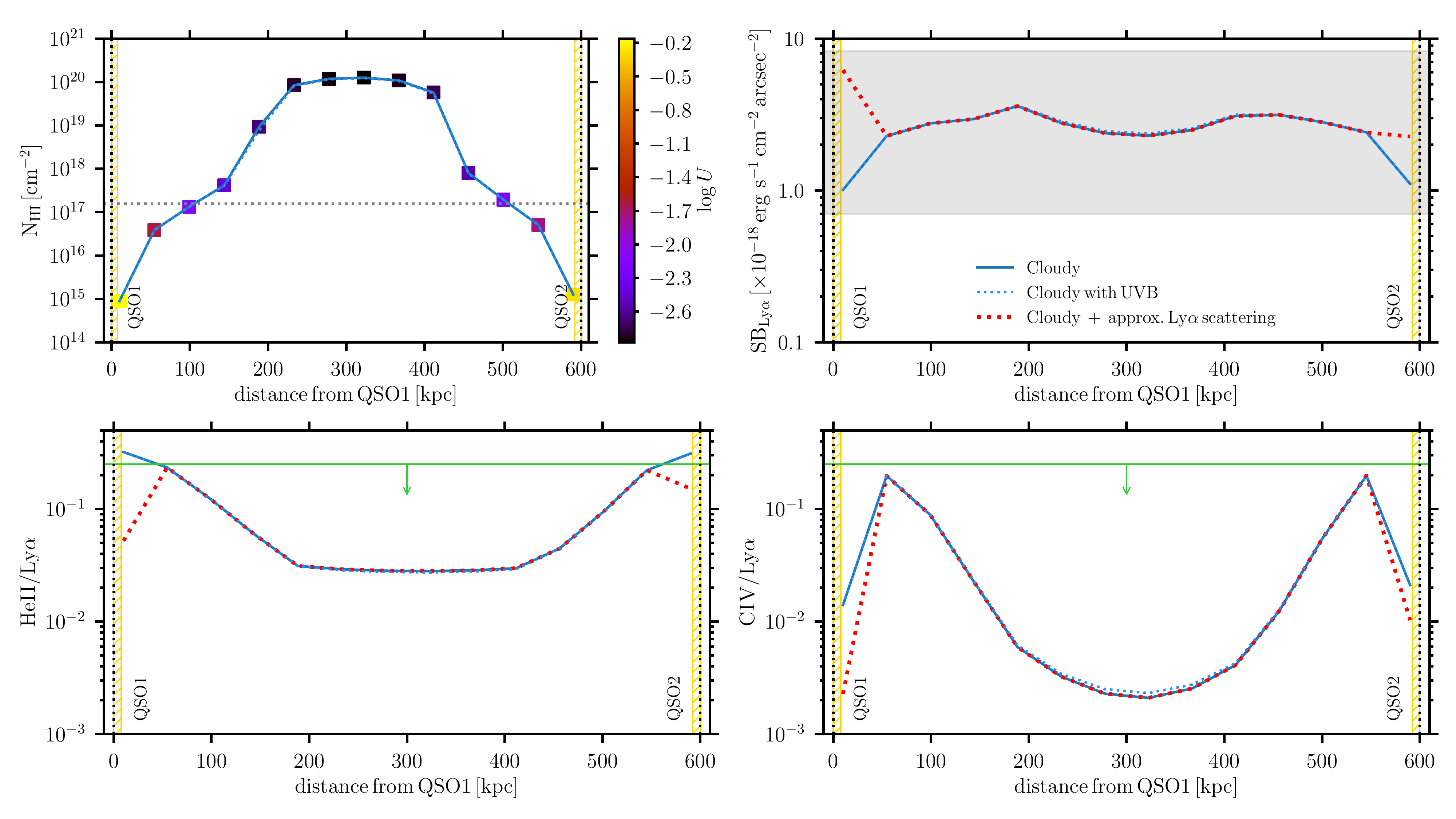}
\caption{Cloudy prediction for plane parallel slabs with total log$(N_{\rm H}/{\rm cm^{-2}})=20.5$ 
and $n_{\rm H}=0.5$~cm$^{-3}$, illuminated by the quasar pair QSO1 and QSO2, placed at an intermediate separation of 600~kpc.
{\it Top left:} column density of \ion{H}{i}, N$_{\ion{H}{i}}$, as a function of distance, color-coded by ionization parameter.
In this scenario, a central region of about 300~kpc is optically thick to the ionizing radiation, 
i.e. N$_{\ion{H}{i}}\gg10^{17.2}$~cm$^{-2}$, while the CGM regions of the two quasars are optically thin.
{\it Top right:}  Predicted SB$_{\rm Ly\alpha}$ for the Cloudy models without (solid blue line) and with (dotted blue line) the UVB. 
The red dotted line indicates the total SB$_{\rm Ly\alpha}$, summing up the Cloudy prediction and the scattering contribution
estimated following Section~\ref{sec:t_scatt}. The predicted SB$_{\rm Ly\alpha}$ is basically flat at a value of SB$_{\rm Ly\alpha}=2.5\times10^{-18}\cgssb$, well
within the observed range (gray shaded region).
{\it Bottom left:} Predicted \ion{He}{ii}/Ly$\alpha$ ratio as function of the distance from the quasars without 
(solid blue line) and with (dotted blue line) the UVB.
{\it Bottom right:} Predicted \ion{C}{iv}/Ly$\alpha$ ratio as function of the distance from the quasars without 
(solid blue line) and with (dotted blue line) the UVB.
In each of the bottom panels, the dotted red lines represent the ratios corrected for the presence of Ly$\alpha$
scattering as modeled in Section~\ref{sec:t_scatt}, while the green horizontal line indicates the local $2\sigma$ upper limit on each ratio (Figure~\ref{fig:Cloudy_projDist}).
The vertical dotted lines in all four panels indicate the position of the two quasars, while the striped yellow 
regions give the zones used to normalize the quasars PSFs.}
\label{fig:Cloudy_intDist}
\end{figure*}

As already discussed in Section~\ref{sec:qpq4}, an interesting configuration places the two quasars at an intermediate distance with respect to the two extremes considered so far.  Specifically, we consider a distance of 600~kpc. 
Indeed, our analytical estimates suggest that optically thick models would be able to reproduce the observed
emission if the two quasars sit at $\sim300$~kpc from the central region of the observed bridges.
As this configuration also stretches considerably the bridges along our line of sight, 
we assume that both quasars shine on the gas.
In this framework the two quasar halos are probably approaching and the velocity shift of the Ly$\alpha$ line 
(Figure~\ref{fig:SB_Vel_map}) should be interpreted as a mixture of complex radiative transfer effects, 
velocities tracing the approaching quasar halos, and extent of the structure along our line of sight.

Our model grid covers distances [10, 590] kpc from each quasar in steps of about 50~kpc. 
We make the same assumptions as in Sections~\ref{sec:proj_dist} and \ref{sec:hub_dist}, 
when computing this grid of models.
We assume (i) a plane parallel geometry, (ii) a fixed volume density $n_{\rm H}=0.5$~cm$^{-3}$, (iii) $Z=0.1\, Z_{\odot}$, and (iv) stop the calculations when $N_{\rm H}=10^{20.5}$~cm$^{-2}$ is reached.
These $N_{\rm H}$, and $Z$ are likely too high for the average cloud in the IGM (e.g., \citealt{meiksin09}), 
but they represent well the properties of absorbing gas seen around high-$z$ quasars 
(e.g., \citealt{QPQ8}). Following the modeling of a single quasar (Section~\ref{sec:sing_qso}), the $n_{\rm H}$ is chosen
large enough to allow for a match of the observed SB$_{\rm Ly\alpha}$. 

We show the predictions of this set of models for $N_{\rm HI}$ (top-left), SB$_{\rm Ly\alpha}$ (top-right), \ion{He}{ii}/Ly$\alpha$ (bottom-left) and \ion{C}{iv}/Ly$\alpha$ (bottom-right) in Figure~\ref{fig:Cloudy_intDist}. 
The color scheme is the same as in Figure~\ref{fig:Cloudy_HubDist}.
As expected from the analytical modeling and from the single source calculation 
in Section~\ref{sec:sing_qso}, the Ly$\alpha$ emission is predicted to be roughly at the same level of 
SB$_{\rm Ly\alpha}\approx2.5\times10^{-18}\cgssb$ throughout all the extent of the bridges. 
This happens even though the model transitions between the optically thin and thick regimes 
at around $R\sim100$~kpc from each quasar. 
At small distances ($R\lesssim50$~kpc), 
our calculation shows that the contribution from scattering could be important.
Furthermore, in this scenario, the contribution of ionizing photons from the UVB of \citet{hm12}, 
is irrelevant (all three curves fall on top of each other for distances larger than 50~kpc).
As already seen in the ``Hubble-flow'' scenario (Section~\ref{sec:hub_dist}), the  \ion{He}{ii} and \ion{C}{iv} line emissions 
are extremely faint, and basically barely observable with current instruments. 
Interestingly, our approximate treatment of scattering creates also in this scenario
a region with peak \ion{He}{ii}/Ly$\alpha$ at a distance of $50-100$~kpc from the quasars. 
Detailed radiative transfer simulations will be able to verify this effect.
The \ion{C}{iv}/Ly$\alpha$ line ratio peaks also at the same location as the \ion{He}{ii}/Ly$\alpha$ one. 
As we discussed in the previous scenarios, this trend is driven by the higher excitation of Carbon on smaller scales, 
and not by the assumption on the Ly$\alpha$ scattering.

As for the ``Hubble-flow'' scenario (Section~\ref{sec:hub_dist}), it is difficult to compare our photoionization models 
with the observations, as complex projection effects (e.g., absorption from the structure itself for both emitted and 
impinging radiation) can drastically change the predicted curves.
For this reason, we do not plot our data in Figure~\ref{fig:Cloudy_HubDist}, but show the available information as done in Figure~\ref{fig:Cloudy_HubDist}. 
Nevertheless, from our models, it is clear that the observed flux can be equally due to emission 
from dense CGM and IGM surrounding the two quasars, with the central region of the bridge possibly 
characterized by optically thick gas.
If this is the case, we would expect to see differences in the Ly$\alpha$ line shape as we move along the bridge, 
with the presence of double peaks or strong asymmetries (e.g., \citealt{Neufeld_1990, Laursen2009}) in its central 
region.
We cannot exclude the presence of these features below the current MUSE spectral resolution 
(FWHM$\approx2.85$~\AA ~or 175~km~s$^{-1}$ at $4870$~\AA).
Deep observations at higher spectral resolution with available IFUs, e.g., MEGARA (\citealt{GildePaz2016}) or KCWI (\citealt{Morrissey2012}), or longslit spectroscopy could 
help to clarify the shape of the Ly$\alpha$ emission,
and assess if optically thick gas is present in this structure.

Finally, we can calculate a rough estimate for the gas mass in this extended structure by assuming 
a cylindrical geometry for each bridge with extent 600~kpc and diameter 35~kpc.
In this case, we can compute the total cool gas mass as 
$M_{\rm cool}=V f_{\rm V} (n_{\rm H}/0.5~{\rm cm^{-3}}) m_{\rm p}/X=8.8\times10^{12} (f_{\rm V}/1.0)$~M$_{\odot}$,
where $V$ is the volume covered by one of the bridges, $m_{\rm p}$ is the proton mass, $X=0.76$ is the Hydrogen mass fraction, and $f_{\rm V}$ is the volume filling factor (e.g. \citealt{qpq4}). 
Multiplying by the number of bridges, we thus obtain a total cool gas mass of $M_{\rm cool}=1.8\times10^{13} (f_{\rm V}/1.0)$~M$_{\odot}$.
As the parcels with high densities $n_{\rm H}=0.5$~cm$^{-3}$
are only the tracer of the structure, i.e. the volume filling factor of parcels with $n_{\rm H}=0.5$~cm$^{-3}$ 
is expected to be much lower than unity for gas on such large scales\footnote{On CGM scales (100~kpc)  the considered densities would imply $f_{\rm V}\sim10^{-2}$ (e.g., \citealt{qpq4}).}, 
this estimate has to be regarded as an upper limit for the total cool gas mass along the observed structure.
Confirming the presence of high densities $n_{\rm H}=0.5$~cm$^{-3}$ within the IGM would imply finding parcels of gas similar to interstellar medium densities 
spread out along filaments. This scenario sounds plausible if we are tracing emission close to faint undetected galaxies, but quite unrealistic at the moment
for ``pure'' IGM ($n_{\rm H}\lesssim10^{-2}$~cm$^{-2}$; e.g. \citealt{meiksin09}).

\section{Modeling the absorbers}
\label{sec:model_abs}

In Section~\ref{sec:abs} we presented the observed properties of absorbers ABS1, ABS2, and ABS3, 
while in this section we discuss their nature in different system configurations.

\subsubsection{ABS1: a metal enriched CGM or IGM absorber}
\label{sec:ABS1_cloudy}

The redshift of ABS1 suggests a link with QSO2. However, peculiar motions could mimic such an association, 
and ABS1 could be related to QSO1, QSO2 or be in the IGM.
Furthermore, the two similar components seen at the \ion{C}{iv} line could be due to different 
structures along the line of sight (Figure~\ref{fig:ABSQSO1}). We constrain the nature of ABS1 by constructing 
photoionization models with Cloudy under three different system configurations. We briefly outline here the models and the results, while we present them in detail in
Appendix~\ref{sec:model_abs_ABS1}.

The photoionization models need to match the observed column densities reported in Table~\ref{Tab:ABS}, 
and the Ly$\alpha$ emission at the location of the absorber, i.e. QSO1. 
The joint constraints from absorption and emission are key in assessing the physical properties of the gas (e.g., $n_{\rm H}$), 
and thus its configuration (\citealt{hennawi+15}). Unfortunately, as our PSF subtraction algorithm is not reliable within the $1$~arcsec$^2$ 
region around QSO1, we can only assume conservative limits for the Ly$\alpha$ emission, i.e. 
below the $5\sigma$ value per channel, which is equivalent to a SB$_{\rm Ly\alpha}$ below $1.75\times10^{-18}\cgssb$. The same applies also to QSO2.
As we show in Appendix~\ref{sec:model_abs_ABS1}, the loose constraint on the Ly$\alpha$ emission does not allow a firm evaluation of the absorber's location.

Specifically, the three system configurations here considered are as follows: (i) ABS1 is only illuminated by the QSO1's radiation, (ii) 
ABS1 sees the radiation from both QSO1 and QSO2, and the two quasars lies at a separation similar to the 
observed projected distance, and (iii) ABS1 is illuminated by both quasars, with QSO1 and QSO2 separated by 600~kpc.
For all the models we consider the presence of the UVB as additional ionizing source.
Importantly, we focus on these three configurations as they are allowed by the modeling of the extended Ly$\alpha$ emission shown in Section~\ref{sec:pow_Lya}.

As explained in detail in Appendix~\ref{sec:model_abs_ABS1}, in all the three configurations we find that ABS1 is a cool ($4.1\lesssim {\rm log}(T/{\rm K}) \lesssim4.4$), metal enriched ($Z>0.3~Z_{\odot}$) absorber, 
located on CGM or IGM scales around the quasar pair. The relatively high metallicity is constrain by the presence of the strong $N_{\ion{N}{V}}$ absorption. Further, our analysis suggests that the location of ABS1 should be characterized by
$-1.7 \lesssim {\rm log}U\lesssim -0.6$. The current data, however, do not allow us to put 
stringent constraints on its precise position due to its loosely constrained $n_{\rm H}$ and $N_{\rm H}$.
Finally, we note that the resulting characteristics of ABS1 are similar to the absorbers usually studied along background sightlines 
piercing the halo of a foreground quasar (e.g., \citealt{QPQ8}). Those absorbers, however, show lower $U$ than ABS1 (${\rm log}U<-1.7$; e.g., Figure~6 in \citealt{QPQ8}).
Indeed, in those cases the quasar pairs are not physically related and the absorbers should not receive much of the radiation from the background quasar.
Observations at higher spectral resolution together with deeper IFU data have the potential to firmly constrain the physical properties of ABS1, and thus its position.

\subsubsection{ABS2 and ABS3: CGM or IGM coherent structures along the quasar pair sight-line}
 
As reported in Section~\ref{sec:abs}, ABS2 (log$(N_{\ion{H}{i}}/{\rm cm^{-2}})\sim14$) and ABS3 (log$(N_{\ion{H}{i}}/{\rm cm^{-2}})=15-17$) appear on both quasars sight-lines, suggesting they trace coherent structures on $\sim100$~kpc (the projected separation between the two quasars). 
At the current depth of the observations, these absorbers are not associated to any continuum
source in the MUSE field-of-view, nor to Ly$\alpha$-emitting galaxies at the absorption redshift.
We evaluate a $5\sigma$ upper limit for the counterpart (if any) in a seeing aperture, finding $L_{\rm Ly\alpha}<3.0\times10^{41}$~erg~s$^{-1}$ ($\sim0.1 L_{\rm Ly\alpha}^{*}$ of \citealt{Ciardullo2012}).
Intriguingly, all these properties are very similar to the absorber detected at $\Delta v=-710$~km~s$^{-1}$ along the line of sight to the quasar pair observed by \citet{Cai2018} with KCWI.

The wider MUSE wavelength range allowed us to detect the presence of strong \ion{C}{iv} 
absorption for ABS3. 
This \ion{C}{iv} detection is only visible along the QSO2 sight-line (log$(N_{\ion{C}{iv}}/{\rm cm^{-2}})>14.7$). 
The presence of this relatively strong high-ionization metal line absorption might indicate 
that this portion of ABS3 is located at a closer distance to QSO2 (or strong ionizing sources, e.g a shock front) 
than the remainder of the structure. 
The absence of absorption at the \ion{N}{v} wavelength might indicate a low metallicity for ABS3. 
The values log$(N_{\ion{H}{i}}/{\rm cm^{-2}})=15-17$ require a relatively large Doppler $b$ parameter ($200-100$~km~s$^{-1}$), which could be due to turbulences in expanding shells around the quasar pair.
Data at higher spectral resolution are needed to explore this occurrence and to firmly constrain the properties of ABS2 and ABS3, 
which are likely CGM or IGM structures coherently extending in front of the quasar pair.

%--------------------------------------------------------------------
\section{Summary and conclusions}
\label{sec:summary}

Recent observations of extended Ly$\alpha$ emission around individual quasars suggest that multiple quasar systems 
are surrounded by more extended and rich structures (\citealt{hennawi+15, FAB2018, FAB2019}).
In an effort to characterize the Ly$\alpha$ emission from CGM and IGM scales, 
we have initiated a ``fast'' survey (45 minutes on source) of $z\sim3$ quasar pairs with MUSE/VLT, 
complementing the work by \citet{Cai2018}.
In this study we focus on the first targeted faint $z\sim3$ quasar pair, SDSS~J113502.03-022110.9 - SDSS~J113502.50-022120.1 ($z=3.020-3.008$; $i=21.84,22.15$), separated by $11.6\arcsec$ (or 89 projected kpc). 

We discovered the presence of filamentary Ly$\alpha$ emission connecting the two quasars 
at an average surface brightness of SB$_{\rm Ly\alpha}=1.8\times10^{-18}\cgssb$.
Using photoionization models constrained with the information on Ly$\alpha$, 
\ion{He}{II}$\lambda$1640, and \ion{C}{iv}$\lambda$1548 line emissions, 
we show that the emitting structures could be explained as intergalactic bridges with an extent between $\sim89$ up to $600$~kpc. 
The faintness of the two quasars and the high levels of Ly$\alpha$ emission seem to rule out 
a $2.9$~Mpc extent for the bridges along our line-of-sight, 
as it could be inferred from the difference between the systemic quasars redshifts.
The intergalactic nature of the emission is also supported by the narrowness of the 
Ly$\alpha$ line ($\sigma_{\rm Ly\alpha} = 162$~km~s$^{-1}$).
At the current spatial resolution and surface brightness limit, 
the projected average width of the bridges is $\sim35$~kpc. 

Additionally, we studied three absorbers found along the two quasar sight-lines.
We detect strong absorption in \ion{H}{i}, \ion{N}{v}, and \ion{C}{iv} along the background quasar
sight-line, which we interpret as due to at least two components of cool ($T\sim10^4$~K), metal enriched ($Z>0.3\, Z_{\odot}$), and relatively ionized circumgalactic or intergalactic gas
characterized by an ionization parameter of $-1.7 \lesssim {\rm log}U \lesssim -0.6$. 
Two additional \ion{H}{i} absorbers are detected along both quasars sight-lines, 
at $\sim -900$~and~$-2800$~km~s$^{-1}$ from the system. 
The \ion{H}{i} absorber at $-2800$~km~s$^{-1}$ has associated \ion{C}{iv} absorption along only the foreground 
quasar sight-line. These two absorbers are not associated to any continuum or Ly$\alpha$ emitters within the MUSE field of view, 
possibly tracing large-scale structures or expanding shells in front of the quasar pair. 

The observations presented in this study confirm that intergalactic bridges can be observed 
even with short exposure times, if peculiar or overdense systems are targeted (e.g. multiple AGN systems).
This is likely due to the presence of dense ($n_{\rm H}\sim0.5$~cm$^{-3}$) gas on large scales coupled with the ionizing radiation
originating from multiple sources. 
Deep high spectral resolution observations of such systems could firmly constrain the physical properties of 
the emitting gas and impinging ionizing continuum,
providing a new leverage to improve current cosmological simulations of structure formation.

\begin{acknowledgements}
We thank Guinevere Kauffmann for providing comments on an early version of this work. 
Based on observations collected at the European Organisation for 
Astronomical Research in the Southern Hemisphere under 
ESO programmes 0100.A-0045(A). 
We acknowledge the role of the ESO staff in providing high-quality service-mode observations, 
which are making this project feasible in a shorter time-scale.
A.O. is funded by the Deutsche Forschungsgemeinschaft (DFG, German Research Foundation) -- MO 2979/1-1.
A.M. is supported by the Dunlap Fellowship through an endowment established by the David Dunlap family and the University of Toronto.
This work made use of matplotlib (\citealt{Hunter2007}).

\end{acknowledgements}

\bibliographystyle{aa} 
\bibliography{allrefs}

\begin{appendix} 
\section{The point spread function of our MUSE data}
\label{app:PSF}

\begin{figure}
\centering
\includegraphics[width=0.8\columnwidth]{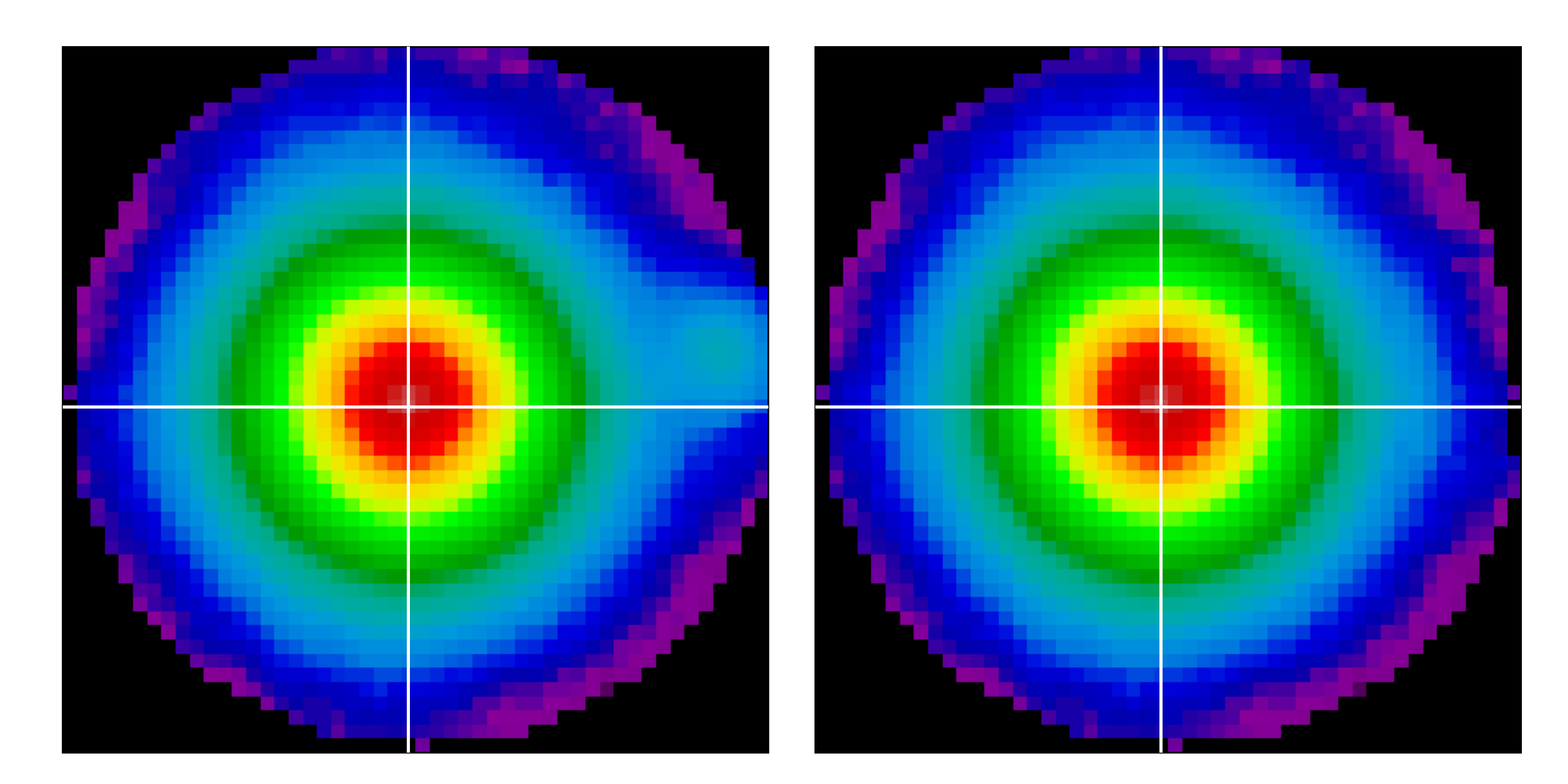}
\caption{Five arcseconds circle cutout of the white-light image of the bright star used as PSF in this work. 
Left: white-light image without post-processing. A faint source on the right part of the star's PSF is clearly evident. 
Right: white-light image after replacing the faint source values with the symmetric portion of the MUSE dataset. 
After this correction, the PSF is well behaved at any wavelength out to five arcseconds 
(Figure~\ref{MoffatPSF} and Appendix~\ref{app:PSF} for details).}
\label{2DimagePSF}
\end{figure}

\begin{figure}
\centering
\includegraphics[width=0.68\columnwidth]{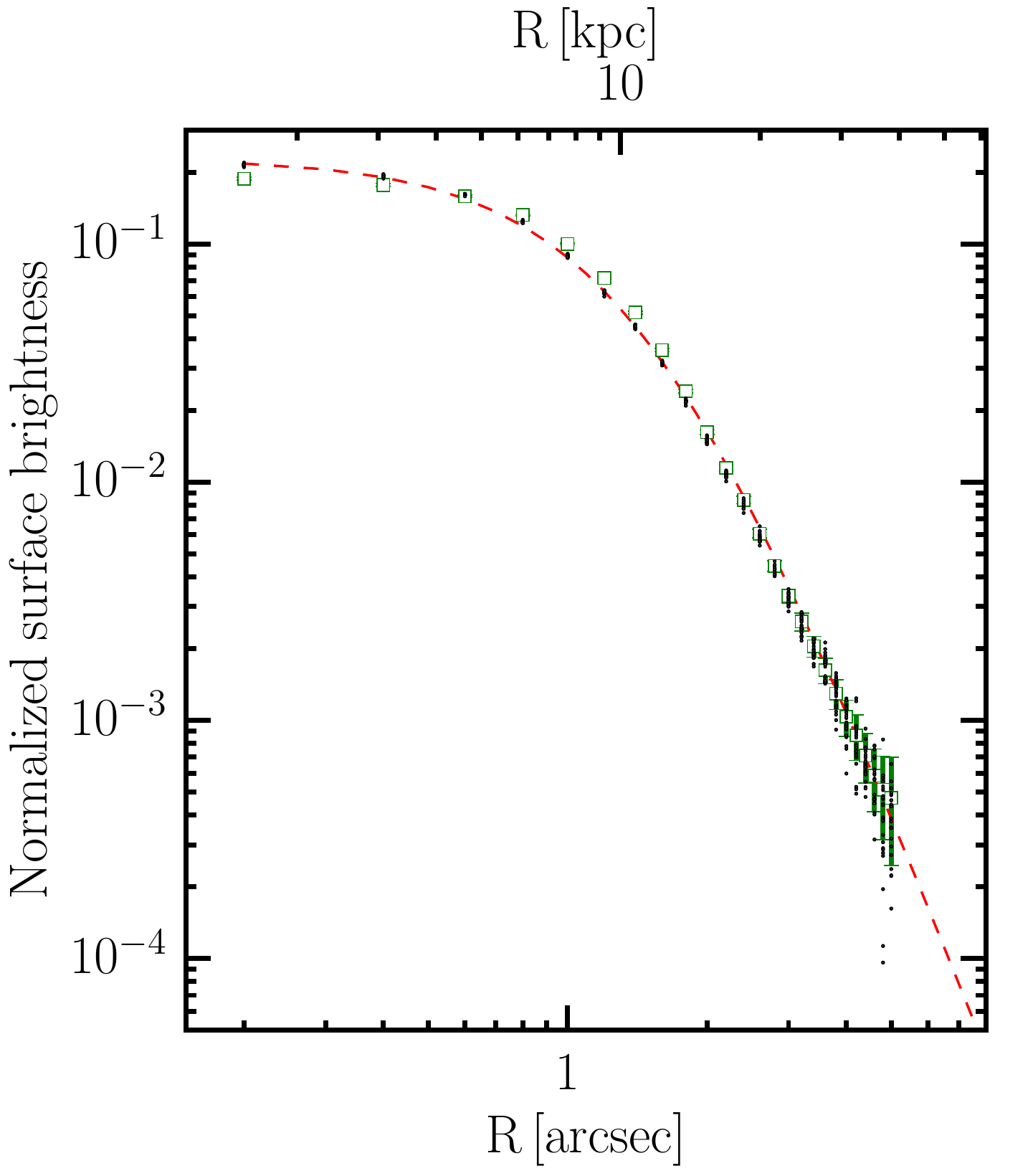}
\caption{Normalized profile of the bright star used as PSF in this work.
The open green squares show the normalized profile derived from the white-light image shown in the right panel of 
Figure~\ref{2DimagePSF}, while the small dots show the normalized profiles for the star within the 17 MUSE layers 
encompassed by the obtained 3D mask for the extended Ly$\alpha$ emission (Section~\ref{sec:PSFsub}). The red dashed line is the best-fit Moffat 
 profile to the white-light image data ($\beta=2.5$ and FWHM$=1.66\arcsec$).
Given the brightness of the used star, there is very good agreement between the profile obtained from the 
white-light image and the individual layers.}
\label{MoffatPSF}
\end{figure}

To model the PSF of our data, which is needed to subtract the unresolved emission from the two quasars (Section~\ref{sec:PSFsub}), 
we rely on the only bright star ($i=16.2$; $r=16.4$) within our observations field of view, 
2MASS J11350307-0220597 (\citealt{Cutri2003}).  
This star has been so far classified as single point source
in all the available catalogues we explored, e.g. the 2MASS All Sky Catalog of point sources (\citealt{Cutri2003}), 
the AllWISE Source Catalog (\citealt{Wright2010}), the 14th Data Release of the Sloan Digital Sky Survey 
(SDSS DR14; \citealt{SDSSXIV}).This star is not saturated in our data.

A faint red source ($r\approx22$) at about 4.4\arcsec\ is present on
the right side of this star, which is clearly visible in the white-light image (Figure~\ref{2DimagePSF}).
Given the red spectrum, this faint source does not contribute significantly at the wavelength of interest for the Ly$\alpha$
emission. We however remove this low-level contaminant by replacing in each layer the values at its
position with the values at the symmetrical position with respect to the star centroid. This is to avoid the introduction of 
any systematic in the PSF subtraction, and in the subsequent extraction of the Ly$\alpha$ signal that we seek. 
The result can be visually inspected in the right panel of
Figure~\ref{2DimagePSF}, while we show the normalized profile of this ``corrected'' star out to five arcseconds in Figure~\ref{MoffatPSF}
(open green squares). 

The star profile is well fitted by a Moffat function with $\beta=2.5$ and FWHM$=1.66\arcsec$. The value for 
$\beta$ is in agreement with the usually assumed value for the MUSE instrument ($\beta=2.8$; e.g., \citealt{Bacon2017}).
For completeness, in the same plot we also show the star profile for each of the 17 layers of the 
3D mask of the extended Ly$\alpha$ emission (small gray dots) built in Section~\ref{sec:PSFsub}. It is clear that the star PSF is defined 
at high S/N out to five arcseconds even in the individual layers, showing a profile consistent with the white-light image.
In our analysis (Section~\ref{sec:PSFsub}) we adopt a normalized version of the star data layer-by-layer
(after removal of the faint source) as empirical PSF.

\section{The superposed galaxy at a lower redshift}
\label{sec:int_gal}

\begin{figure}
\centering
\includegraphics[width=0.63\columnwidth]{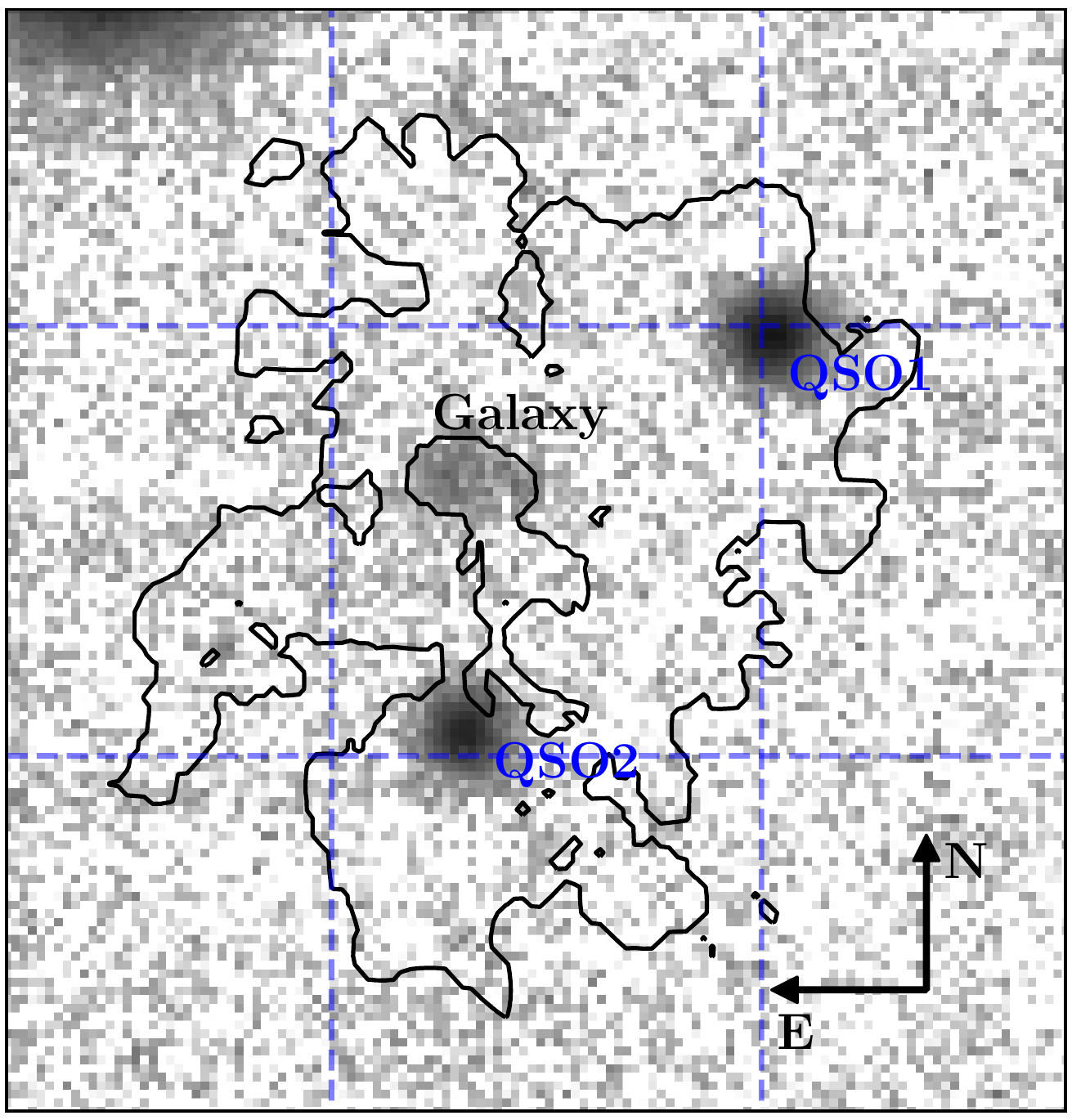}
\caption{SDSS $i$-band image extracted from the final MUSE datacube for the same field-of-view shown in Figure~\ref{fig:SB_Vel_map}.
We indicate the position of QSO1, QSO2 and of a faint galaxy unveiled in projection between the targeted quasar pair.
This galaxy is an interloper at a different redshift, tentatively at $z=0.457\pm0.001$ (Appendix~\ref{sec:int_gal} and Figure~\ref{gal_spec}). For comparison purposes, we also show the $2\sigma$ isophote for the
extended Ly$\alpha$ emission (black). To guide the eye we overlay a grid spaced by $10\arcsec$ (or $77$~kpc), as done in Figure~\ref{fig:SB_Vel_map}.}
\label{gal_fov}
\end{figure}

\begin{figure}
\centering
\includegraphics[width=0.8\columnwidth]{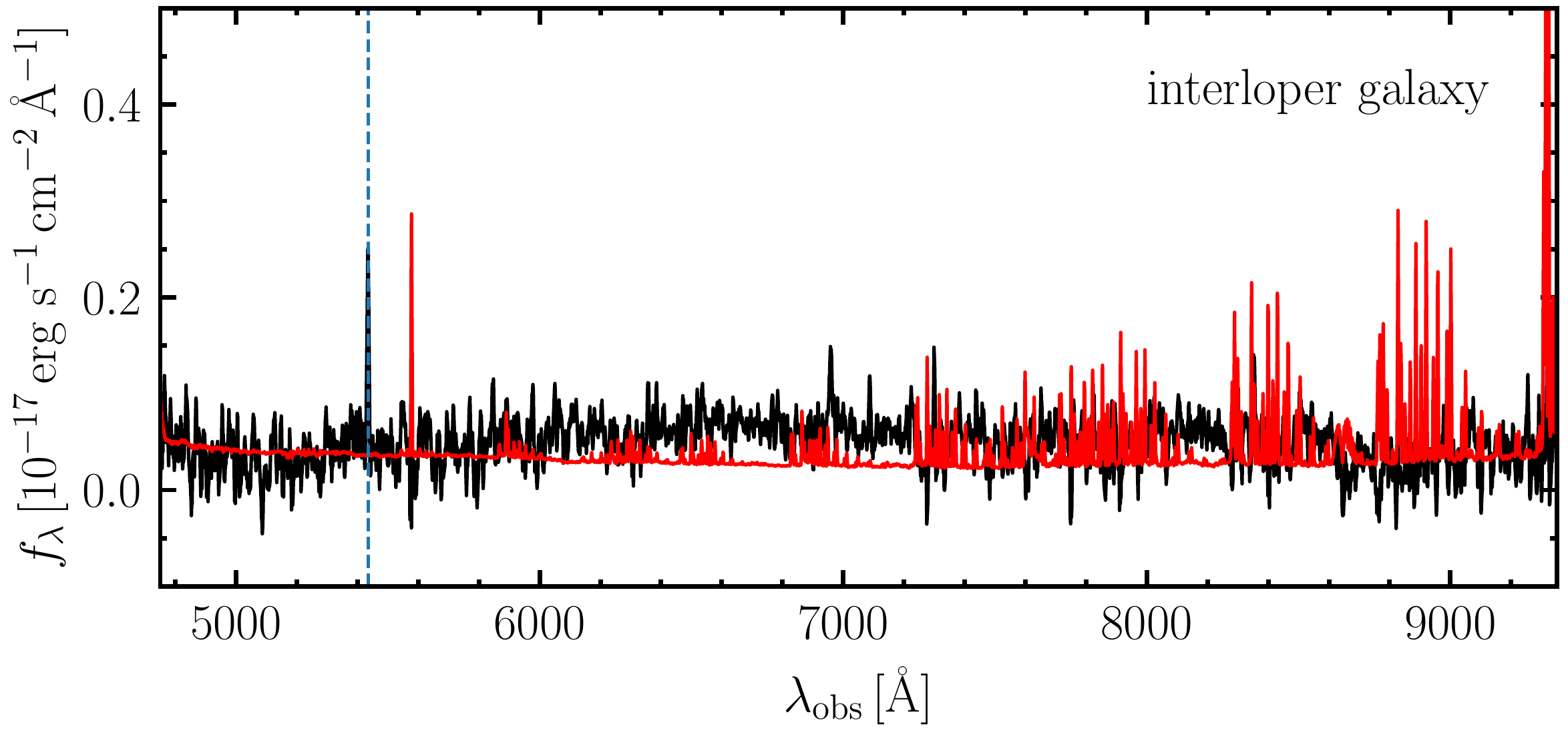}
\caption{1D spectrum (black) of the faint galaxy shown in Figure~\ref{gal_fov}, as extracted using a circular aperture with radius 1\arcsec\ from the MUSE data.
The red spectrum indicates the error vector. The vertical dashed blue line indicates the position of the only detected line emission ($\lambda=5435\AA$).
This galaxy is not associated to the quasar pair as there are no known strong line emissions at a rest-frame 
wavelength of $\sim1800$~\AA. If this line is [\ion{O}{ii}]$\lambda3729$, this galaxy would be at $z=0.457\pm0.001$.
}
\label{gal_spec}
\end{figure}

The MUSE observations unveil the presence of a faint galaxy located in projection between the quasar pair.
In Figure~\ref{gal_fov} we show the $i$-band image extracted from the MUSE datacube using the transmission curve of the corresponding SDSS filter (\citealt{Fukugita1996}). 
The image encompasses the same field of view of Figure~\ref{fig:SB_Vel_map}. 
We indicate the position of the two quasars, of the faint galaxy and of the $2\sigma$ isophote of the Ly$\alpha$ emission. 
The galaxy has a $i$ magnitude of $i=23.82\pm0.03$ when extracted in a circle with radius of 1\arcsec.

We show the spectrum of this faint galaxy in Figure~\ref{gal_spec}.
It is evident a relatively strong line emission at $\lambda=5435$~\AA, $F=(2.1\pm0.1)\times10^{-17}$~erg~s$^{-1}$~cm$^{-2}$. Given the 
absence of any other signature useful to identify the galaxy redshift, we cannot firmly place this galaxy in 
a cosmological context. Its redshift, however, is surely not close to the quasar pair as there are no known strong line emissions at a rest-frame 
wavelength of $\sim1800$~\AA. Further, the galaxy morphology seems resolved even with the large seeing of these observations, possibly hinting at a
low-redshift nature for this object. For reference, we compute its redshift by assuming the line emission to be [\ion{O}{ii}]$\lambda3729$. We find $z=0.457\pm0.001$.
If this galaxy is indeed a foreground object, its dust and gas could absorb the higher redshift Ly$\alpha$ photons of interest to us.
Deeper spectroscopy could unveil the nature of this galaxy and quantify its effect on the extended Ly$\alpha$ emission.

\section{Narrow-band and $\chi$ maps of the Ly$\alpha$ bridge}
\label{app:NBandChi}

\begin{figure}
\centering
\includegraphics[width=0.88\columnwidth]{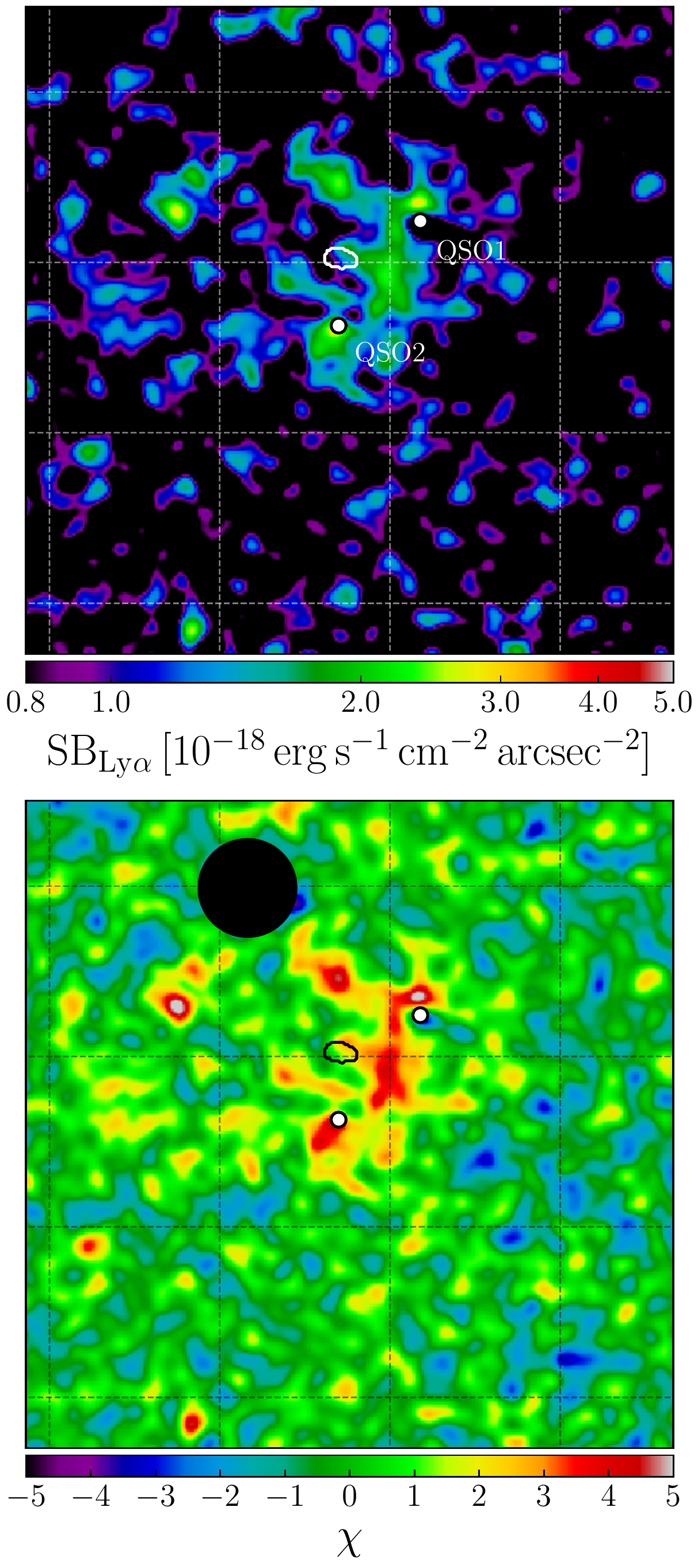}
\caption{Top: Pseudo 6.25\AA\ (5 layers) narrow-band map centered at the central wavelength (4872.7\AA) of the 3D mask of the Ly$\alpha$ emission obtained in Section~\ref{sec:PSFsub}. The map, obtained after PSF and continuum subtraction, shows a $57\arcsec \times 57 \arcsec$ (or 438~kpc~$\times$~438~kpc) FoV and it is color coded following the Ly$\alpha$ surface brightness. Bottom: $\chi_{\rm smth}$ map for the same wavelength range as in the left panel, and obtained using a Gaussian kernel with FWHM = 1.66\arcsec (i.e. similar to the seeing), as explained in Appendix~\ref{app:NBandChi}. 
To guide the eye, in both panels we overlay a grid spaced by 15\arcsec (or 115~kpc) and we indicate the position of QSO1 and QSO2 prior to their PSF subtraction. In both panels, the interloper galaxy ``G'' is indicated with its contour.}
\label{NB_chi_Lya}
\end{figure}

In Section~\ref{sec:em_res} we show the optimally extracted map of the extended Ly$\alpha$ emission connecting the quasar pair.
For completeness and comparison purposes, we present here also a pseudo narrow-band image.
Specifically, we collapsed the five layers (or $6.25$~\AA) of the final MUSE datacube centered at the wavelength of 4872.7~\AA. This wavelength
corresponds to the central layer of the 3D mask obtained in Section~\ref{sec:em_res}.
To avoid introducing too large of a sky noise, the wavelength range of the pseudo narrow-band is chosen to be small and comparable to the width of the Ly$\alpha$
line in the central part of the observed structure. We caution that the chosen width does not encompass the whole velocity range spanned by the aforementioned 3D mask.
The top panel of Figure~\ref{NB_chi_Lya} shows the SB map obtained in this way after a smoothing with a Gaussian kernel with FWHM~$=1.66\arcsec$ (i.e. the seeing of the observations). 
The extended Ly$\alpha$ emission connecting the two bridges is readily visible.

Further, we visualize the noise properties of this map and the significance of the detection by constructing a smoothed $\chi$ image of the same dataset following the recipe in 
\citet{qpq4} and \citet{fab+15a}, for a Gaussian kernel with FWHM~$=1.66\arcsec$. 
The smoothed $\chi$ image is obtained by dividing the smoothed data shown in the left panel of Figure~\ref{NB_chi_Lya}, $I_{\rm smth}$, by the smoothed sigma image
$\sigma_{\rm smth}$ computed by propagating the variance image of the unsmoothed data (details in \citealt{fab+15a}).
The bottom panel of Figure~\ref{NB_chi_Lya} shows this smoothed $\chi$ image after masking a circular region of radius 4\arcsec\ around the bright star 2MASS J11350307-0220597. 
This map reveals that the extended Ly$\alpha$ emission is detected at relatively high significance, and that the noise behaves quite well throughout all the field of view.

\section{Spectrum of box 4 along the pseudoslit}
\label{app:box2}

\begin{figure}
\centering
\includegraphics[width=0.8\columnwidth]{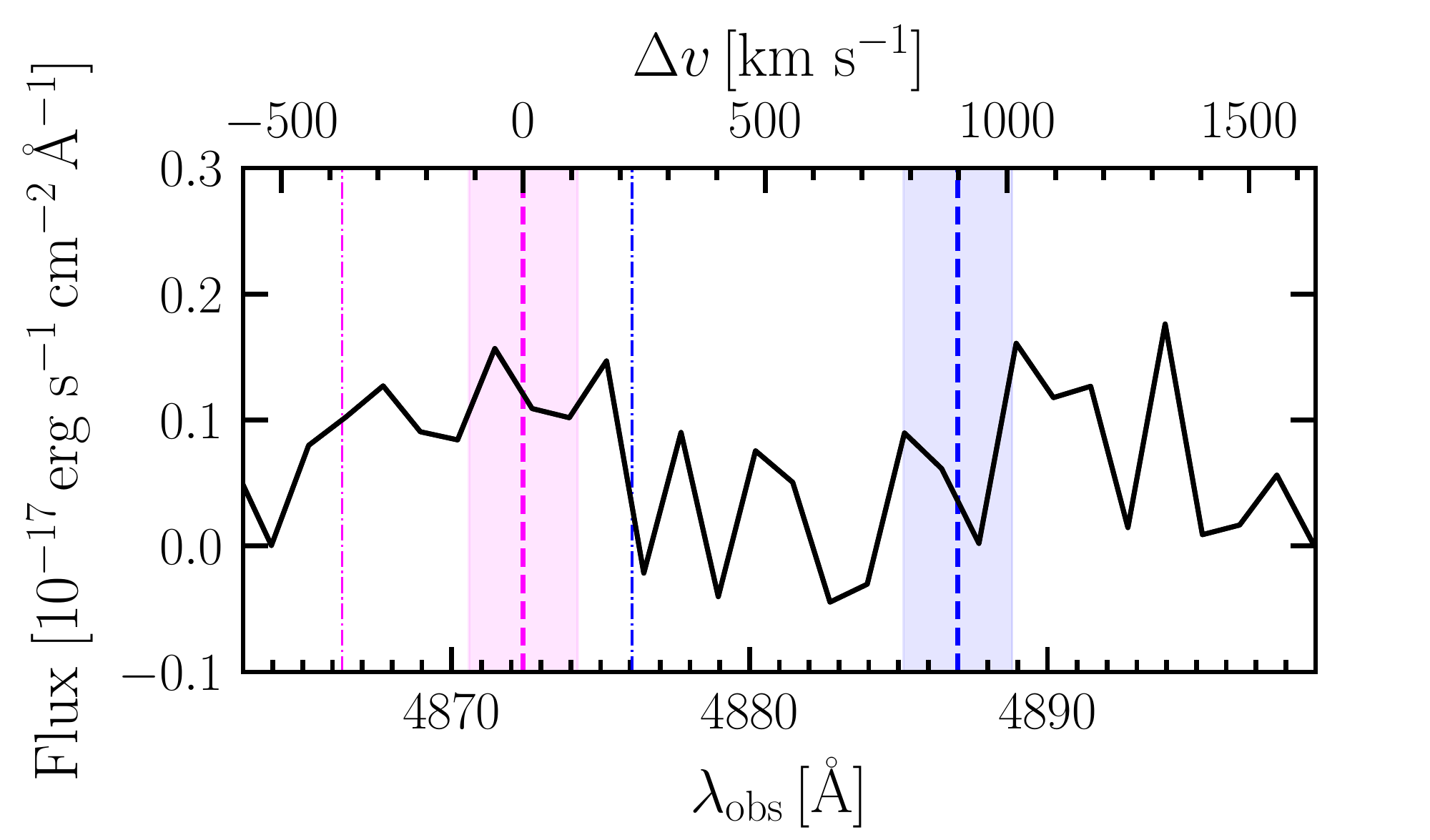}
\caption{Spectrum of box 4 along the pseudoslit used in the analysis of Figure~\ref{fig:Pseudoslit}. The dashed (dotted-dashed) vertical lines show the systemic (peak of the Ly$\alpha$) redshifts for QSO1 (blue) and QSO2 (magenta). The respective shaded regions indicate the error on the redshift as estimated by SDSS.}
\label{Box2_spectrum}
\end{figure}

Here we present the spectrum of box 4 along the pseudoslit used in Section~\ref{sec:em_res} (Figure~\ref{fig:Pseudoslit}).
Figure~\ref{Box2_spectrum} shows this spectrum in physical units. We omitted these data from Figure~\ref{fig:Pseudoslit} as it would have made that normalized plot harder to read. 
The faint level of emission at this location is in agreement with the optimally extracted map presented in Section~\ref{sec:em_res}.

\section{$\chi$ maps at the \ion{C}{iv} and \ion{He}{ii} wavelengths}
\label{app:chiCIVHeII}

In Section~\ref{sec:em_res} we quoted upper limits for the \ion{C}{iv} and \ion{He}{ii} extended line emissions.
Here we show a cut of the final MUSE datacube at their expected observed wavelengths given the flux-weighted center of the Ly$\alpha$ emission,  6208.8~\AA\ and 6573.5~\AA\, respectively.
In particular, we construct smoothed $\chi$ images following the method described in Appendix~\ref{NB_chi_Lya}. These maps have the potential of better visualizing the presence of extended emission.

\begin{figure}
\centering
\includegraphics[width=0.82\columnwidth]{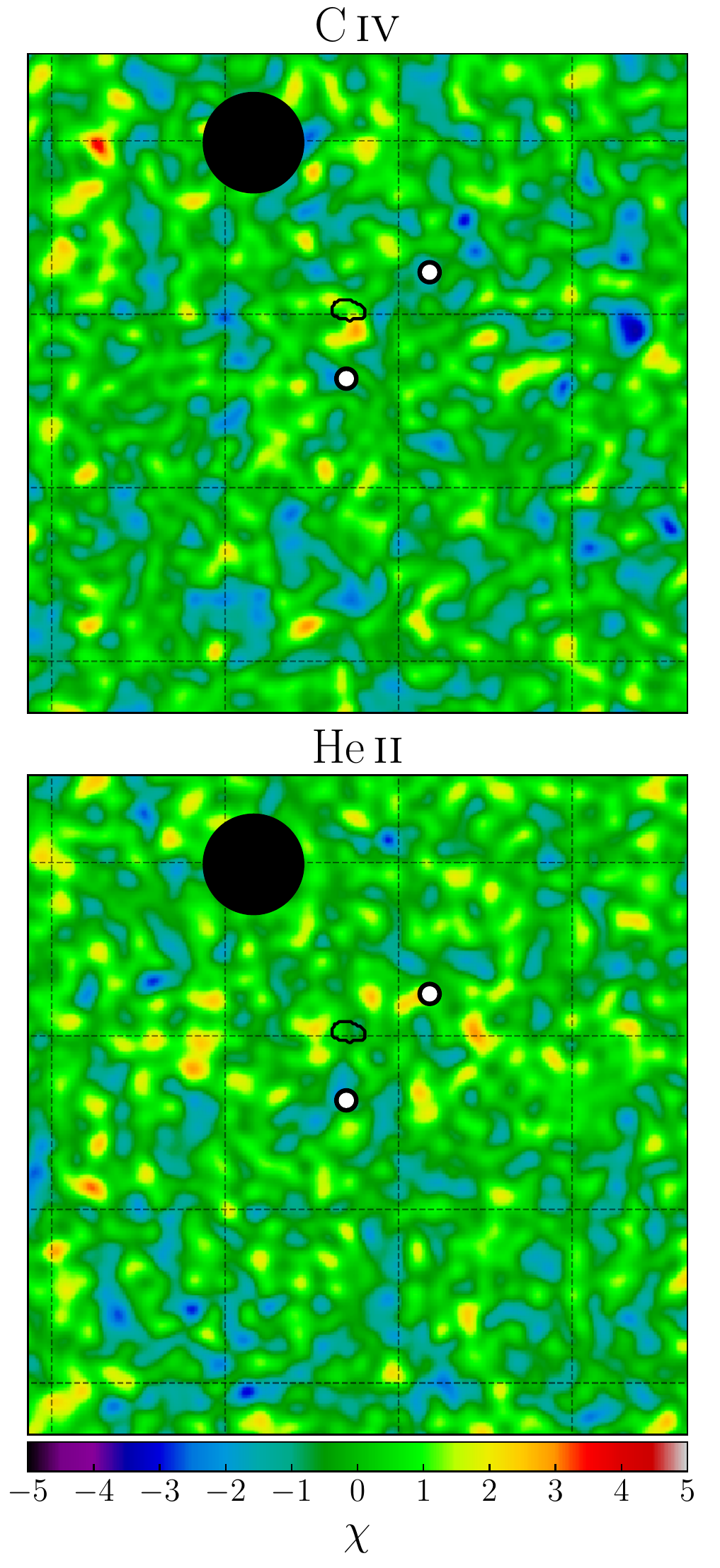}
\caption{Top: $\chi_{\rm smth}$ map of $57\arcsec \times 57 \arcsec$ (or 438~kpc~$\times$~438~kpc) FoV for 6.25\AA\ (5 layers) centered at the wavelength expected for the \ion{C}{iv} line emission (6208.8~\AA) given the center of the 3D mask for the Ly$\alpha$ emission. The map is obtained after PSF and continuum subtraction using a Gaussian kernel with FWHM~$= 1.66$\arcsec (i.e. similar to the seeing) as explained in Section~\ref{app:NBandChi}.   
Bottom: Same as the left panel, but centered at the wavelength expected for the \ion{He}{ii} line emission (6573.5~\AA). 
Both maps, covering the corresponding velocity range of Figure~\ref{NB_chi_Lya}, do not reveal the presence of any clear detection of extended \ion{C}{iv} or \ion{He}{ii} line emission associated with the extended Ly$\alpha$ emission.
In both panels, we overlay a grid spaced by 15\arcsec (or 115~kpc) and we indicate with circles the position of QSO1 and QSO2 prior to their PSF subtraction.  In both panels, the interloper galaxy ``G'' is indicated with its contour.}
\label{chi_CIV_HeII}
\end{figure}

Figure~\ref{chi_CIV_HeII} presents the two smoothed $\chi$ maps obtained using a Gaussian kernel with FWHM~$= 1.66$\arcsec. The white circles indicate the position of the two quasars prior to their PSF subtraction. 
We mask a circular region of radius 4\arcsec\ around the bright star 2MASS J11350307-0220597.
As mentioned in Section~\ref{sec:em_res}, there is no evidence for extended emission at these wavelengths.

\section{Modeling the absorber ABS1}
\label{sec:model_abs_ABS1}

\begin{figure*}
\centering
\includegraphics[width=0.95\textwidth]{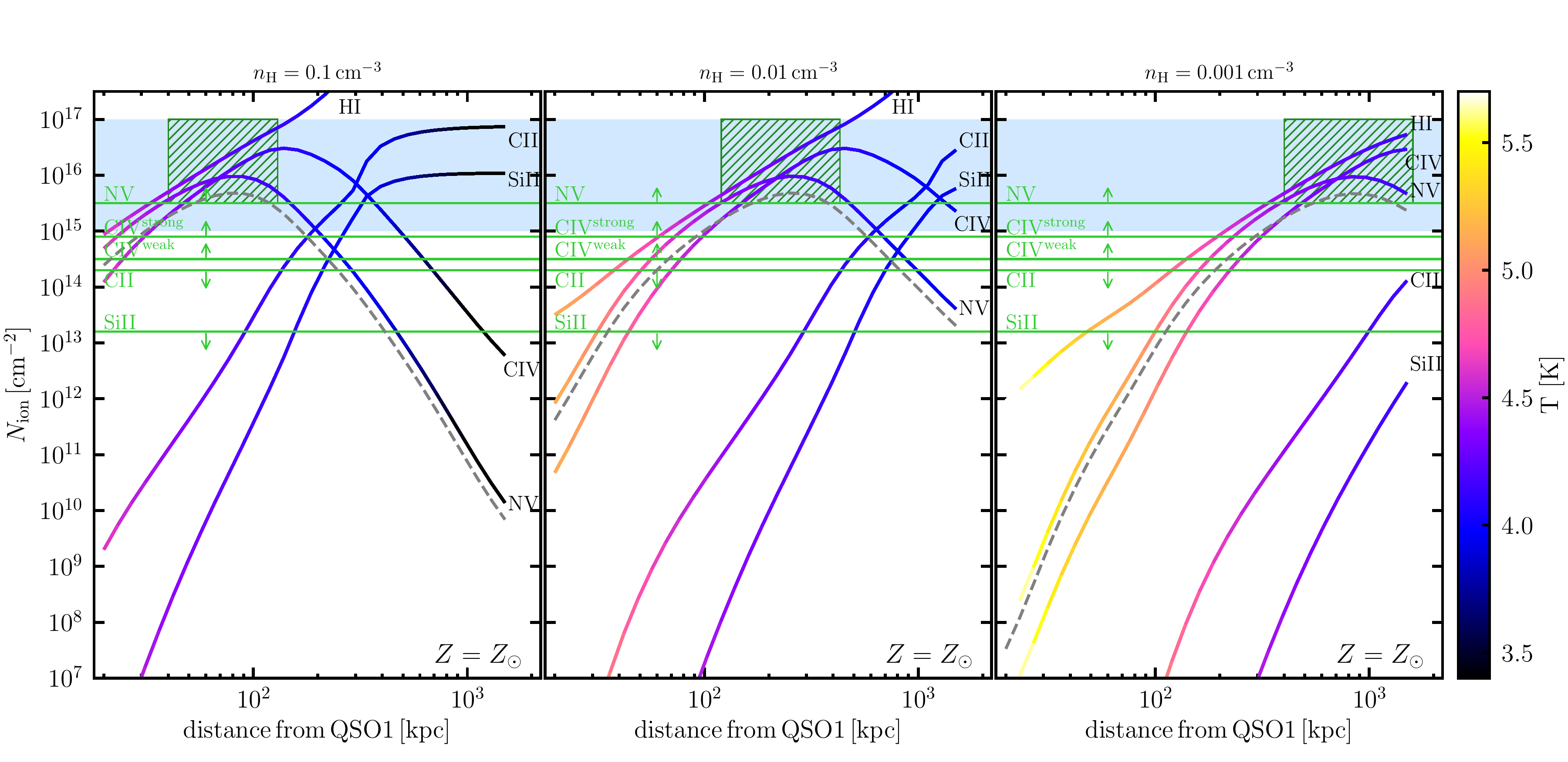}
\caption{Predictions of photoionization models for the absorber ABS1 with $Z=Z_{\odot}$, 
in the case it is illuminated by QSO1 and the UVB (see Appendix~\ref{sec:model_abs_ABS1} for details on the models assumptions).
The predicted column densities for \ion{H}{i}, \ion{C}{iv}, \ion{N}{v}, \ion{C}{ii}, and  \ion{Si}{ii} 
are plotted as a function of distance from QSO1. The horizontal green lines with arrows indicate the observational limits
for the same metal ions, while the blue shaded regions show the observational limits for \ion{H}{i}. 
The green hatched boxes indicate the regions where the models matched the observations. 
The model curves are color-coded following the predicted temperature. The grey dashed lines represent the curves
for \ion{N}{v} for $Z=0.5\,Z_{\odot}$. The models in agreement with the observations are characterized by $4.1\lesssim{\rm log}(T/{\rm K})\lesssim4.4$ and $-1.7\lesssim{\rm log}U\lesssim-0.7$. 
To match the \ion{N}{v} absorption, the metallicity should be $Z>0.3\,Z_{\odot}$.}
\label{fig:photo_abs_QSO1}
\end{figure*}

In Section~\ref{sec:ABS1_cloudy} we summarize the results of our photoionization models concerning ABS1 in three different system configurations.
In this appendix we present in detail the assumptions and the predictions of these calculations.

For simplicity, we assume the following for all the Cloudy models here discussed:
(i) a plane parallel geometry, (ii) three values of fixed volume density $n_{\rm H}=0.1,0.01,0.001$~cm$^{-3}$, 
(iii) three values of fixed metallicity $Z=0.1, 0.5, 1~Z_{\odot}$, 
and (iv) a column density stopping criteria ($N_{\rm H}=10^{20.5}$~cm$^{-2}$).
We do not consider higher values for $n_{\rm H}$ as these would
result in higher SB$_{\rm Ly\alpha}$ than the assumed upper limit for the emission (e.g. Figure~\ref{fig:Cloudy_QSO1}). 
Therefore, all models are already in agreement with the limits on the emission at the absorber position (SB$_{\rm Ly\alpha}<1.75\times10^{-18}\cgssb$; Section~\ref{sec:ABS1_cloudy}). 
We further note that all the models presented in this section include the presence of the UVB at $z\sim3$ (\citealt{hm12}). 
The three system configurations probed are as follows.

\begin{figure*}
\centering
\includegraphics[width=0.95\textwidth]{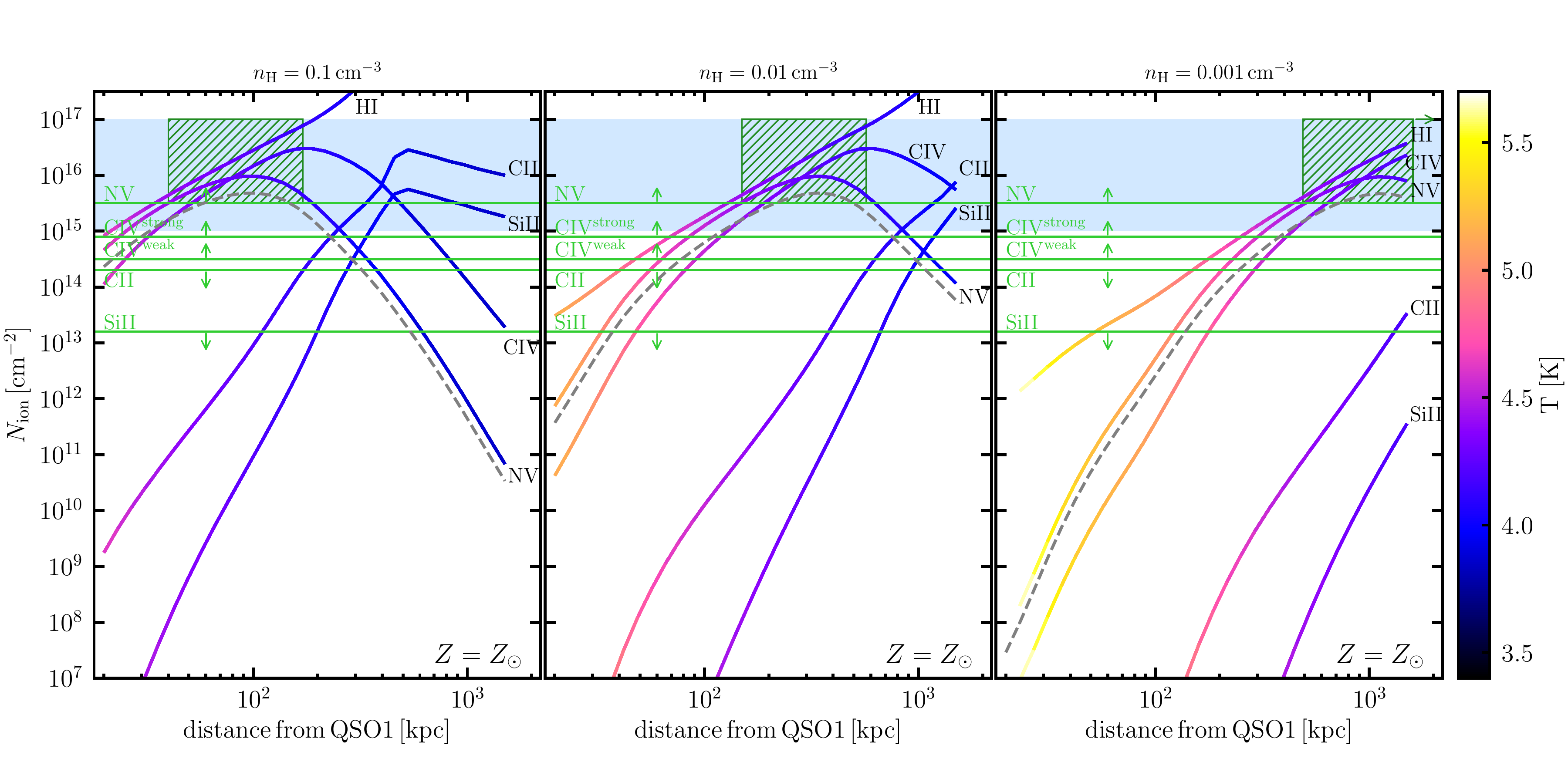}
\caption{Predictions of photoionization models for the absorber ABS1 with $Z=Z_{\odot}$, in the case it is illuminated by 
QSO1, QSO2 (placed at a projected distance of 89~kpc) and the UVB (details on 
the models assumptions in Appendix~\ref{sec:model_abs_ABS1}).
All symbols and colors are explained in the caption of Figure~\ref{fig:photo_abs_QSO1}.
Similarly to Figure~\ref{fig:photo_abs_QSO1}, the models in agreement with the observations are characterized by 
$4.1\lesssim{\rm log}(T/{\rm K})\lesssim4.4$, $-1.7\lesssim{\rm log}U\lesssim-0.7$, 
and $Z>0.3\,Z_{\odot}$.}
\label{fig:photo_abs_QSO1_QSO2_ProjDist}
\end{figure*}

First, as ABS1 is only seen along the QSO1 sight-line, we assume the absorber to be illuminated only by QSO1. 
In this framework, QSO2 is obscured in the direction of ABS1, i.e. ABS1 is not within the ionizing ``cones'' of QSO2.
We thus run Cloudy models assuming as input only the continuum of QSO1 and the UVB, and consider distances 
in the range [20, 1500] kpc.

Figure~\ref{fig:photo_abs_QSO1} shows how the column densities for the different ions change as a function of distance from QSO1 in the grid of models at solar metallicity. 
From left to right, we show the results for $n_{\rm H}=0.1,0.01,0.001$~cm$^{-3}$, respectively. We note that the decrease in $n_{\rm H}$ (and thus increase in $U$) causes the predicted curves 
to be shifted towards larger distances (as photoionization models are self-similar in $U$).
This shift follows equation~\ref{eq_R} 
so that e.g. the curves for 
$n_{\rm H}=0.01$~cm$^{-3}$ are at $\sim3.2$ times larger distances than the ones for 
$n_{\rm H}=0.1$~cm$^{-3}$.  
Allowing for higher $n_{\rm H}$ values ($>0.1\rm cm^{\rm -3}$) would require 
the absorber to be at small distances ($<40$~kpc) from QSO1. 
This seems to be ruled out not only by the Ly$\alpha$ levels implied by higher $n_{\rm H}$, but 
also by the relatively quiescent kinematics of the metal absorptions (Table~\ref{Tab:ABS}).

\begin{figure*}
\centering
\includegraphics[width=0.95\textwidth]{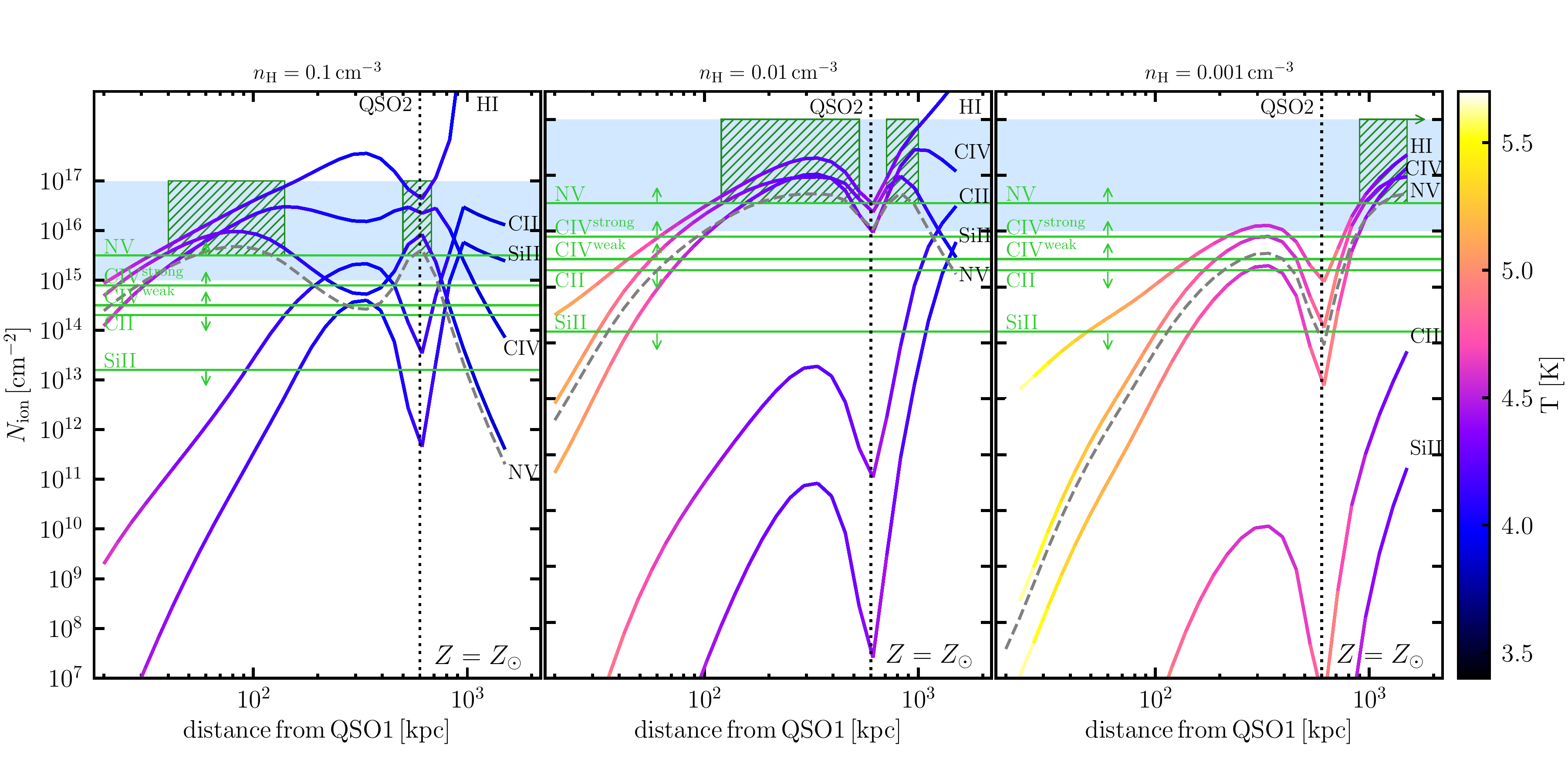}
\caption{Predictions of photoionization models for the absorber ABS1 with $Z=Z_{\odot}$, 
in the case where it is illuminated by QSO1, QSO2 (placed at 600~kpc from QSO1) and the UVB  (details on 
the models assumptions in Appendix~\ref{sec:model_abs_ABS1}). 
All symbols and colors are explained in the caption of Figure~\ref{fig:photo_abs_QSO1}.
Similarly to Figures~\ref{fig:photo_abs_QSO1} and \ref{fig:photo_abs_QSO1_QSO2_600}, 
the models in agreement with the observations are characterized by $4.1\lesssim{\rm log}(T/{\rm K})\lesssim4.4$ and $-1.7\lesssim{\rm log}U\lesssim-0.7$ 
(in Appendix~\ref{sec:model_abs_ABS1} we report the detailed ranges for each plot). The metallicity is constrained to be $Z>0.3\,Z_{\odot}$.}
\label{fig:photo_abs_QSO1_QSO2_600}
\end{figure*}

In each panel, the curves are color-coded by their temperature, and 
we indicate the observed limits on the metal ions column densities as horizontal green lines with arrows, and 
the limits on Hydrogen as blue shaded regions.
The green hatched boxes indicate the regions where the models match the observations. 
It is evident that these hatched regions encompass models with the same temperature in all three panels of Figure~\ref{fig:photo_abs_QSO1}, 
i.e. $4.1\lesssim{\rm log}(T/{\rm K})\lesssim4.4$. These also translate to the same ionizing parameters $-1.7\lesssim{\rm log}U\lesssim-0.7$.
Furthermore, the observed lower limit on $N_{\ion{N}{v}}$ requires the models to be relatively 
metal enriched ($Z>0.3\, Z_{\odot}$). As example, 
we show the predictions for $N_{\ion{N}{v}}$ at $Z=0.5\, Z_{\odot}$ as a dashed grey line in each panel.
These photoionization models thus predict that ABS1 is a cool ($4.1\lesssim{\rm log}(T/{\rm K})\lesssim4.4$) absorber, 
already enriched, and located at a distance 
$40\sqrt{0.1{\rm cm^{-3}}/n_{\rm H}}$~kpc~$\lesssim R \lesssim 130\sqrt{0.1{\rm cm^{-3}}/n_{\rm H}}$~kpc from QSO1,
where the ionization parameter is constrained to be $-1.7\lesssim{\rm log}U\lesssim-0.7$.
It is thus clear that, in this configuration, ABS1 could be located from the CGM of QSO1 out to the IGM 
(even at Mpc distances from QSO1).

As we found good agreement between our models and the observed Ly$\alpha$ emission for a configuration in which the two quasars sit at their projected distance (Section~\ref{sec:proj_dist}),
we assume that the distance along the line-of-sight between QSO1 and QSO2 is negligible and that both illuminate ABS1.
Therefore, in the next step we run Cloudy models assuming as input the continua of both QSO1 and QSO2, scaled accordingly to their distance from the absorber. In particular, we consider distances 
in the range [20,1500] kpc from QSO1, and distances $d_{\rm QSO2}=\sqrt{d_{\rm QSO1}^2+89^2}$~kpc from QSO2.

Figure~\ref{fig:photo_abs_QSO1_QSO2_ProjDist} shows how the column densities for the different ions change as a function of distance 
from QSO1 in the grid of models at solar metallicity. The addition of QSO2 at a small projected distance from QSO1 only slightly 
changes the predictions of the previously considered configuration, with the absorber now positioned at slightly larger distances 
from QSO1.
The location of the hatched boxes is shifted roughly following equation~\ref{eq_R}. 
Indeed, on scales comparable to the distance between the two quasars, equation~\ref{eq_R} is no longer strictly valid.
Specifically, for $n_{\rm H}=0.1$~cm$^{-3}$ ABS1 sits at $40$~kpc $\lesssim R_{\rm QSO1}\lesssim 170$~kpc and $98$~kpc $\lesssim R_{\rm QSO2}\lesssim 190$~kpc. For $n_{\rm H}=0.01$~cm$^{-3}$, we find $150$~kpc $\lesssim R_{\rm QSO1}\lesssim 570$~kpc and $170$~kpc $\lesssim R_{\rm QSO2}\lesssim 580$~kpc, while for $n_{\rm H}=0.001$~cm$^{-3}$, we get $490$~kpc $\lesssim R_{\rm QSO1}\lesssim 1600$~kpc and $500$~kpc $\lesssim R_{\rm QSO2}\lesssim 1610$~kpc. For the lowest-density grid, larger distances 
are also allowed. All the selected ranges where the models agree with the observations correspond to temperatures 
$4.1\lesssim{\rm log}(T/{\rm K})\lesssim4.4$ and ionizing parameters $-1.7\lesssim{\rm log}U \lesssim -0.7$.
We again find that the models require metallicities $Z>0.3\, Z_{\odot}$ in order to match the observed absorptions 
in the metal ions, especially $N_{\ion{N}{v}}$.
For this configuration, ABS1 could be located from the CGM of the system comprising QSO1 and QSO2 out to the IGM.

Last, a configuration in which the two quasars sit at an intermediate distance of 
$\sim600$~kpc, can also explain the observed levels of Ly$\alpha$ emission (Section~\ref{sec:int_dist}). 
Therefore, we also model this case for the illumination of ABS1. 
In particular, we run Cloudy models assuming as input the continua of both QSO1 and QSO2, 
considering distances $20\leqslant d_{\rm QSO1}\leqslant1500$~kpc, and accordingly $R_{\rm QSO2}=\sqrt{(d_{\rm QSO1}-600)^2+89^2}$~kpc.

Figure~\ref{fig:photo_abs_QSO1_QSO2_600} shows how the column densities for the different ions change depending on the distance from QSO1, in the grid of models at solar metallicity. 
In each panel we indicate the locations of QSO2, at 600~kpc 
from QSO1. However, the absorber is never closer than 89~kpc from QSO2, as per the formula above.
From left to right, we show the results for $n_{\rm H}=0.1,0.01,0.001$~cm$^{-3}$, respectively. 
It is clear that the presence of QSO2 at 600~kpc creates a more complex behavior of the curves with respect to the previous two configurations. 
A high ionized region can now be seen around QSO2 as well.
For this reason, there is not a straightforward formula to derive the distance at which the absorber is located depending on its 
density. We can derive $d_{\rm QSO1}$ by looking for the position where the same $U$ is achieved  in all three different density cases.
For $n_{\rm H}=0.1$~cm$^{\rm -3}$, ABS1 would be located at $40$~kpc $\lesssim R_{\rm QSO1}\lesssim 140$~kpc 
and $470$~kpc $\lesssim R_{\rm QSO2}\lesssim 570$~kpc, or at $500$~kpc $\lesssim R_{\rm QSO1}\lesssim 680$~kpc and 
$ 90$~kpc $\lesssim R_{\rm QSO2}\lesssim 140$~kpc. These distances correspond to  $4.1\lesssim {\rm log}(T/{\rm K}) \lesssim4.4$ and 
$-1.7 \lesssim {\rm log}U\lesssim -0.7$, or  $4.1\lesssim {\rm log}(T/{\rm K}) \lesssim4.2$ and $-1.7 \lesssim {\rm log}U\lesssim -1.5$.
For $n_{\rm H}=0.01$~cm$^{\rm -3}$, ABS1 would be located at $120$~kpc $\lesssim R_{\rm QSO1}\lesssim 530$~kpc 
and $114$~kpc $\lesssim R_{\rm QSO2}\lesssim 488$~kpc, or at $710$~kpc $\lesssim R_{\rm QSO1}\lesssim 1000$~kpc and 
$ 145$~kpc $\lesssim R_{\rm QSO2}\lesssim 500$~kpc.  These locations result in  $4.2\lesssim {\rm log}(T/{\rm K}) \lesssim4.4$ 
and $-1.2 \lesssim {\rm log}U\lesssim -0.6$, or  $4.1\lesssim {\rm log}(T/{\rm K}) \lesssim4.3$ and $-1.8 {\rm log}U\lesssim -0.9$.
Finally, for $n_{\rm H}=0.001$~cm$^{\rm -3}$, ABS1 would be located 
at $900$~kpc $\lesssim R_{\rm QSO1}\lesssim 1500$~kpc and $400$~kpc $\lesssim R_{\rm QSO2}\lesssim 900$~kpc. 
Larger distances (not modelled here) are allowed in the lowest density case.
All the aforementioned models give similar ranges for the temperature $4.1\lesssim {\rm log}(T/{\rm K}) \lesssim4.4$, and 
ionization parameter $-1.2 \lesssim {\rm log}U\lesssim -0.6$.
For each panel, the selected distances thus reflect similar $T$ and $U$ as the two configurations 
previously discussed.
It is thus clear that, in this configuration, ABS1 could be located from the CGM of QSO1 or QSO2 out to the IGM.  
As in the two previous cases, the observed lower limit on $N_{\ion{N}{v}}$ requires the models to be relatively enriched, 
with metallicities $Z>0.3\, Z_{\odot}$.

\end{appendix}

\end{document}